\documentclass[journal ]{new-aiaa}
\usepackage[utf8]{inputenc}
\usepackage{textcomp}

\usepackage{graphicx}
\usepackage{amsmath}
\usepackage[version=4]{mhchem}
\usepackage{siunitx}
\usepackage{longtable,tabularx}
\usepackage{multirow}
\usepackage{subcaption} % for subfigures

\title{Sensitivities of Free-Air RANS and DDES Methods on the High-Lift NASA CRM}

\author{Zauner M.\footnote{Aerospace Project Research Associate, JAXA Aviation Technology Directorate, Aircraft Lifecycle Innovation Hub. Corresponding author, zauner.markus@jaxa.jp.}, 
Sansica A.\footnote{Researcher, JAXA Aviation Technology Directorate, Aircraft Lifecycle Innovation Hub, sansica.andrea@jaxa.jp.}, 
Matsuzaki T.\footnote{Senior Researcher, FMIC R\&D INC., Engineering Department, matsuzaki.tomoaki2@jaxa.jp.},
Lusher D. J.\footnote{Researcher, JAXA Aviation Technology Directorate, Aircraft Lifecycle Innovation Hub, lusher.david@jaxa.jp.},
and Hashimoto A.\footnote{Manager, JAXA Aviation Technology Directorate, Aircraft Lifecycle Innovation Hub, hashimoto.atsushi@jaxa.jp.}.}
\affil{Japan Aerospace Exploration Agency, Chofu, Tokyo, 182-8522 Japan}
\affil{FMIC R\&D INC., Musashino, Tokyo, 180-0023 Japan}

\begin{document}

\maketitle

\begin{abstract}
To reduce the time-to-market of future aircraft, it is crucial to predict the flight envelope accurately before building prototypes for flight tests. The High-Lift Prediction Workshop (HLPW) series aims to assess the numerical prediction capability of current CFD technology considering NASA's high-lift version of the Common Research Model (CRM-HL).
The present work contributes to these collaborative efforts, quantifying sensitivities for RANS-based steady, unsteady, and hybrid RANS/LES scale-resolving approaches. Uncertainties associated with the choice of turbulence model, initialization strategies, grid resolution, and iterative convergence at free-air conditions are covered.
Near stall%($C_{L,max}\approx2.5$)
, a large spread of RANS results was observed for different turbulence models %($\Delta C_L=0.27$) 
and initialization strategies%($\Delta C_L=0.24$)
, while iterative convergence appeared less crucial for the present simulations.
Steady and unsteady RANS simulations were unable to predict the correct flow physics near $C_{L,max}$, even for large grids. 
Delayed Detached Eddy Simulations (DDES), however, showed good accuracy compared with wind-tunnel experiments and predicted $C_{L,max}$ with an error of around $5\%$. Compared to steady RANS, the computational cost of DDES was a factor of ten higher. Lessons learned and potential best-practice strategies are shared to aid future studies. While warm-started RANS simulations using SA models are recommended at lower angles of attack, scale-resolving methods %- such as hybrid RANS/LES - 
are required near stall.
\end{abstract}

\section*{Nomenclature}

\noindent(Nomenclature entries should have the units identified)

{\renewcommand\arraystretch{1.0}
\noindent\begin{longtable*}{@{}l @{\quad=\quad} l@{}}
$\alpha$ & Angle of Attack (AoA) \\
$C_p$ & Wall-pressure coefficient \\
$C_f$, $C_{f,x}$ & Skin-friction coefficient magnitude and x-component \\
$C_L$, $C_D$, $C_M$ & Lift-, drag-, and pitching-moment coefficient \\
$\overline{q}$ & Time-averaged quantities \\
$C_{L,max}$ & Lift coefficient near stall\\
$M=U^{*}_{\infty}/a^{*}_{\infty}$ & Mach number \\
$\Delta \tau= \Delta \tau^{*} c^{*}_{mac} / U^{*}_{\infty}$ & time step \\
$St$  & Strouhal number, $f^{*}c^{*}_{mac} / U^{*}_{\infty}$ \\
$CTU$ & Convective time unit, $c^{*}_{mac} / U^{*}_{\infty}$
\end{longtable*}}

\section{Introduction}

\lettrine{T}{he} majority of commercial airplane accidents happen during the low-speed phases of the flight envelope (24\% during final approach, 30\% during landing, and 18\% during take-off and initial climb \citep{Boeing2022}). 
It is therefore crucial to guarantee that sufficient lift force is generated during these low-altitude flight phases, even at off-design conditions with high-lift devices engaged, which increase the geometrical complexity of the aircraft significantly. The lift generation at low speeds is typically limited by the maximum lift coefficient ($C_{L,\textrm{max}}$) and its corresponding critical angle of attack, beyond which stall-related phenomena detrimentally impact the aerodynamic performance and handling of the aircraft. 
At present, around two thirds of total test flights for certification are performed at low speeds. With the aim of reducing the amount of expensive flight testing required, it is essential to demonstrate that numerical tools are reliable enough to predict the aerodynamic performance at low speeds and increased angles of attack. 
% As a measure of accuracy, \citet{Clark2020} suggests deviations of aerodynamic coefficients of $\Delta C_L \le 0.03$ and $\Delta C_D \le 0.001$ near $C_{L,max}$. 
As a measure of accuracy, \citet{Clark2020} suggest that deviations of aerodynamic coefficients between Computational Fluid Dynamics (CFD) and experiment should satisfy $\Delta C_L \le 0.03$ and $\Delta C_D \le 0.001$ near $C_{L,max}$ for acceptable agreement.
%While this is surely an outstanding challenge, the assessment of computational fluid dynamics (CFD) methods is an area of active academic and industrial research. 
While meeting these criteria is certainly a considerable challenge for current CFD solvers, substantial efforts are being made in both academic and industrial settings to assess and improve the accuracy of state-of-the-art CFD methods.
The National Aeronautics and Space Administration (NASA) and Boeing jointly proposed the CFD Vision 2030 \cite{SKADGLM2014,CCDGKKNS2021} and Certification by Analysis (CbA) \cite{MACLMMMSS2021} reports and the AIAA Recommended Practice for supporting aircraft certification through simulations \cite{GKMRSW2021}, where areas of focus to improve state-of-the-art CFD solvers are documented. While CFD tools are well calibrated in very small regions of the flight envelope, their extension to other design and off-design conditions is key for maximizing the effectiveness of numerical simulations during design and certification phases.

One example of an international collaboration bringing together numerous world-leading aerospace research institutions, universities, and commercial companies for accurate CFD-based prediction of the low-speed flight regime is the NASA led AIAA CFD High Lift Prediction Workshop (HLPW) [\url{https://hiliftpw.larc.nasa.gov/}]. Across all workshop editions, widely-used Reynolds-Averaged Navier-Stokes (RANS) methods have been demonstrated to fail near $C_{L,\textrm{max}}$ \cite{RSW2022}. Furthermore, due to experimental unknowns and uncertainties as well as simplifications in the numerical setup (i.e. selection of inflow/outflow and wall boundary conditions), the inclusion of wind-tunnel walls \cite{DHFMK2022} was also shown not to improve the quality of the RANS results. 

Recently, \citet{Kiris2022} performed an extensive and broad parametric study, employing both RANS methods and higher-fidelity approaches, such as wall-modeled Large Eddy Simulations (WM-LES), which were shown $11-12$ times more expensive than conventional RANS methods.
% Despite some WM-LES cases being only 11-12 times more expensive than conventional RANS methods, the characterization of the full flight envelope might still be prohibitively expensive.
Theoretically, hybrid RANS/Large Eddy Simulation (HRLES) allow significantly larger time steps compared to WM-LES and could deliver results of similar accuracy at cheaper costs. 
%In search of cheaper alternatives to WM-LES, the extensive computational study of \citet{Browne2022} assessed high-order hybrid RANS/Large Eddy Simulation (HRLES) approaches.
In the extensive computational study of \citet{Browne2022}, %While 
it was clearly shown that HRLES methods overcome some of the RANS shortcomings. However, the authors also showed that HRLES can be rather sensitive to grid resolution, which often leads to spurious flow separation reminiscent of conventional RANS solutions. 
Grids consisting of approximately 400 million points (denoted as Level C grids at HLPW4) are of great interest for practical application, but were shown to be insufficient for the structured over-set grids used by \cite{Browne2022}. While HRLES performed on meshes containing 571 million points (Level D grids) yielded very promising results, we have to acknowledge that their grid size can exceed those required by WM-LES.
As scale resolving simulations, such as HRLES, are sparse for complex configurations like the CRM-HL, it is important to gain more experience to improve best practises. Eventually, this will help to better understand grid-requirements and numerical settings that lead to reduced costs and uncertainties.   
%While time-step values can be around 100 times larger in the case of HRLES for efficient time-integration, \citet{Kiris2022} estimated the total costs of HRLES using Level D grids to be approximately twice as high as those of WM-LES. 

%In the present contribution, we extend the sensitivity study of \citet{Browne2022} for RANS, URANS, and HRLES methods. 
% In the present contribution, we extend the sensitivity study of \citet{Browne2022} for RANS, URANS, and HRLES methods over the widest range of parameters and sensitivities to date using unstructured meshes, including  respectively four and six versions of Spalart-Allmaras and Menter's turbulence models.
% As well as understanding the effects of grid resolution and initialization strategy, particular effort is focused on determining the sensitivity to different variants of the Spalart-Allmaras (SA) and Shear Stress Transport (SST) turbulence models.
\citet{Mavriplis2022} performed RANS and HRLES for the present test case considering an adaptively refined Cartesian mesh. The latter approach was limited to only two angles of attack.
The extensive work of \citet{DHFMK2022} and \citet{Browne2022} considered structured over-set grids for RANS, URANS, and HRLES methods. In the present contribution, we extend their study using unstructured meshes. We are particularly interested to understand, if we encounter for Delayed Detached-Eddy Simulations (DDES) using unstructured meshes similar limitations as \cite{Browne2022} for Zonal Detached-Eddy Simulations (ZDES) using over-set grids. 
While \cite{Browne2022} considered Spalart-Allmaras (SA) turbulence models only, we also test a Shear Stress Transport (SST) model. 
Since previous work considered time histories below $80$ Convective Time Units (CTUs), we increase the simulation time for selected cases to $>160$ CTUs, which allows us to better characterize low-frequency phenomena.
For RANS and URANS simulations, we assess turbulence models and initialisation strategies over the widest range of parameters and sensitivities to date. Eventually, the present study and learned lessons allow us to define recommended practises for high-lift predictions of aircraft models.

The paper is structured as follows: 
geometry, flow conditions, and experimental datasets are briefly outlined in section~\ref{sec:hlpw4_exp}. 
The numerical solver, methods, and setups are described in section~\ref{sec:num_method}, where an iterative convergence criterion as well as different boundary condition and initialization strategies are reported. Results obtained by RANS/URANS and HRLES simulations are examined in section~\ref{sec:results} and best-practices/lessons learned are discussed in section~\ref{sec:results_bestpractices}. Final conclusions are summarized in section~\ref{sec:conclusions}. 
A comprehensive description of the database for RANS and URANS is given in the Appendix.

\section{Geometry, Flow Conditions and Experimental Dataset \label{sec:hlpw4_exp}}

\begin{figure}[hbt!]
  \centering
  \begin{tabular}{ll}
    a) & b) \\
    \includegraphics[width=0.45\textwidth,trim={1mm 1mm 1mm 1mm},clip]{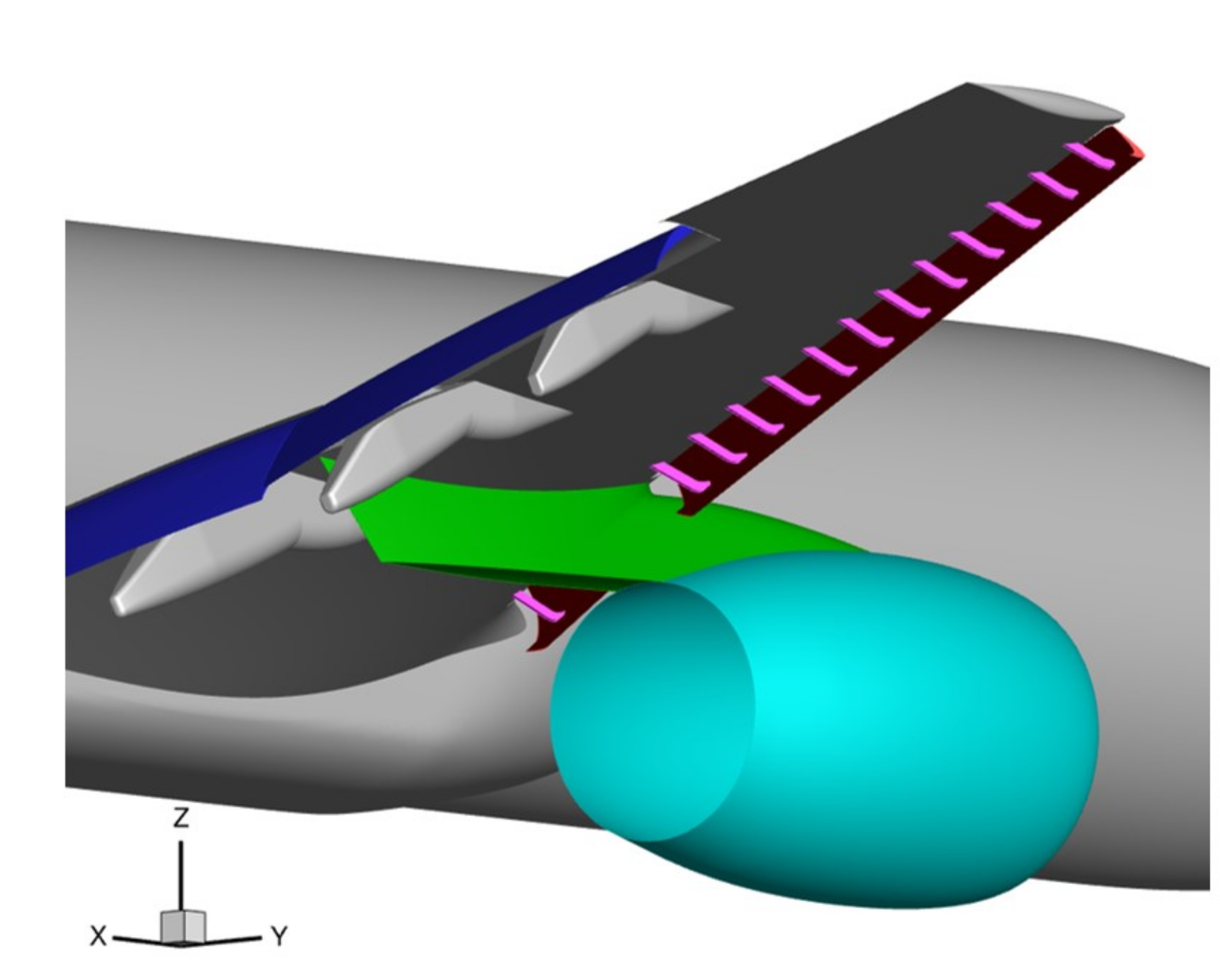} &
    \includegraphics[width=0.45\textwidth,trim={1mm 1mm 1mm 1mm},clip]{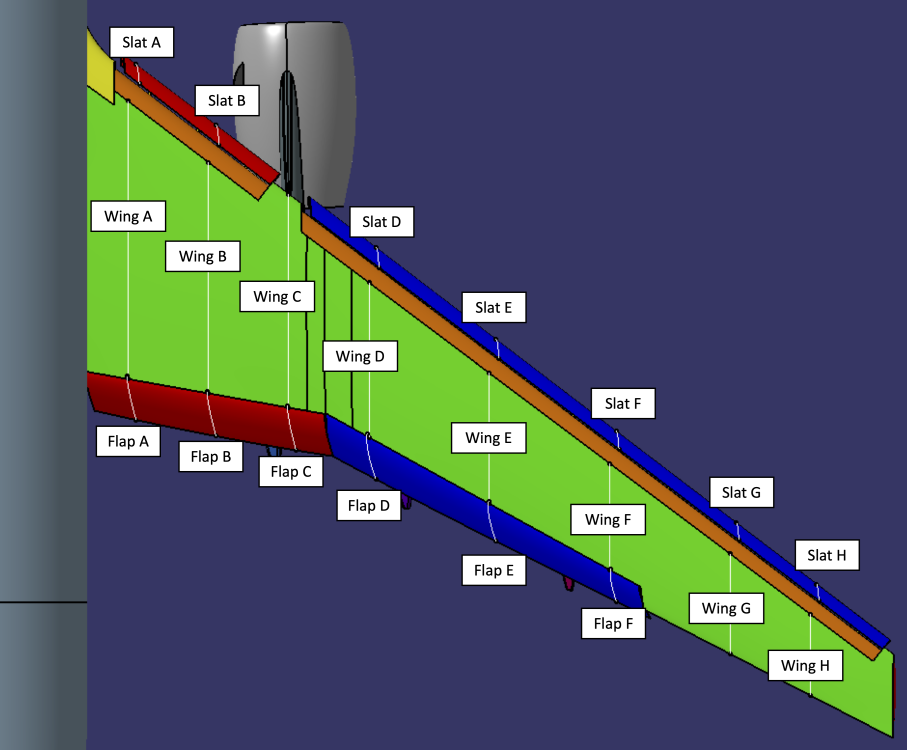} \\
  \end{tabular}
  \caption{
  %(a) Side-view of the High-Lift configuration of NASA's Common Research Model (CRM-HL). The deflected slats and flaps are coloured magenta and blue, respectively. Pylon and nacelle are respectively shown in green and cyan. (b) Top-down view with indication of cross-sections equipped with pressure tabs taken from the HLPW-4 web-page.
  (a) Side-view of the CRM-HL configuration. (b) Top-down view with indication of cross-sections equipped with pressure tabs taken from the HLPW-4 webpage.
  }
  \label{fig:Intro}
\end{figure}

The geometry and flow conditions of interest are based on the $4^{th}$ AIAA CFD High-Lift Prediction Workshop (\url{https://hiliftpw.larc.nasa.gov/index-workshop4.html}). The aim of this workshop was to assess the capabilities of state-of-the-art CFD solvers for medium-to-high-aspect ratio wings on full-aircraft configurations at landing/take-off (high-lift) conditions. The High-Lift configuration of the NASA Common Research Model (CRM-HL) is analysed at representative corrected (accounting for wind-tunnel confinement effects) angles of attack of $\alpha=7.05^{\circ}$, $17.05^{\circ}$, $19.57^{\circ}$, and $21.47^{\circ}$. Reference Mach and Reynolds numbers (based on mean aerodynamic chord and free-stream quantities) are $M_{\infty}=0.2$ and $Re=5.49$ million, respectively. The main components of the CRM-HL model considered here are shown in Fig~\ref{fig:Intro}(a) and consist of: a semi-span wing with extracted flaps and slats, flap track fairings and slat brackets, and a nacelle/pylon equipped with a chine. Only nominal $40^\circ/37^\circ$ inboard/outboard trailing-edge flap and $30^\circ$ slats deflections are considered. A large experimental database is provided for comparisons by \citet{ELSR2020}, which includes polars of aerodynamic forces, oil flow visualizations, and sectional pressure distributions at several spanwise wing cross-sections denoted by letters $A$-$H$ in Fig~\ref{fig:Intro}(b).

\section{Numerical Methodology \label{sec:num_method}}

\subsection{Simulation Setup \label{sec:sim_setup}}

For all simulations presented in this work, JAXA's unstructured flow solver FaSTAR \cite{Hetal2012,IHOAT2017} is employed. The solver applies a node-centered finite volume approach, with the Harten-Lax-van Leer-Einfeldt-Wada (HLLEW) scheme\cite{O1995} being used to accurately compute advective terms. Additionally, a Green–Gauss/Weighted-Least-Squares (GLSQ) method\citep{GLSQ} is used for gradient computations. For stabilisation, a Hishida (van Leer-type) slope limiter and U-MUSCL scheme \citep{Burg2005} are applied. Steady solutions are obtained by using the Lower/Upper Symmetric Gauss-Seidel (LU-SGS) \cite{Setail1998} time-stepping scheme with a Courant–Friedrichs–Lewy (CFL)-fixed local time step. All calculations were performed with a CFL number of 10. For the unsteady calculations, dual-time stepping \cite{Vetal2000} is used to improve the accuracy of the implicit time integration method. The LU-SGS scheme is used for the pseudo-time sub-iterations, with the physical time derivatives being approximated by a three-point backward difference. 
For hybrid RANS/LES, a non-dimensional time step of $\Delta \tau= \Delta \tau^{*} c^{*}_{mac} / U^{*}_{\infty} = 0.00036$ CTUs has been used for all simulations, where $U^{*}_{\infty}$ and $c^{*}_{mac}$ denote the free-stream velocity and aerodynamic mean chord, respectively. For URANS computations, the time-step was set to $\Delta \tau= \Delta \tau^{*} c^{*}_{mac} / U^{*}_{\infty} = 0.0002$ CTUs to ensure numerical robustness on the given RANS grid. Details will be follow later.
FaSTAR has been validated over numerous years and on many different flow configurations, and has shown excellent agreement with other comparable industrial solvers \cite{MHIUTI2013,jaxadpw62018,SMIHF2023,SH2023}. Both RANS and hybrid RANS/LES approaches are used to assess the solver's capabilities for different levels of fidelity. The main modeling features of both approaches are described as follows:

\begin{itemize}
\item RANS and Unsteady RANS:
\item[] Several turbulence models are available in FaSTAR to close the averaged Reynolds stresses. In addition to the classic Spalart-Allmaras turbulence model with zero $f_{t2}$ tripping term (\textit{SA-noft2}) \cite{SA1992}, variants considering rotation (\textit{SA-noft2-R}) \cite{sar1995} and rotation-curvature (\textit{SA-noft2-RC}) \cite{SSTS2000} corrections have also been assessed in this work. The effects of the quadratic constitutive relation 2000 version applied to the SA-noft2-R model (\textit{SA-noft2-R-QCR2000}) \cite{S2000} have also been investigated. 
The two-equation $k-\omega$ Shear Stress Transport turbulence model in its standard form (\textit{SST}), as well as the 2003 updated version (\textit{SST-2003}) \cite{MKL2003} with controlled decay (\textit{SST-2003sust}) \cite{RS2007} or with additional vorticity source terms (\textit{SST-V}) have also been considered.

\item Hybrid RANS/LES:
\item[] In the context of this work, the RANS/URANS framework described above is referred to as a "low-fidelity" method, in contrast to the higher fidelity offered by the hybrid RANS/LES approach. In the present study, the SA-noft2-R based Delayed Detached Eddy Simulation (DDES)\citep{Spalart2006} hybrid method is used. The turbulence length scales are estimated by using the wall distance in the RANS branch and the local grid scale in the LES branch. The transition from the RANS region to the LES region is determined based on the local grid scale, wall distance, and the local state of the degree of turbulence through the eddy viscosity and the deformation tensor.

\end{itemize}

\subsection{Numerical Domains and Grids \label{sec:sim_setup_grid}}

Different sets of numerical grids have been used to target low (RANS/URANS) and high (DDES) fidelity calculations. Categorized by the fidelity of the method, the numerical grids are available on the HLPW-4 website (\url{https://hiliftpw.larc.nasa.gov/Workshop4/grids_downloads.html}) and are further described below:

\begin{itemize}
\item RANS/URANS grids: the JAXA committee-provided C-level grid (240-JAXA-unstructured/240-Config\_CRM-HL\_40-37\_Nominal/240.C) consists of $86$ million nodes and $208$ million cells. It is selected as our `Standard RANS grid'. To perform a refinement study, the standard RANS grid was uniformly refined by reducing the grid spacing by $\approx 33\%$ in each direction, resulting in a $D$-level grid comprised of $182$ million nodes.
Both unstructured hybrid grids are composed of hexahedra, prisms, pyramids, and tetrahedra and were generated using an unstructured mesh generator named Mixed-Element Grid Generator in 3 Dimensions (MEGG3D)\cite{Ito2002, Ito2013, Ito2013b}.
\item Hybrid RANS/LES: The C-level ANSA grid family for free-air simulations (103-ANSA-Unstructured-hiA-Yplus1/103-Config\_CRM-HL\_40-37\_Nominal/103.C) has been selected. The unstructured grid uses a quad dominant surface mesh with hexa layers and hexa-dominant in the farfield. Constructed for hybrid RANS/LES calculations, the grid has a $\Delta y^{+}$ at the wall of unity and wake refinement up to angles of attack around $20^\circ$. In total, the mesh contains $218$ million nodes and $276$ million cells.
\end{itemize}

\subsection{Boundary Conditions and Initialization Strategies \label{sec:results_freeair_setup}}

No-slip velocity and adiabatic temperature conditions are enforced on the aircraft walls. Far-field boundary conditions are employed at the lateral boundaries, and the angle of attack $\alpha$ is applied at the inflow of the numerical domain. Only half of the aircraft is simulated to reduce computational cost, with a symmetry condition applied on the longitudinal center plane. Regarding the initialization of the simulations, we can distinguish between:
\begin{enumerate}
\item \textbf{Cold-start:} Initializing the flow field with uniform free-stream conditions, performing linear blending to enforce no-slip conditions at the wall. Initializing the simulation in this manner may require for some cases reduced CFL numbers or time steps for the first $\approx10,\!000$ iterations, before resuming to the target set up described in \S \ref{sec:sim_setup}.
\item \textbf{Warm-start:} In this approach, the simulation is restarted from a previous RANS solution at a lower angle of attack.
\end{enumerate}
The increments for the angles of attack to be considered are selected according to the experimental data provided for HLPW4, which are summarised in table \ref{tab:angles of attack}. 
\begin{table}[h]
\centering
%\begin{tabular}{cc|c|c|c|c|c|c|c|c}
\begin{tabular}{cccccccccc}
$\alpha$-step & & 1 & 2 & 3 & 4 & 5 & 6 & 7 & 8 \\
\hline
angle of attack & $\alpha$ & $0.0^{\circ}$ & $2.78^{\circ}$ & $7.05^{\circ}$ & $11.29^{\circ}$ & $17.05^{\circ}$ & $19.57^{\circ}$ & $20.55^{\circ}$ & $21.47^{\circ}$ \\
\hline
experiment \cite{ELSR2020} & $C_L$ & 1.16 & 1.36 & 1.78 & 2.15 & 2.50 & 2.51 & 2.46 & 2.32\\
							& $C_D$ & 0.12 & 0.13 & 0.19 & 0.25 & 0.33 & 0.36 & 0.37 & 0.44\\
							& $C_M$ & -0.40 & -0.39 & -0.37 & -0.34 & -0.29 & -0.24 & -0.23 & -0.30 \\
\end{tabular}
\caption{Angles of attack considered for advancing warm-started RANS simulations.}
\label{tab:angles of attack}
\end{table}
It has been shown in previous studies (e.g. \cite{Lacy2016}) that warm-started simulations tend to deliver increased lift, as flow separation can be delayed by restarting the simulation from solutions containing attached flow. 
Starting from a low angle of attack of $7.05^{\circ}$, it has been shown to be sufficient to run simulations for only $10,\!000$ iterations to obtain an initial solution for a higher angle of attack (see figure 3 in \cite{Zauner2023}). Interestingly, \citet{Zauner2023} showed that further increasing the number of iterations before raising the angle of attack can again lead to pre-mature flow separation and decreased lift, highlighting the sensitivity of RANS-based solvers and the need for assessment and publication of best practices in this field.

% In the present contribution, initializing the flow field with uniform free-stream conditions is denoted as `cold-start'. Linear blending is performed at the wall to enforce no-slip conditions. 
% Initializing simulation may require for some cases reduced CFL numbers or time steps for the first $\approx10,\!000$ iterations, before resuming to the target set up described in \S \ref{sec:sim_setup}.
% Cold-starting from a low-angle of attack, solutions after $10,\!000$ iterations are considered to `warm-start' simulations at increased angles of attack. 

\subsection{Iterative Convergence of the RANS Solutions \label{sec:sim_setup_convparam}}

Residuals of the conservative variables are often used as a measure of the iterative convergence for RANS solutions. The maximum of the $L2$-norm in the entire domain for density, momentum, total energy, and turbulence-model specific conservative variables can be tracked individually at each iteration and used to determine when a solution has sufficiently converged and the simulation is completed. Ideally, we consider RANS solutions to be iteratively converged when all residuals reach machine-precision zeros. In practice, however, it can be extremely difficult to reach sufficiently low residuals when complex geometries such as the CRM-HL are considered. Although increased grid resolution and quality of the CAD models, longer simulation run-time, and improved convergence methods can help in lowering the residuals, we may still not reach sufficient convergence due to modelling limitations and errors (e.g. flow unsteadiness). For industrial applications, however, machine-zero precision convergence may not be needed as long as residuals are reasonably low in key-regions of interest and aerodynamic coefficients exhibit minimal variation. Alternative more relaxed criteria can therefore be decided a priori to determine a satisfactory level of iterative convergence for the solution.

For the present CRM-HL test case, we are mainly interested in the aerodynamic performance of the entire aircraft as a whole.
The present measure for convergence is inspired by NASA's approach of using the Root Mean Squared ($rms$) values of $C_D$ fluctuations\citep{DHFMK2022}, but is extended to also include both $C_L$ and $C_M$ normalised by their mean values. It is defined as
\begin{equation}\label{convp}
CPA = \sqrt{ \left( \frac{C_{L,rms}}{\overline{C_L}} \right) ^2 + \left( \frac{C_{D,rms}}{\overline{C_D}} \right) ^2 + \left( \frac{C_{M,rms}}{\overline{C_M}} \right) ^2},
\end{equation}
where mean values of the considered aerodynamic coefficients are assumed to be non-zero. This coefficient is calculated over the final $10,000$ iterations of each simulation and is used as a measure of the iterative convergence of the RANS solution. For completeness, we note that we typically use the logarithm of this parameter, denoted as $\log CPA$. FaSTAR typically provides histories of aerodynamic coefficients up to a precision of 5 significant digits, which is sufficient for present simulations targeting $CPA<10^{-4}$.

\section{Results \label{sec:results}}

% We will now discuss steady and unsteady RANS simulation as well as hybrid RANS/LES results.
\subsection{Sensitivities of Steady RANS Methods \label{sec:results_freeair_rans}}
In this section RANS simulations are performed to first assess sensitivities to turbulence modelling, before studying the influence of different initialisation strategies. Comments are also made on convergence properties for different test cases, and grid sensitivities are reported at the end of this sub-section.
%%%%%%%%%%%%%%%%%%%%%%%%%%%%%%%%%%%%%%%%%%%%%%%%%%%%%%%%%%%%%%%%%%%%%%%%%%%%%%%%%%%%
\subsubsection{Sensitivity to the Choice of Turbulence Model}
% Turbulence model
\begin{figure}
    \centering
    \begin{subfigure}[b]{0.33\textwidth}
        a) \\
        \includegraphics[width=\textwidth, trim=10 10 50 50, clip]{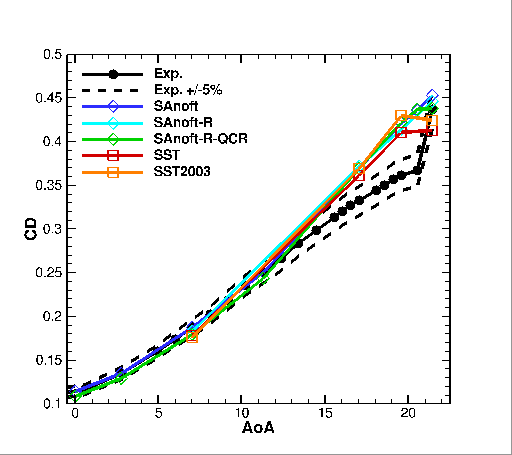}
        %\caption{Subplot 1}
    \end{subfigure}
    \begin{subfigure}[b]{0.33\textwidth}
        b) \\
        \includegraphics[width=\textwidth, trim=10 10 50 50, clip]{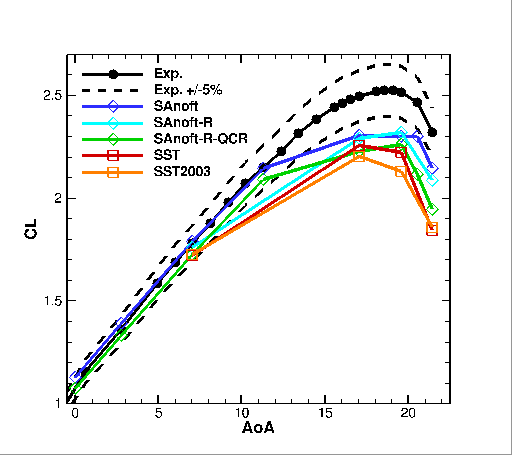}
        %\caption{Subplot 2}
    \end{subfigure}
    \begin{subfigure}[b]{0.33\textwidth}
        c) \\
        \includegraphics[width=\textwidth, trim=10 10 50 50, clip]{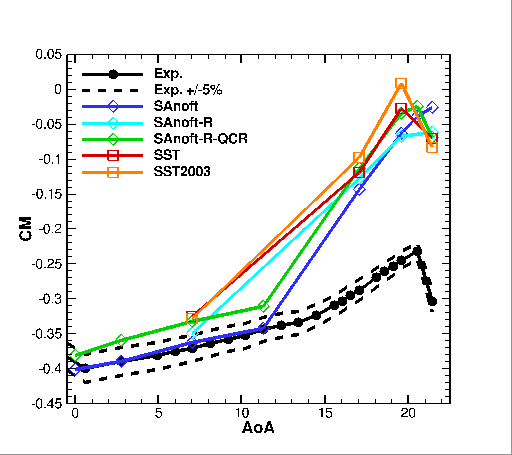}
        %\caption{Subplot 3}
    \end{subfigure}
    \caption{(a) $C_D$, (b) $C_L$, and (c) $C_M$ over $\alpha$ for selected RANS turbulence models. Black symbols and curves respectively denote experimental measurements and $\pm 5\%$ errors.}\label{fig:FA_RANS_Coeff}
\end{figure}
Fig~\ref{fig:FA_RANS_Coeff} shows (a) drag, (b) lift, and (c) moment coefficients for selected turbulence models as a function of the angle of attack. Black symbols denote experimental measurements and black dashed lines correspond to $\pm 5\%$ error margins. For $\alpha \le 7.05^{\circ}$, all considered turbulence models deliver consistent results, where lift and drag stay within $5\%$ error compared to experiments. Most turbulence models overestimate the moment coefficient already at reduced angles of attack. Further increasing the angle of attack leads to all turbulence models significantly over-predicting moment and drag coefficients, while lift is consistently under-predicted.
\begin{figure}
    \centering
    \begin{subfigure}[b]{0.48\textwidth}
        a) \\
        \includegraphics[width=\textwidth, trim=10 10 50 250, clip]{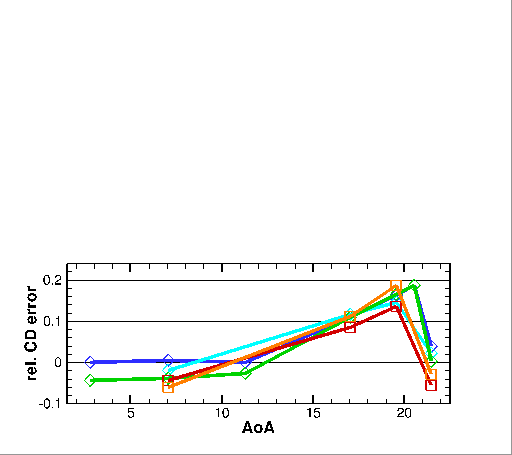}
        %\caption{Subplot 1}
    \end{subfigure}
    \begin{subfigure}[b]{0.48\textwidth}
        b) \\
        \includegraphics[width=\textwidth, trim=10 10 50 250, clip]{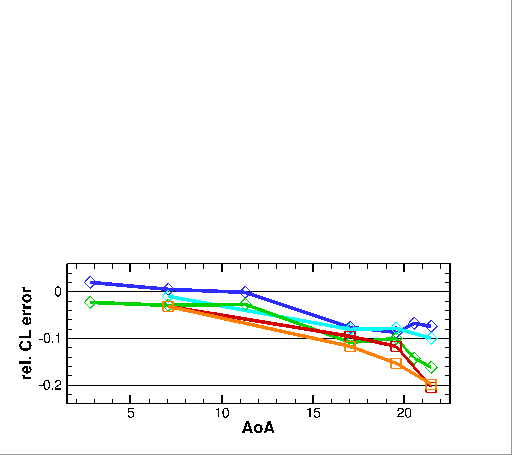}
        %\caption{Subplot 2}
    \end{subfigure}
    \vspace{0.5cm} % adjust vertical spacing between rows of subplots
    \begin{subfigure}[b]{0.48\textwidth}
        c) \\
        \includegraphics[width=\textwidth, trim=10 10 50 250, clip]{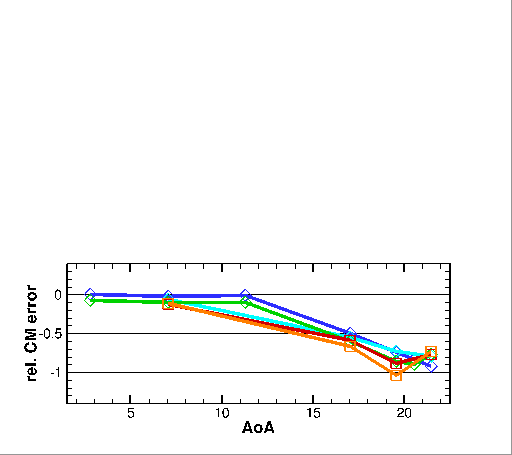}
        %\caption{Subplot 3}
    \end{subfigure}
    \begin{subfigure}[b]{0.48\textwidth}
        d) \\
        \includegraphics[width=\textwidth, trim=10 10 50 250, clip]{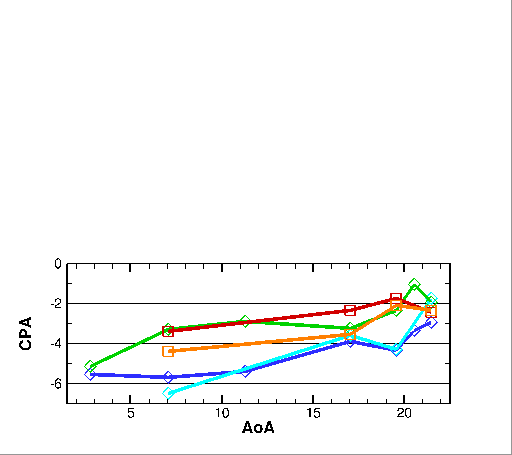}
        %\caption{Subplot 4}
    \end{subfigure}
    \caption{
    %Relative errors comparing RANS results with respect to experimental data \citep{ELSR2020}. (a) $C_D$, (b) $C_L$, and (c) $C_m$ as well as (d) convergence parameters (CPA) are shown as a function of angle of attack, where colours correspond to lines in Fig~\ref{fig:FA_RANS_Coeff}.
    Relative errors and (CPA) comparing RANS results with respect to experimental data \citep{ELSR2020}. Colours correspond to lines in Fig~\ref{fig:FA_RANS_Coeff}.
    }\label{fig:FA_RANS_Err}
\end{figure}
Corresponding relative errors are quantified in Fig~\ref{fig:FA_RANS_Err} for selected turbulence models. Cyan, blue, and green lines correspond to simulations using Spalart-Allmaras (SA) models with and without rotation correction, and with rotation and QCR corrections, respectively. Red and orange curves denote our baseline and 2003 versions of Menter's Shear-Stress Transport (SST) models, respectively. We can see peaks in relative errors with respect to experimental measurements near $C_{L,max}$ for all aerodynamic coefficients shown in sub-figures (a-c). The convergence parameter shown in Fig~\ref{fig:FA_RANS_Err}(d) indicates convergence problems with increasing angels of attack. We remind the reader that based on previous experience $\log CPA<-4$ would be favourable (\S \ref{sec:sim_setup_convparam}). This is satisfied at reduced angles of attack, where $CPA$ values compare well with the standard deviation of $C_D$ divided by their mean value, reported in \cite{DHFMK2022}. However, for such complex geometries grid-convergence (i.e. obtaining low $CPA$ values) is challenging. In particular, SST models and the SA models with \verb|QCR| corrections show problems with iterative convergence. We will return to discussions on iterative convergence later in this sub-section. First, we want to look at some flow visualisations to understand the origin of differences in aerodynamic coefficients.

%%%%%%%%%%%%%%%%%%%%%%%%%%%%%%%%%%%%%%%%%%%%%%%%%%%%%%%%%%%%%%%%%%%%%%%%%%%%%%%%%%%%
\begin{figure}
    \centering
    \begin{subfigure}[b]{0.33\textwidth}
    	a) SA-noft2\\
        \includegraphics[width=\textwidth, trim=10 10 10 10, clip]{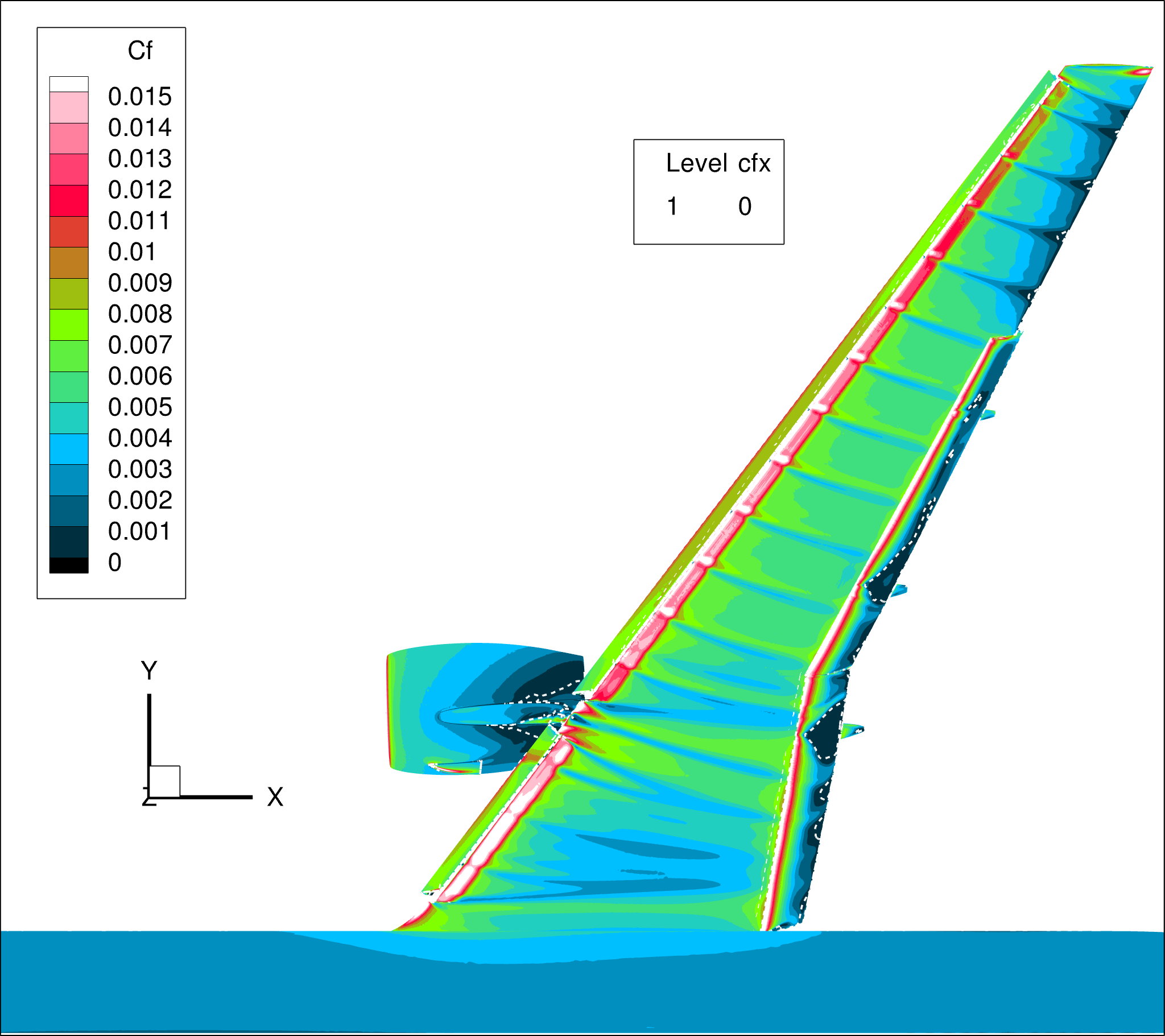}
        %\caption{Subplot 1}
    \end{subfigure}
    \begin{subfigure}[b]{0.33\textwidth}
    	b)\\
        \includegraphics[width=\textwidth, trim=10 10 10 10, clip]{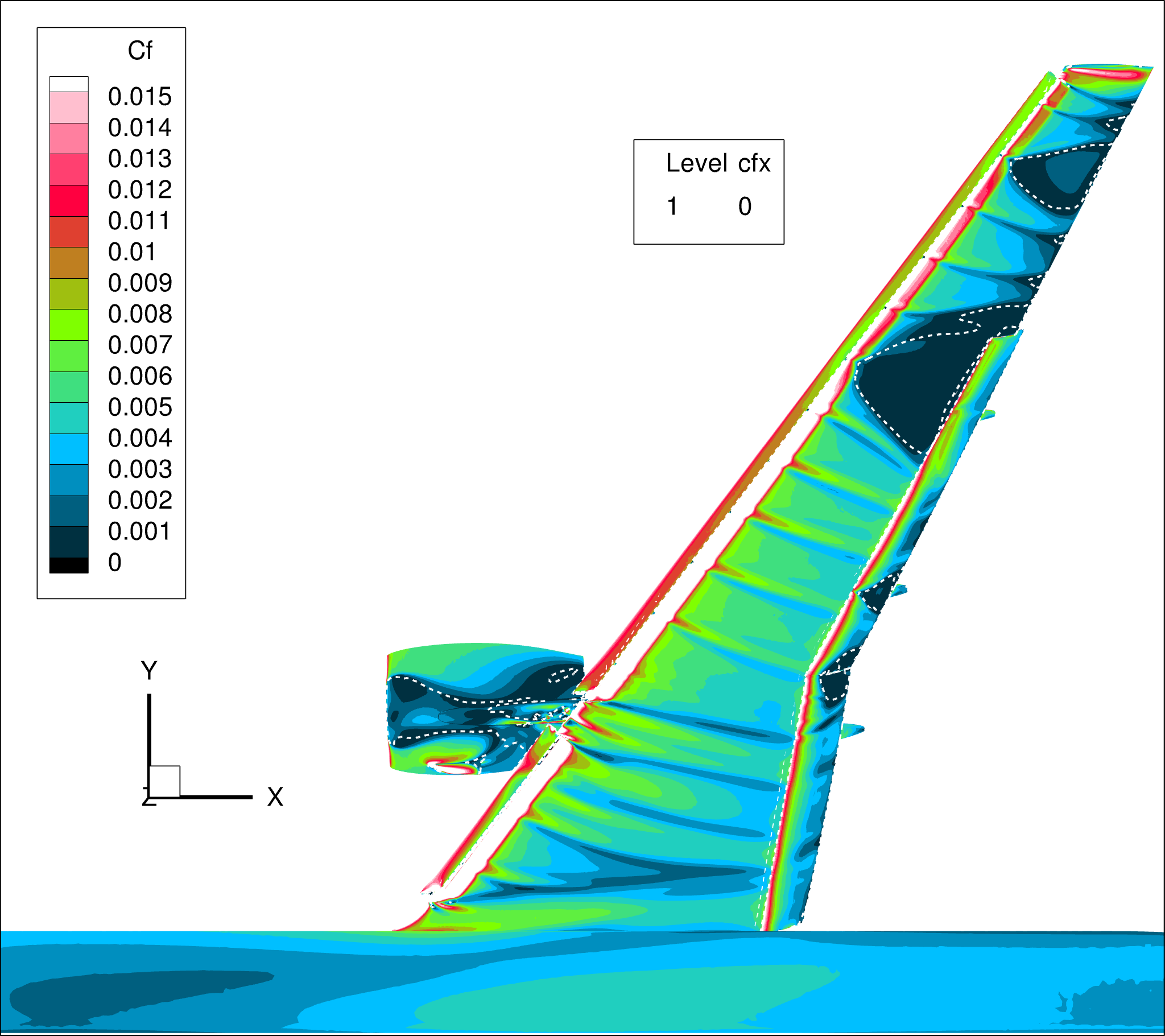}
        %\caption{Subplot 2}
    \end{subfigure}
    \begin{subfigure}[b]{0.33\textwidth}
    	c)\\
        \includegraphics[width=\textwidth, trim=10 10 10 10, clip]{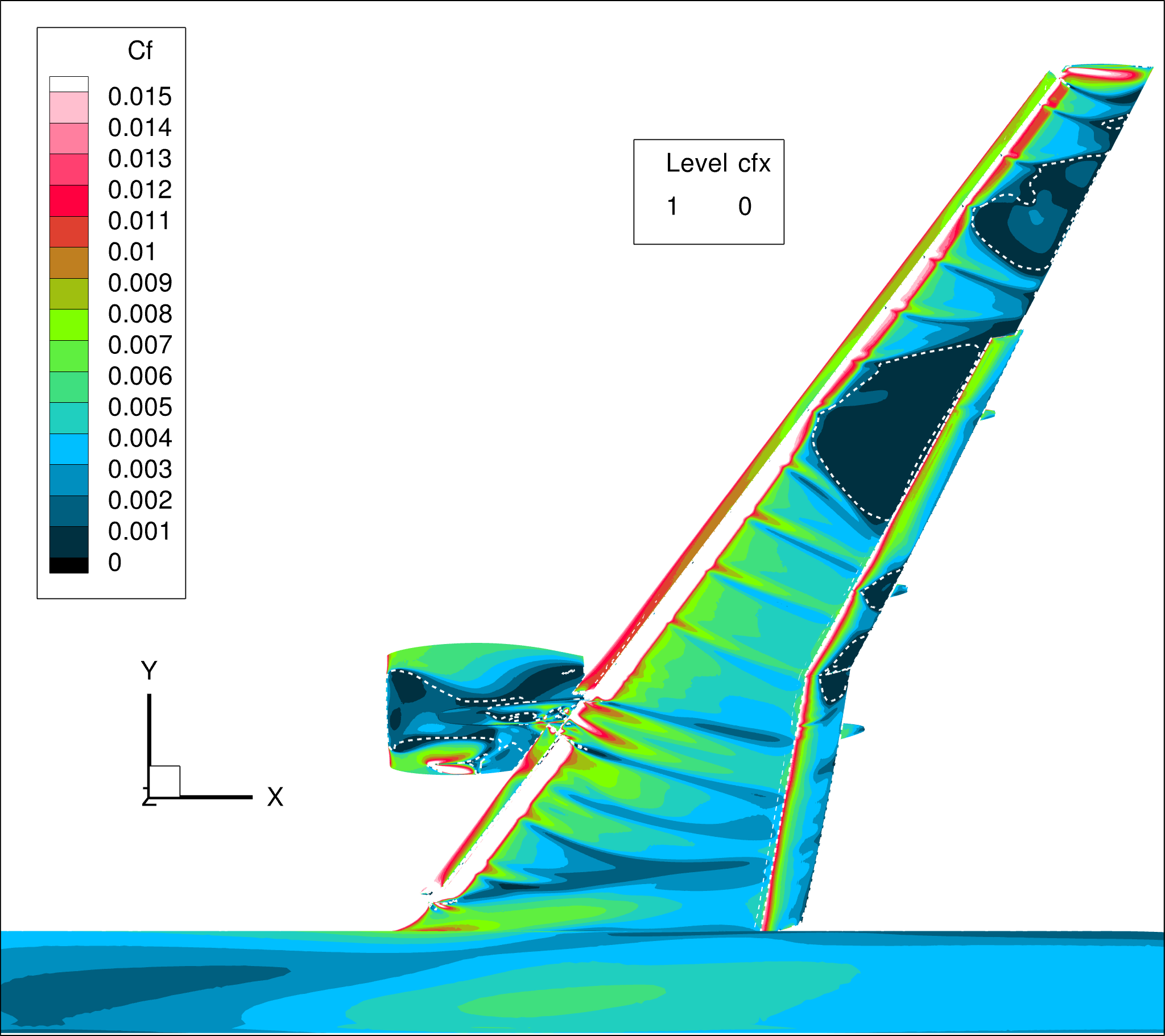}
        %\caption{Subplot 3}
    \end{subfigure}\\
    \vspace{0.1cm} % adjust vertical spacing between rows of subplots
%%%%%%%%%%%%%%%%%%%%%%%%%%%%%%%%%%%%%%%%%%%%%%%%%%%%%%%%%%%%%%%%%%%%%%%%%%%%%%%%%%%%%%%%%%%%%%%%%%
    \begin{subfigure}[b]{0.33\textwidth}
    	d) SA-noft2-R-QCR\\
        \includegraphics[width=\textwidth, trim=10 10 10 10, clip]{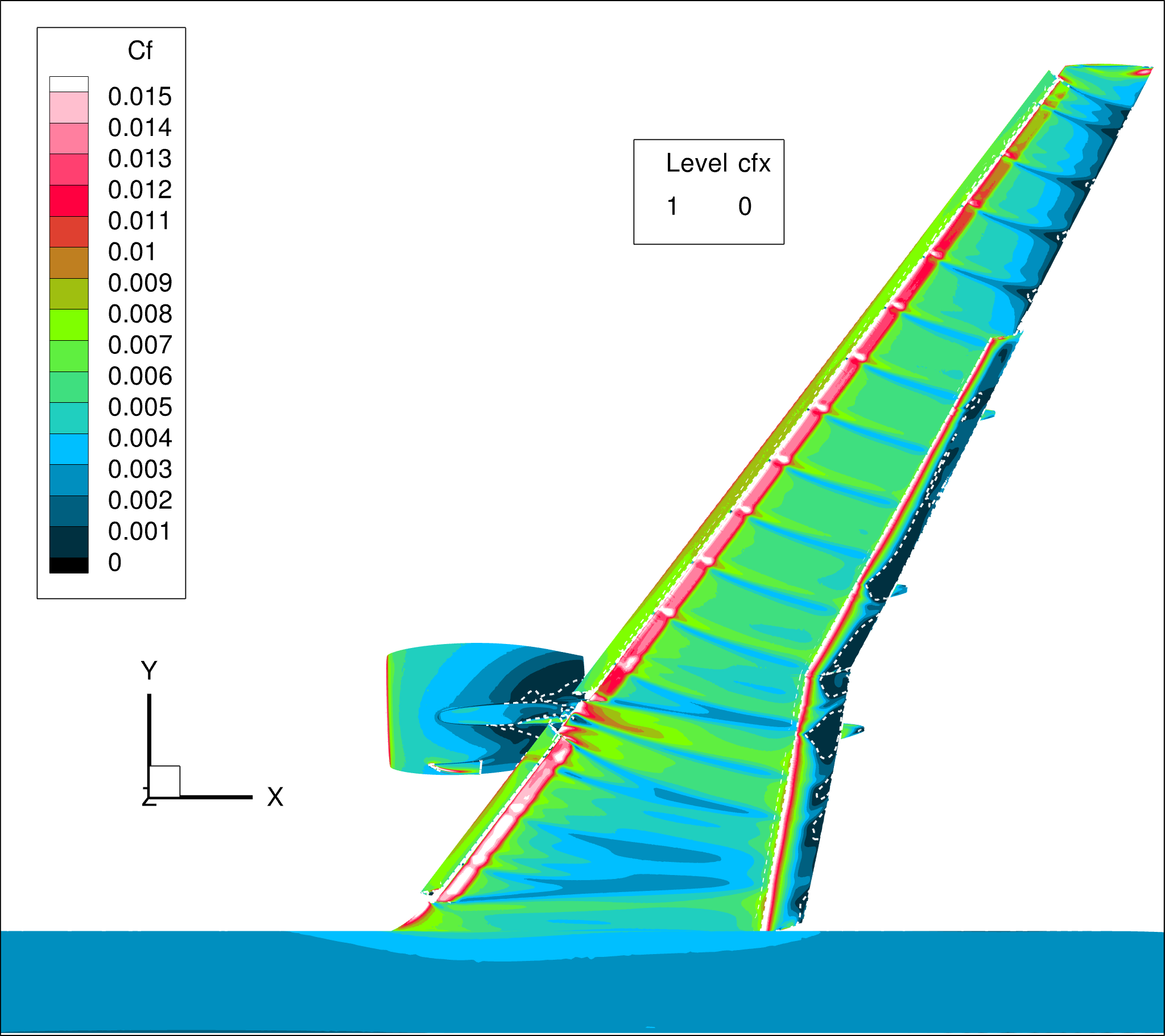}
        %\caption{Subplot 1}
    \end{subfigure}
    \begin{subfigure}[b]{0.33\textwidth}
    	e)\\
        \includegraphics[width=\textwidth, trim=10 10 10 10, clip]{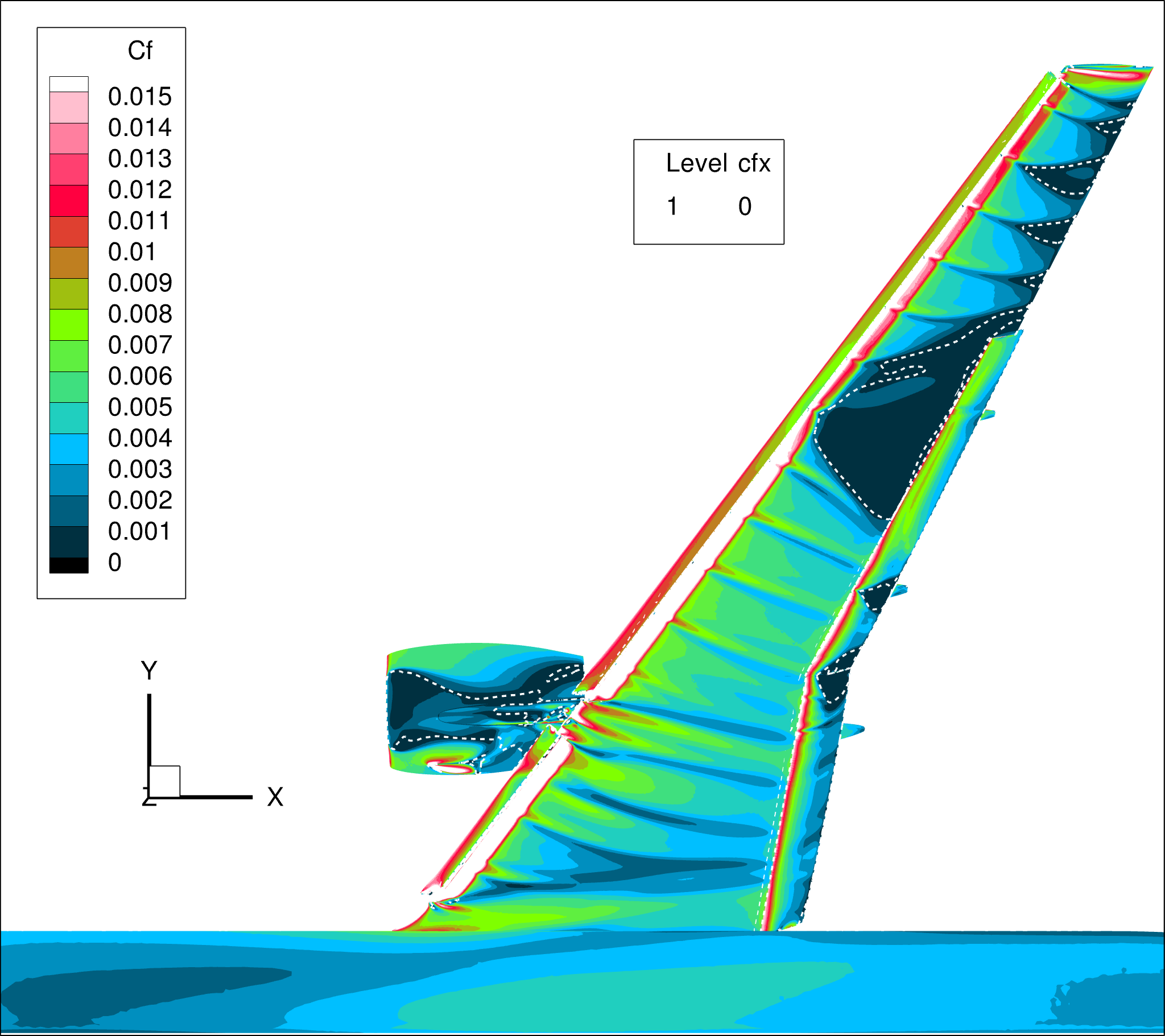}
        %\caption{Subplot 2}
    \end{subfigure}
    \begin{subfigure}[b]{0.33\textwidth}
    	f)\\
        \includegraphics[width=\textwidth, trim=10 10 10 10, clip]{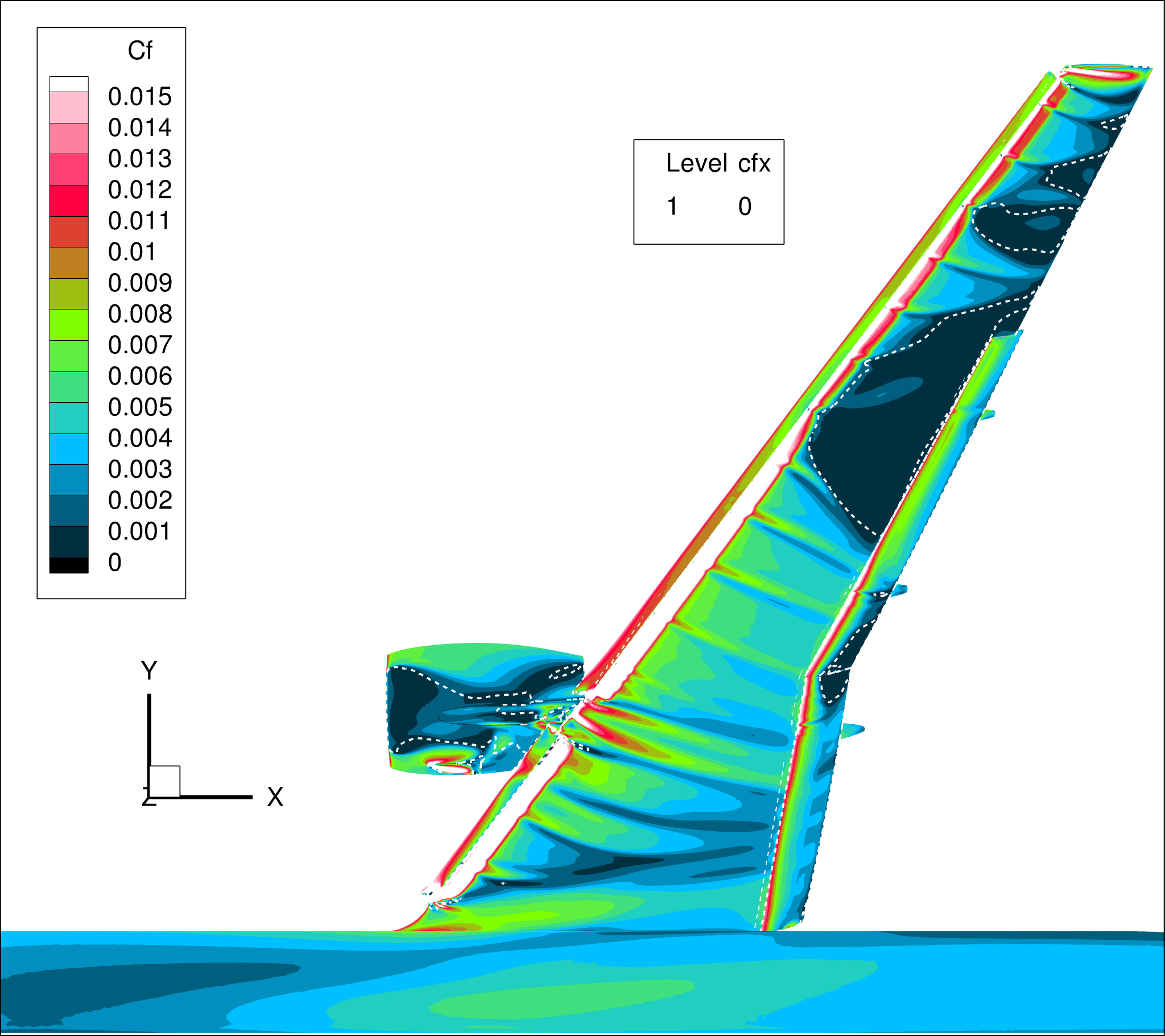}
        %\caption{Subplot 3}
    \end{subfigure}\\
    \vspace{0.1cm} % adjust vertical spacing between rows of subplots
%%%%%%%%%%%%%%%%%%%%%%%%%%%%%%%%%%%%%%%%%%%%%%%%%%%%%%%%%%%%%%%%%%%%%%%%%%%%%%%%%%%%%%%%%%%%%%%%%%
%    \begin{subfigure}[b]{0.33\textwidth}
%    	g)\\
%        \includegraphics[width=\textwidth, trim=10 10 10 10, clip]{Figures_FA_RANS/Cf/SST20_cold_07p05.png}
%        %\caption{Subplot 1}
%    \end{subfigure}
%    \begin{subfigure}[b]{0.33\textwidth}
%    	h)\\
%        \includegraphics[width=\textwidth, trim=10 10 10 10, clip]{Figures_FA_RANS/Cf/SST20_cold_17p05.png}
%        %\caption{Subplot 2}
%    \end{subfigure}
%    \begin{subfigure}[b]{0.33\textwidth}
%    	i)\\
%        \includegraphics[width=\textwidth, trim=10 10 10 10, clip]{Figures_FA_RANS/Cf/SST20_cold_19p57.png}
%        %\caption{Subplot 3}
%    \end{subfigure}\\
%    \vspace{0.1cm} % adjust vertical spacing between rows of subplots
%%%%%%%%%%%%%%%%%%%%%%%%%%%%%%%%%%%%%%%%%%%%%%%%%%%%%%%%%%%%%%%%%%%%%%%%%%%%%%%%%%%%%%%%%%%%%%%%%%
    \begin{subfigure}[b]{0.33\textwidth}
    	j) SST-2003\\
        \includegraphics[width=\textwidth, trim=10 10 10 10, clip]{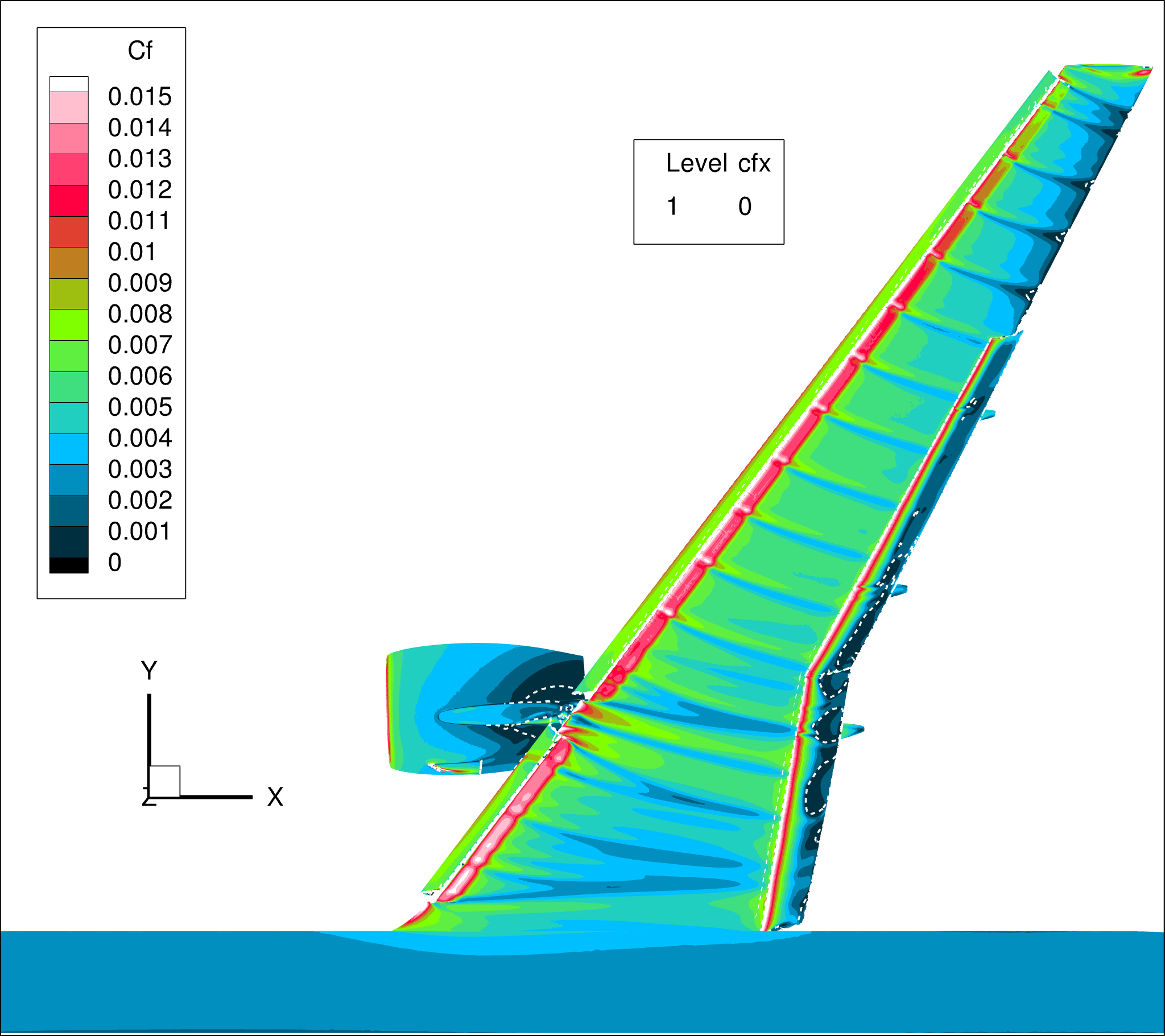}
        %\caption{Subplot 1}
    \end{subfigure}
    \begin{subfigure}[b]{0.33\textwidth}
    	k)\\
        \includegraphics[width=\textwidth, trim=10 10 10 10, clip]{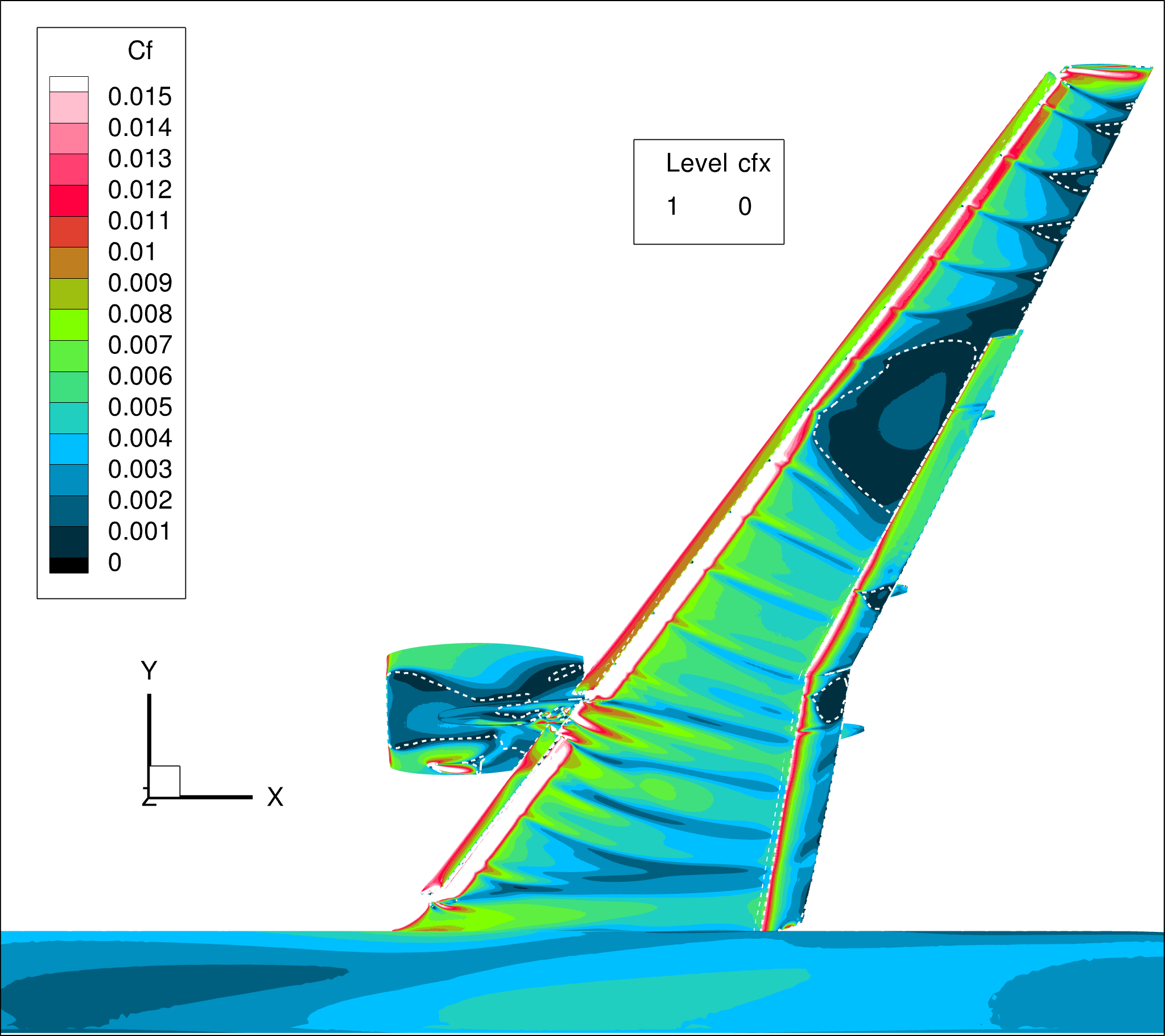}
        %\caption{Subplot 2}
    \end{subfigure}
    \begin{subfigure}[b]{0.33\textwidth}
    	l)\\
        \includegraphics[width=\textwidth, trim=10 10 10 10, clip]{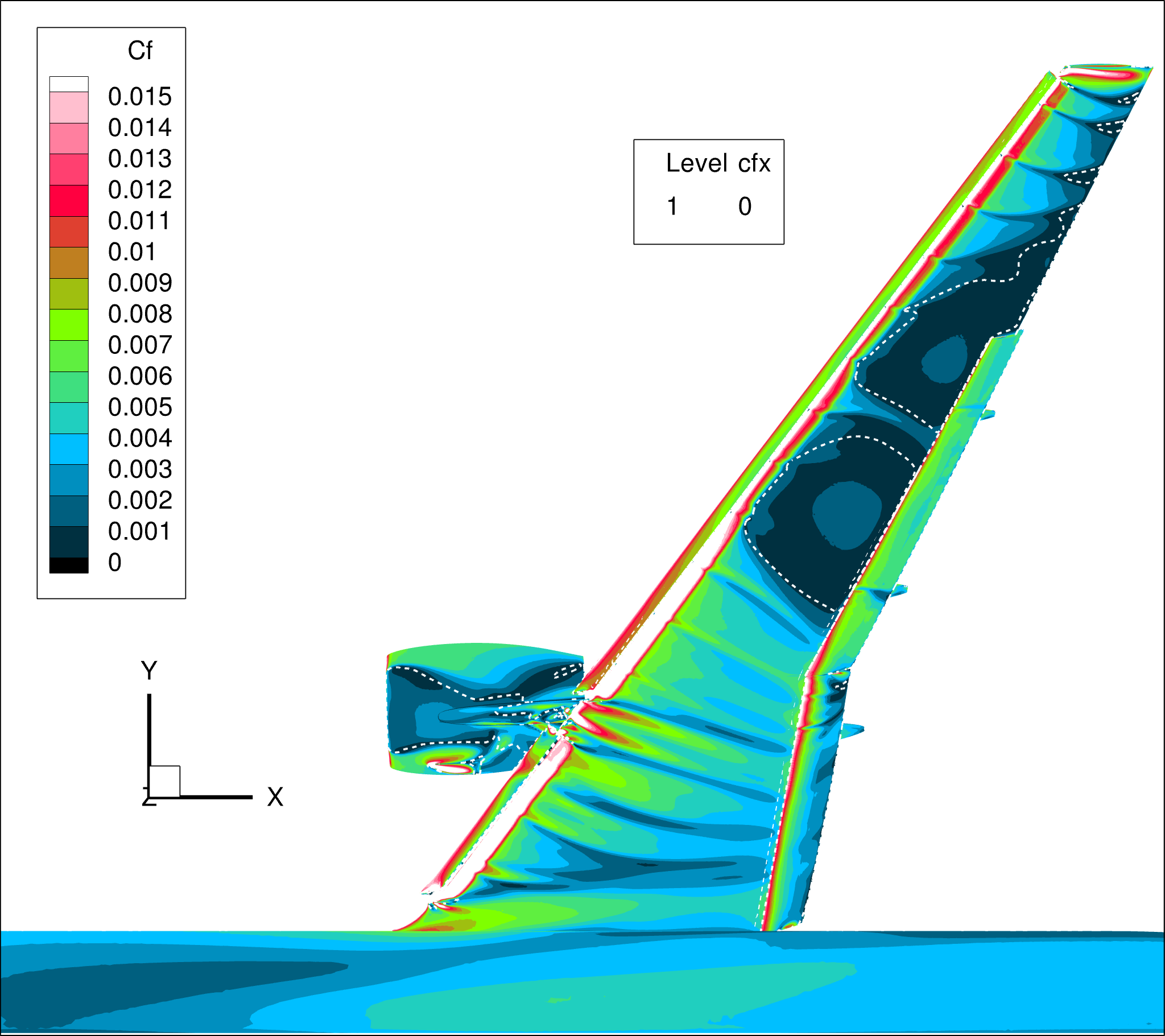}
        %\caption{Subplot 3}
    \end{subfigure}\\
%    \vspace{0.1cm} % adjust vertical spacing between rows of subplots
%%%%%%%%%%%%%%%%%%%%%%%%%%%%%%%%%%%%%%%%%%%%%%%%%%%%%%%%%%%%%%%%%%%%%%%%%%%%%%%%%%%%%%%%%%%%%%%%%%
    \caption{
    % Contours showing skin-friction coefficient $C_f$ for cold-started RANS simulations using (a-c) SA-noft2, (d-f) SA-noft2-R-QCR, and (g-i) SST-2003 turbulence models. Columns correspond to angles of attack of $7.05^{\circ}$ (left-hand-side), $17.05^{\circ}$ (middle), and $19.57^{\circ}$ (right-hand-side). White-dashed iso-curves denote zero stream-wise skin-friction components.
    $C_f$-contours for cold-started RANS using (a-c) SA-noft2, (d-f) SA-noft2-R-QCR, and (g-i) SST-2003 turbulence models. Columns (left-to-right) correspond to $\alpha=7.05^{\circ}$, $17.05^{\circ}$, and $19.57^{\circ}$. Iso-curves denote $C_{f,x}=0$.
    }\label{fig:FA_RANS_Cf}
\end{figure}
Fig~\ref{fig:FA_RANS_Cf} shows skin-friction contours for cold-started RANS simulations using (a-c) SA-noft2, (d-f) SA-noft2-R-QCR, and (g-i) SST-2003 turbulence models. Columns correspond to angles of attack of $7.05^{\circ}$ (left-hand-side), $17.05^{\circ}$ (middle), and $19.57^{\circ}$ (right-hand-side), where $C_{L,max}$ is expected just before the latter of these angles of attack. 
While cold-started simulations indeed suggest maximum lift around these angles, we can already observe for $\alpha=17.05^{\circ}$ significant regions of separated flow that are not present in the experimental measurements. While such non-physical large coherent separation regions appear mainly at the mid-chord location for $\alpha = 17.05^{\circ}$, they can also be observed for the baseline SA model near the wing tip. At $\alpha=19.57^{\circ}$, flow separation appears for all turbulence models also in the outboard wing region. Regarding the observed differences in the moment coefficient, it is also important to note significant regions of flow separation on the wing flaps even at low angles of attack ($\alpha=7.05^{\circ}$). It should be noted that due to the strong flap deflection with respect to the free-stream, the flaps experience high aerodynamic loads. At increased angles of attack, flow separation along the flaps weakens and is mainly pronounced near the wing bend.
Near the wing root, all turbulence models appear to produce consistent behavior, while the SST model predicts reduced flow separation on top of the nacelle. As we will see when warm-starting RANS simulations by incrementally increasing the angle of attack, significant changes in these regions of separated flow occur, which also impact the aerodynamic coefficients.
%%%%%%%%%%%%%%%%%%%%%%%%%%%%%%%%%%%%%%%%%%%%%%%%%%%%%%%%%%%%%%%%%%%%%%%%%%%%%%%%%%%%
\subsubsection{Sensitivity to Initialisation Strategies}
For the sake of conciseness, a subset of three representative turbulence models are selected here to show sensitivities to initialization strategies: the baseline turbulence model SA-noft2, which predicted $C_{L,max}$ closest to experimental measurements, and the SA-noft2-R-QCR and SST-2003 models (which showed rather poor agreement for cold starts). A complete list of aerodynamic coefficients including warm-started simulations using other turbulence models can be found in the appendix in tables \ref{tab:Sum_SA} and \ref{tab:Sum_SST}.
% Initialisation
\begin{figure}
    \centering
    \begin{subfigure}[b]{0.33\textwidth}
        a) \\
        \includegraphics[width=\textwidth, trim=10 10 50 50, clip]{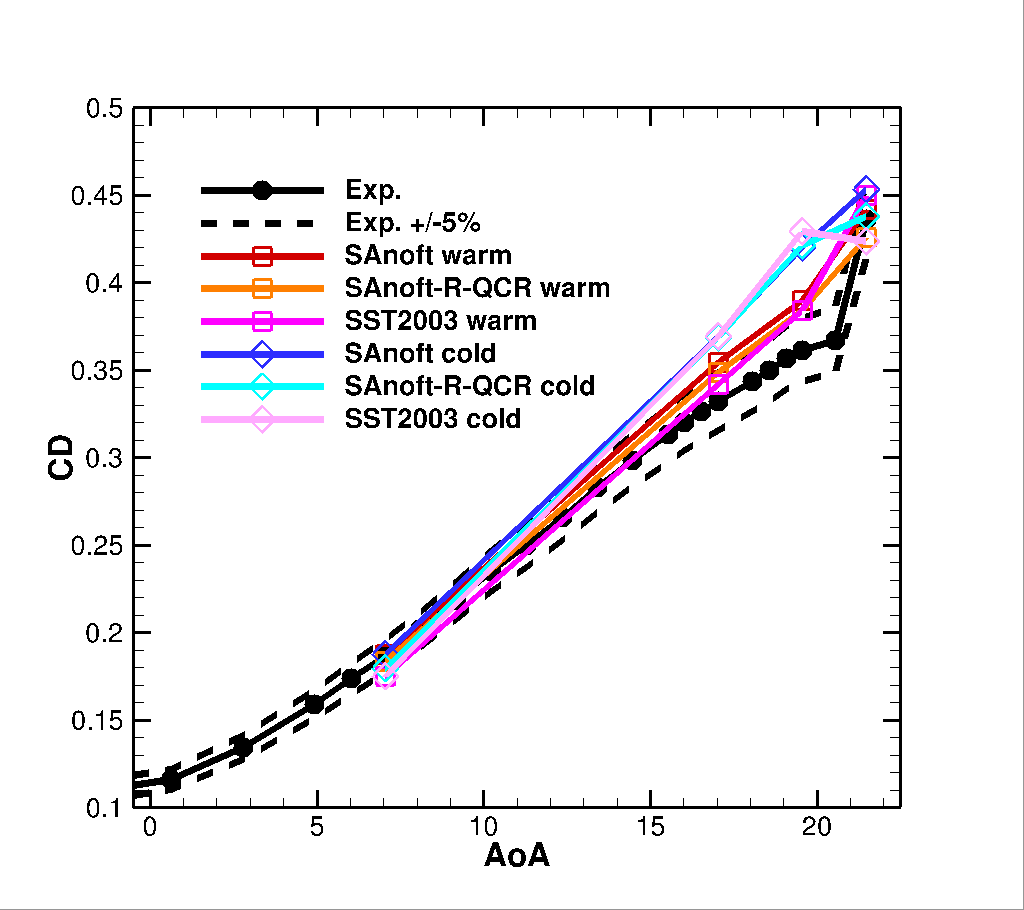}
        %\caption{Subplot 1}
    \end{subfigure}
    \begin{subfigure}[b]{0.33\textwidth}
        b) \\
        \includegraphics[width=\textwidth, trim=10 10 50 50, clip]{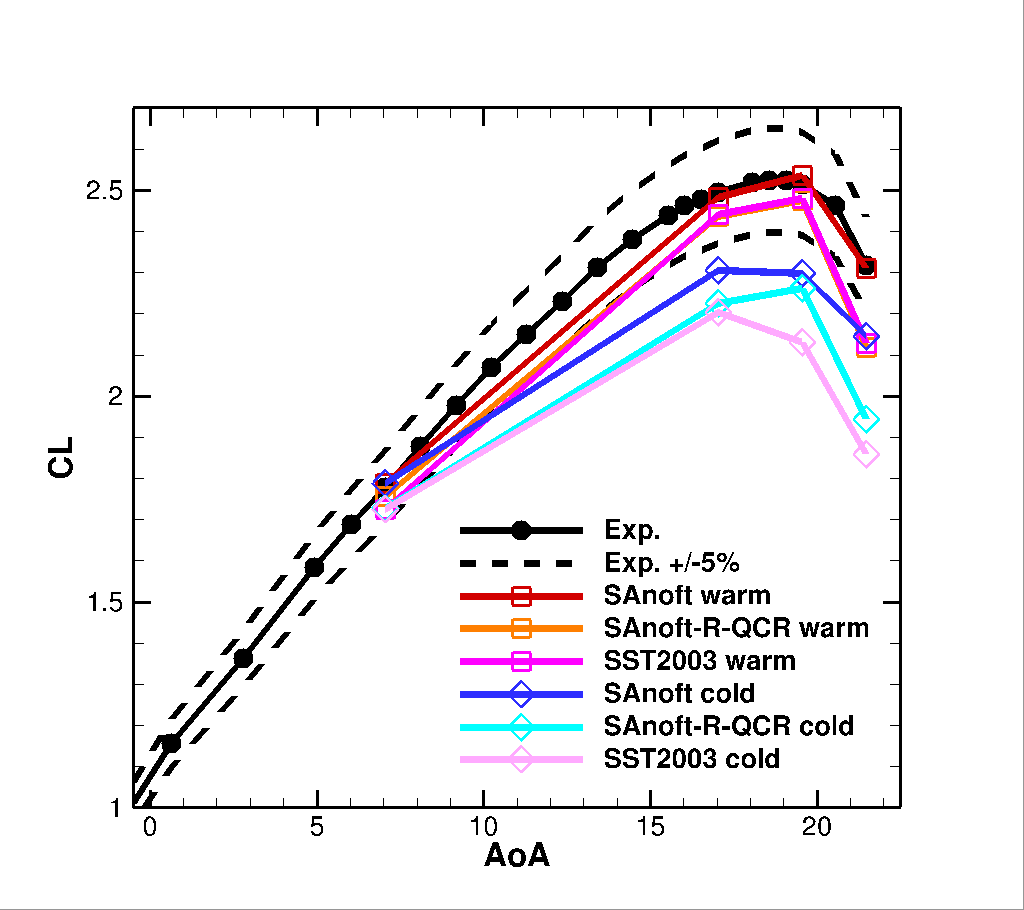}
        %\caption{Subplot 2}
    \end{subfigure}
    \begin{subfigure}[b]{0.33\textwidth}
        c) \\
        \includegraphics[width=\textwidth, trim=10 10 50 50, clip]{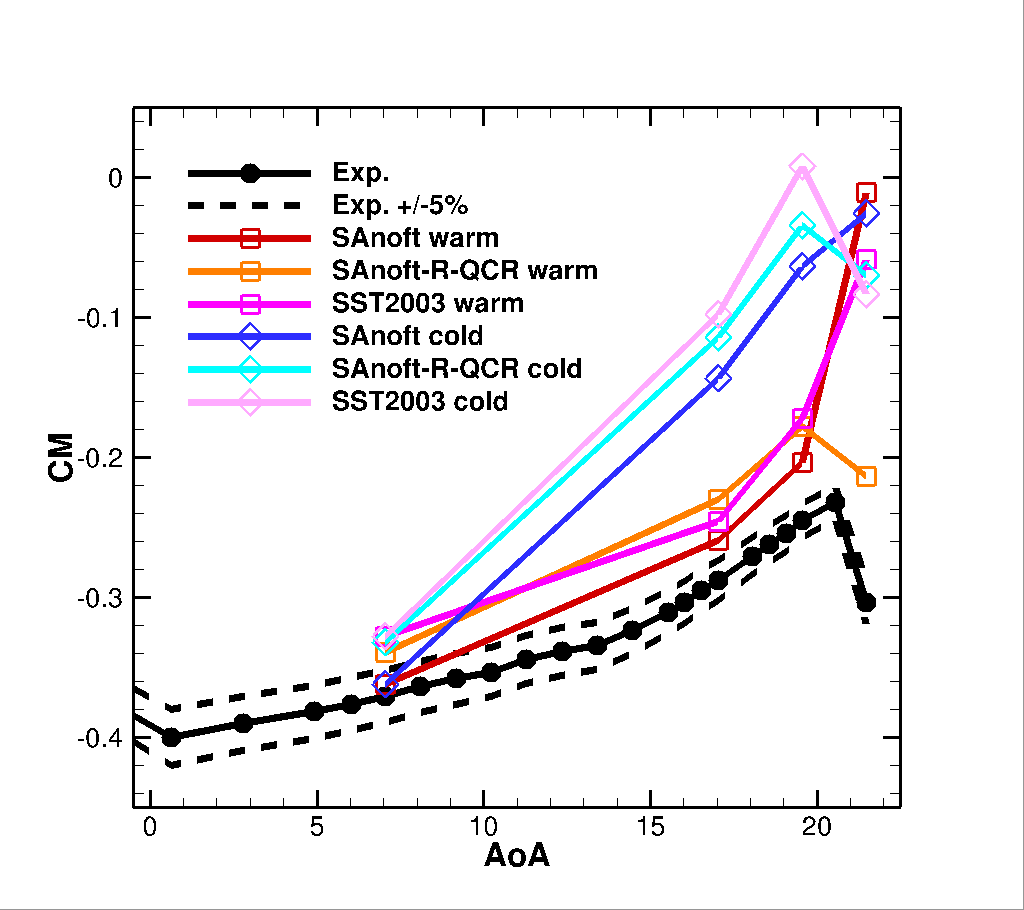}
        %\caption{Subplot 3}
    \end{subfigure}
    \caption{
    % (a) $C_D$, (b) $C_L$, and (c) $C_M$ for warm- and cold-started RANS simulations as a function of the angle of attack. Black symbols denote experimental measurements and black dashed lines correspond to $\pm 5\%$ error margins.
    (a) $C_D$, (b) $C_L$, and (c) $C_M$ over $\alpha$ for warm- and cold-started RANS simulations. Black symbols and curves respectively denote experimental measurements and $\pm 5\%$ errors.
    }\label{fig:FA_RANS_Coeff_Init}
\end{figure}
Fig~\ref{fig:FA_RANS_Coeff_Init} shows (a) drag-, (b) lift-, and (c) moment-coefficients as functions of $\alpha$ for the selected cold- (blue, cyan, and pink curves) and warm-started (red, orange, and magenta curves) simulations. 
Experimental measurements and the $\pm 5\%$ error margin are denoted by black solid and dashed curves, respectively. For $7.05^{\circ} < \alpha < 21.47^{\circ}$ warm-started simulations using SA-noft2 (red curves) and SA-noft2-R-QCR (orange curves) turbulence models show significant improvement for the main aerodynamic quantities. Also the warm-started simulations using the SST-2003 model perform well in terms of aerodynamic coefficients. 
This tendency can be confirmed in figures \ref{fig:FA_RANS_Err_Init} (a-c), showing relative errors with respect to experimental measurements. However, at post-stall conditions, we experience significant convergence problems for the warm-started simulation using SA-noft2 (denoted by the red curve), which leads to absence of the pitch-break in Fig~\ref{fig:FA_RANS_Coeff_Init}(c). The SA-noft2-R-QCR turbulence model shows slightly better convergence characteristics and leads to aerodynamic coefficients which are in fair agreement with experiments. However, as we shall see later, these promising trends are rather misleading, as they are a result of error-cancellation.
\begin{figure}
    \centering
    \begin{subfigure}[b]{0.45\textwidth}
        a) \\
        \includegraphics[width=\textwidth, trim=20 20 100 500, clip]{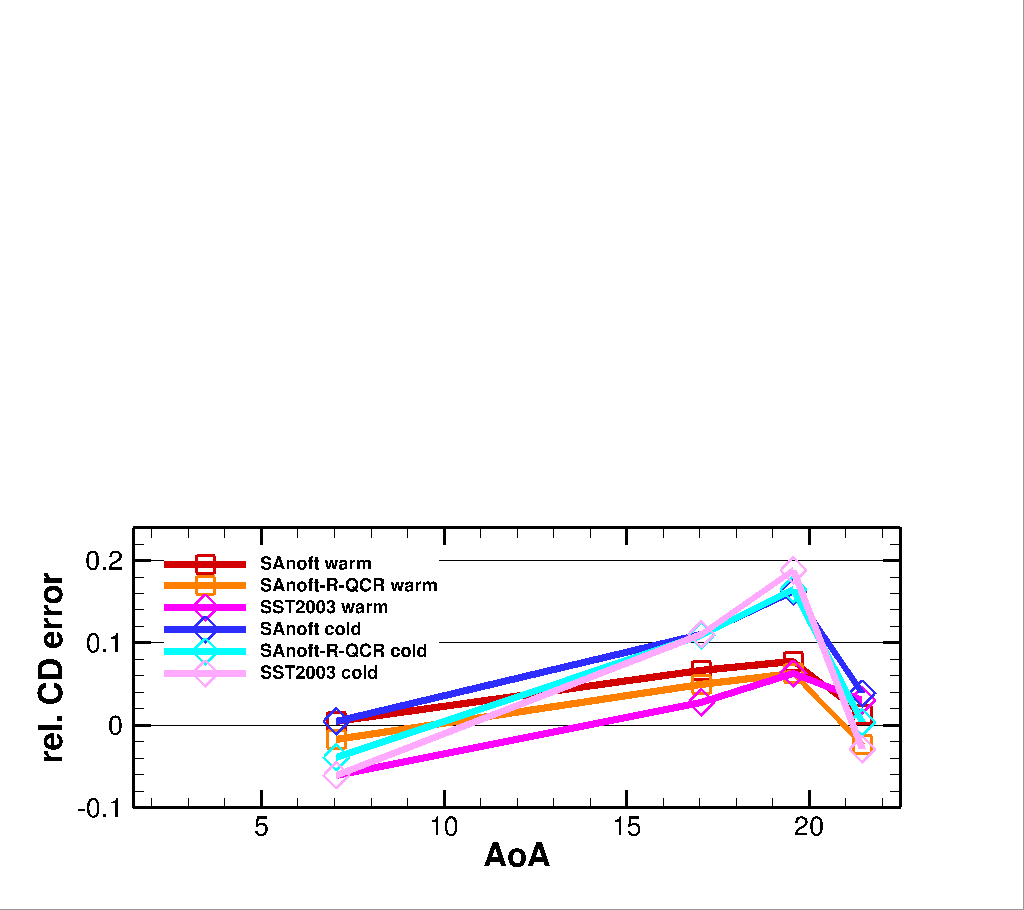}
        %\caption{Subplot 1}
    \end{subfigure}
    \begin{subfigure}[b]{0.45\textwidth}
        b) \\
        \includegraphics[width=\textwidth, trim=20 20 100 500, clip]{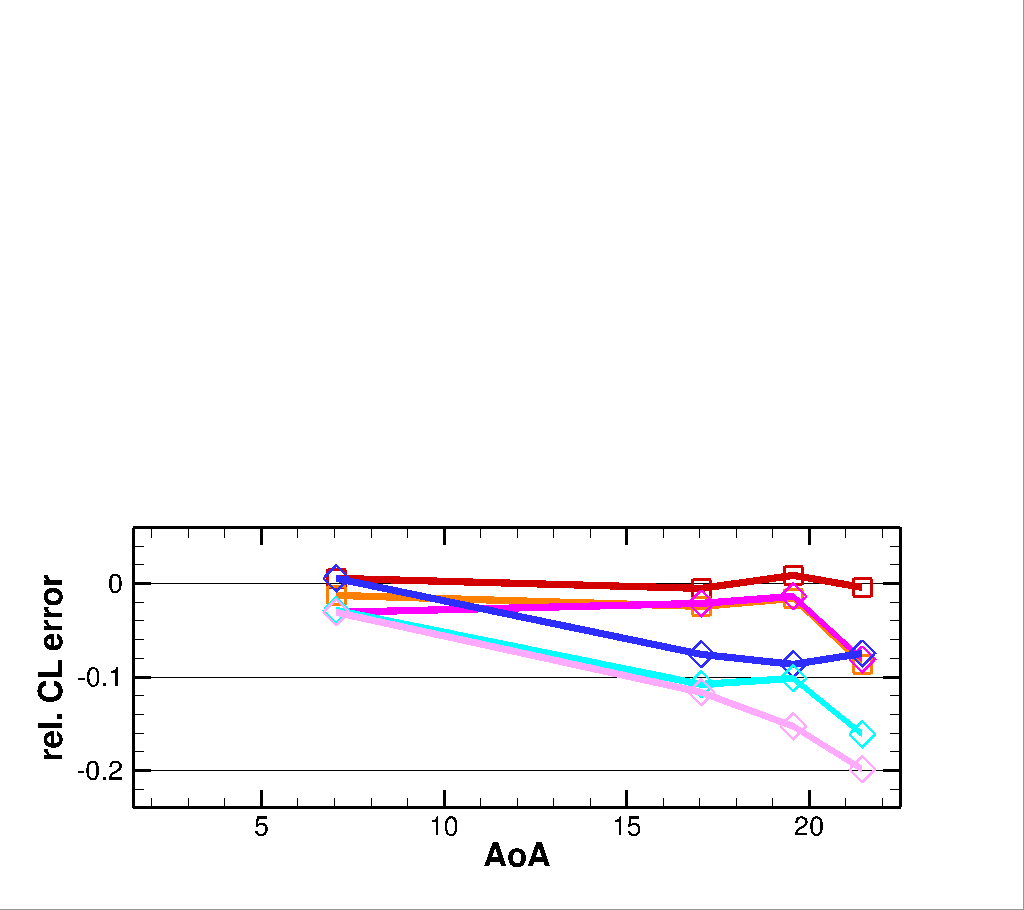}
        %\caption{Subplot 2}
    \end{subfigure}
    \vspace{0.5cm} % adjust vertical spacing between rows of subplots
    \begin{subfigure}[b]{0.45\textwidth}
        c) \\
        \includegraphics[width=\textwidth, trim=20 20 100 500, clip]{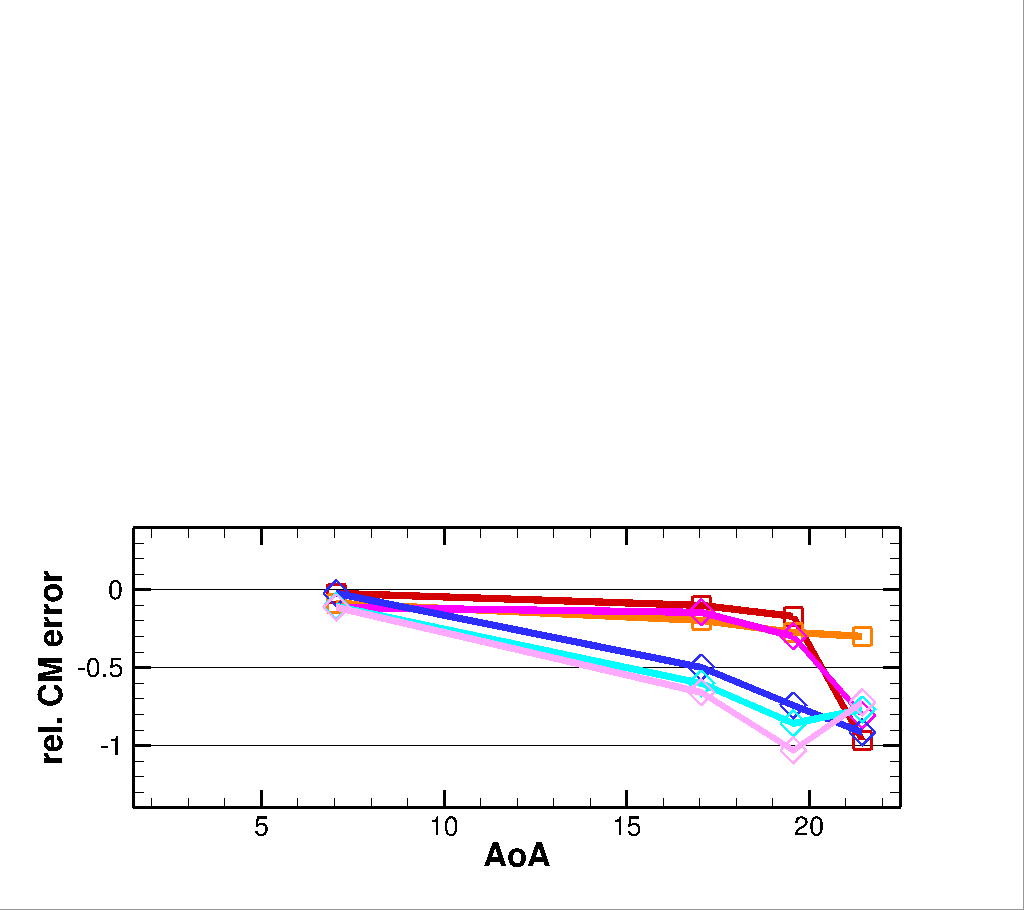}
        %\caption{Subplot 3}
    \end{subfigure}
    \begin{subfigure}[b]{0.45\textwidth}
        d) \\
        \includegraphics[width=\textwidth, trim=20 20 100 500, clip]{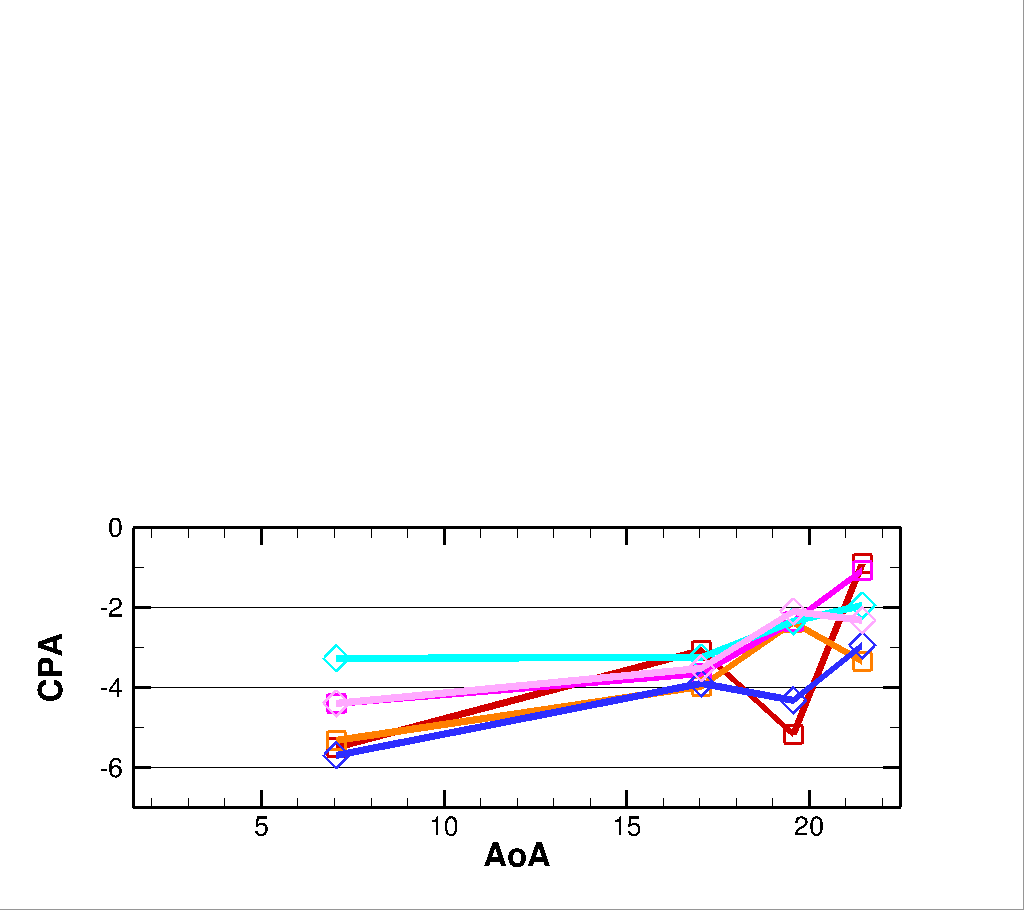}
        %\caption{Subplot 4}
    \end{subfigure}
    \caption{
    %Relative errors comparing cold- and warm-started RANS results with respect to experimental data \citep{ELSR2020}. (a) $C_D$, (b) $C_L$, and (c) $C_m$ as well as (d) convergence parameters (CPA) are shown as a function of angle of attack, where colours correspond to lines in Fig~\ref{fig:FA_RANS_Coeff_Init}
    Relative errors and (CPA) comparing RANS results with respect to experimental data \citep{ELSR2020}. Colours correspond to lines in Fig~\ref{fig:FA_RANS_Coeff_Init}
    }\label{fig:FA_RANS_Err_Init}
\end{figure}

\begin{figure}
    \centering
    % \begin{subfigure}[b]{0.33\textwidth}
    % 	a) \\
    %     \includegraphics[width=\textwidth, trim=10 10 10 10, clip]{Figures_FA_RANS/Cf/SA10_warm_07p05.png}
    %     %\caption{Subplot 1}
    % \end{subfigure}
    % \begin{subfigure}[b]{0.33\textwidth}
    % 	b)\\
    %     \includegraphics[width=\textwidth, trim=10 10 10 10, clip]{Figures_FA_RANS/Cf/SA12_warm_07p05.png}
    %     %\caption{Subplot 2}
    % \end{subfigure}
    % \begin{subfigure}[b]{0.33\textwidth}
    % 	c)\\
    %     \includegraphics[width=\textwidth, trim=10 10 10 10, clip]{Figures_FA_RANS/Cf/SST22_cold_07p05.png}
    %     %\caption{Subplot 3}
    % \end{subfigure}\\
    % \vspace{0.1cm} % adjust vertical spacing between rows of subplots
%%%%%%%%%%%%%%%%%%%%%%%%%%%%%%%%%%%%%%%%%%%%%%%%%%%%%%%%%%%%%%%%%%%%%%%%%%%%%%%%%%%%%%%%%%%%%%%%%%  
    \begin{subfigure}[b]{0.33\textwidth}
    	a) SA-noft2\\
        \includegraphics[width=\textwidth, trim=10 10 10 10, clip]{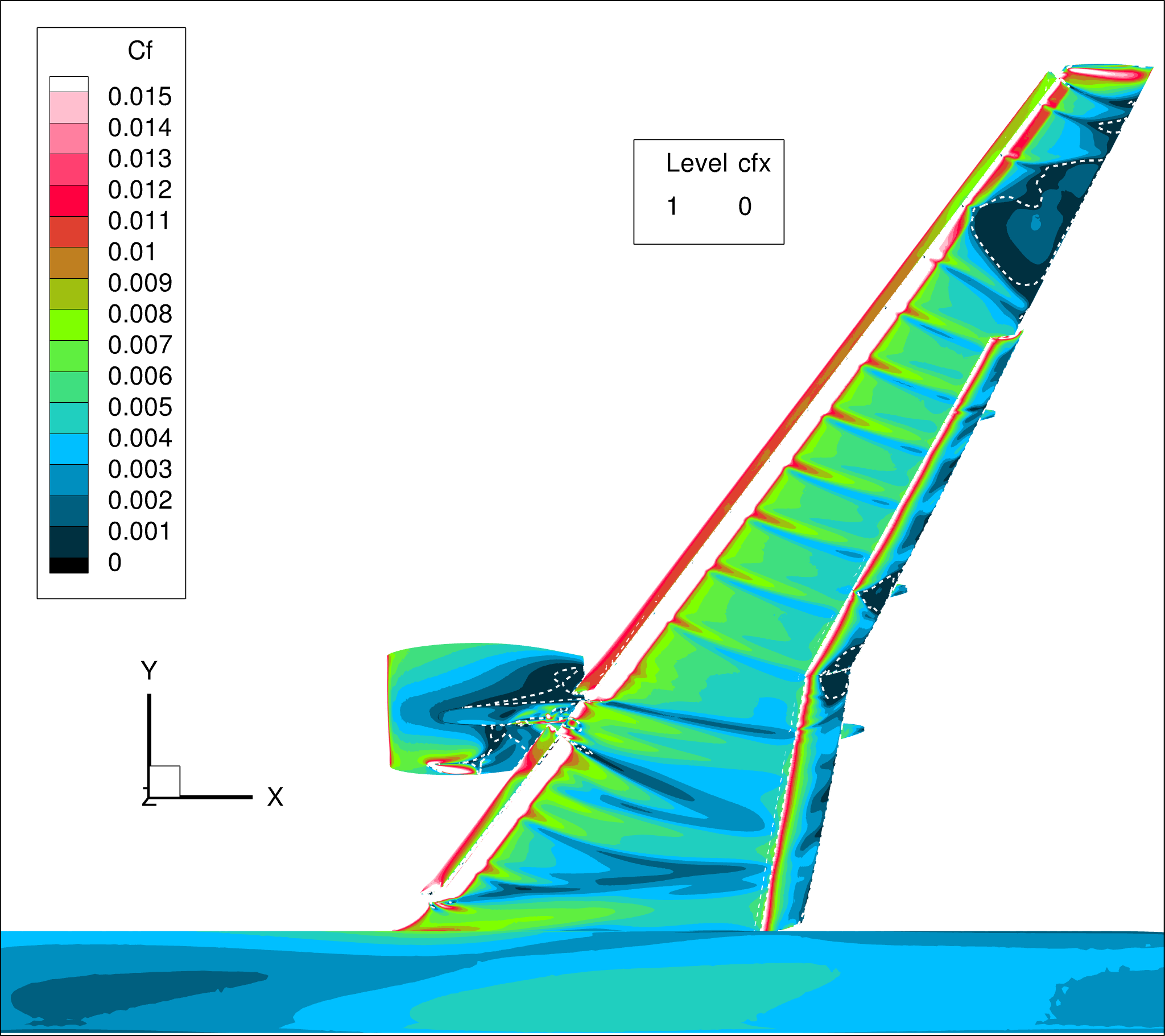}
        %\caption{Subplot 1}
    \end{subfigure}
    \begin{subfigure}[b]{0.33\textwidth}
    	b)\\
         \includegraphics[width=\textwidth, trim=10 10 10 10, clip]{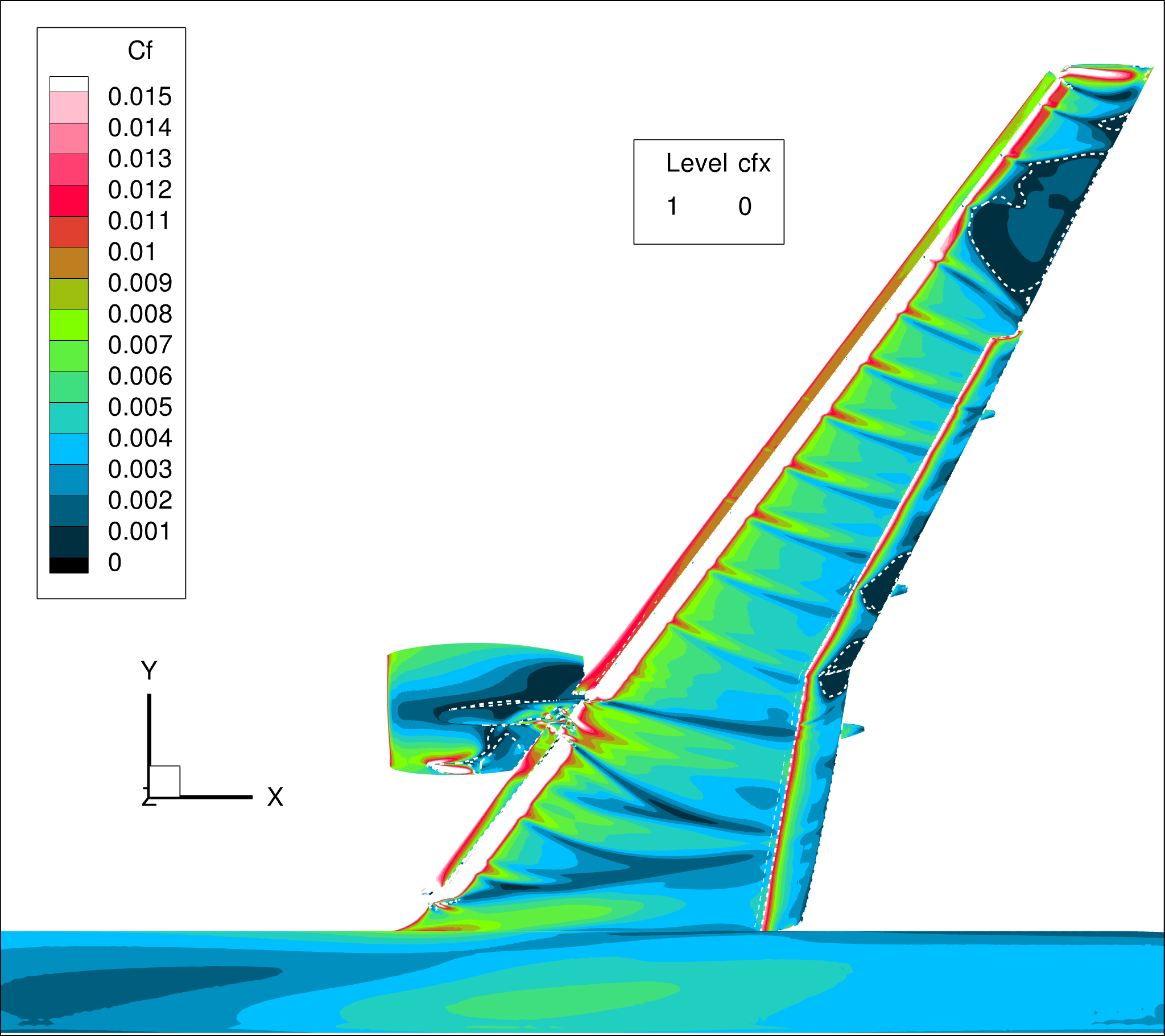}
        %\caption{Subplot 2}
    \end{subfigure}
    \begin{subfigure}[b]{0.33\textwidth}
    	c)\\
        \includegraphics[width=\textwidth, trim=10 10 10 10, clip]{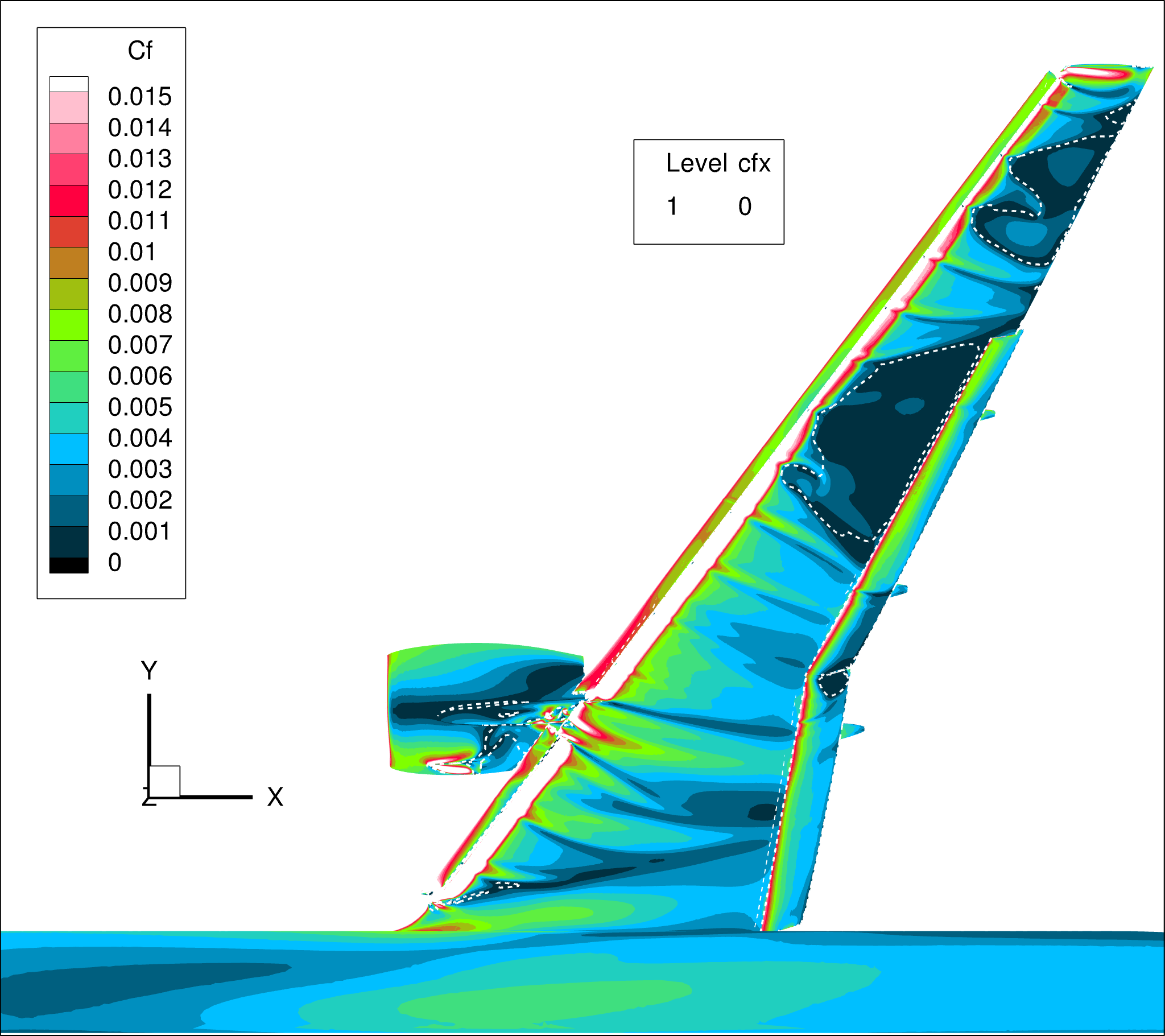}
        %\caption{Subplot 3}
    \end{subfigure}\\
    \vspace{0.1cm} % adjust vertical spacing between rows of subplots
%%%%%%%%%%%%%%%%%%%%%%%%%%%%%%%%%%%%%%%%%%%%%%%%%%%%%%%%%%%%%%%%%%%%%%%%%%%%%%%%%%%%%%%%%%%%%%%%%%  
    \begin{subfigure}[b]{0.33\textwidth}
    	d) SA-noft2-R-QCR\\
        \includegraphics[width=\textwidth, trim=10 10 10 10, clip]{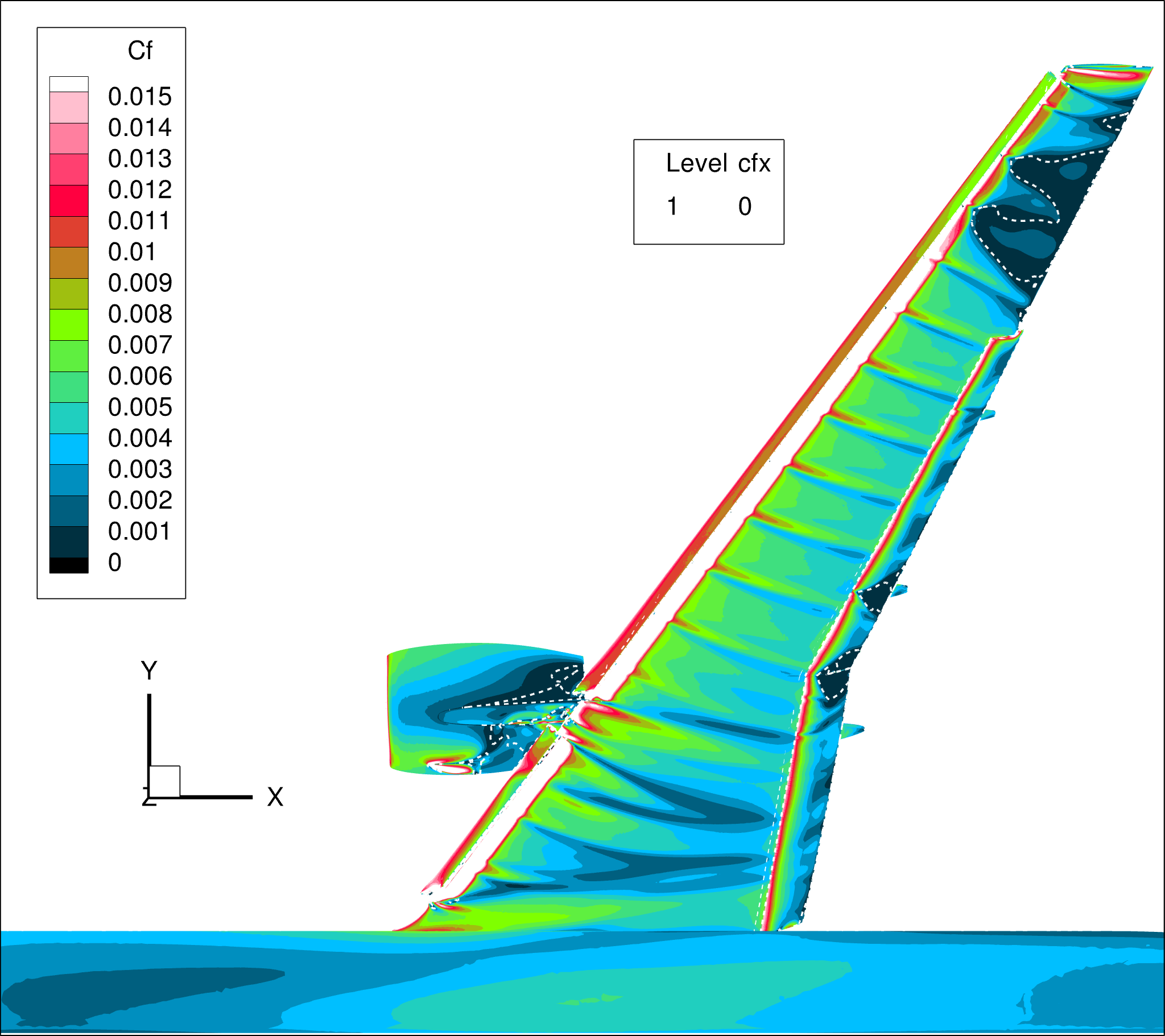}
        %\caption{Subplot 1}
    \end{subfigure}
    \begin{subfigure}[b]{0.33\textwidth}
    	e)\\
        \includegraphics[width=\textwidth, trim=10 10 10 10, clip]{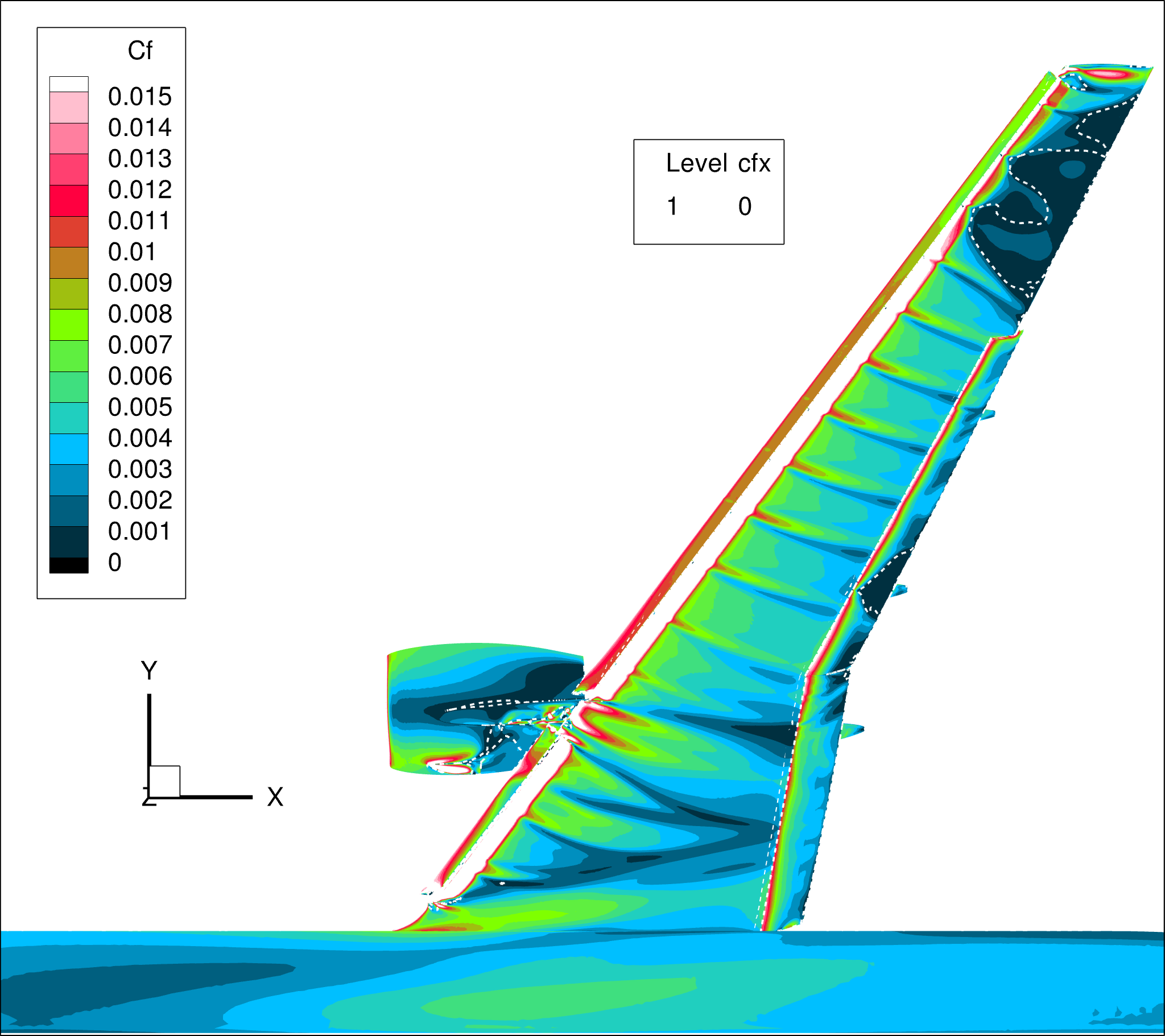}
        %\caption{Subplot 2}
    \end{subfigure}
    \begin{subfigure}[b]{0.33\textwidth}
    	f)\\
        \includegraphics[width=\textwidth, trim=10 10 10 10, clip]{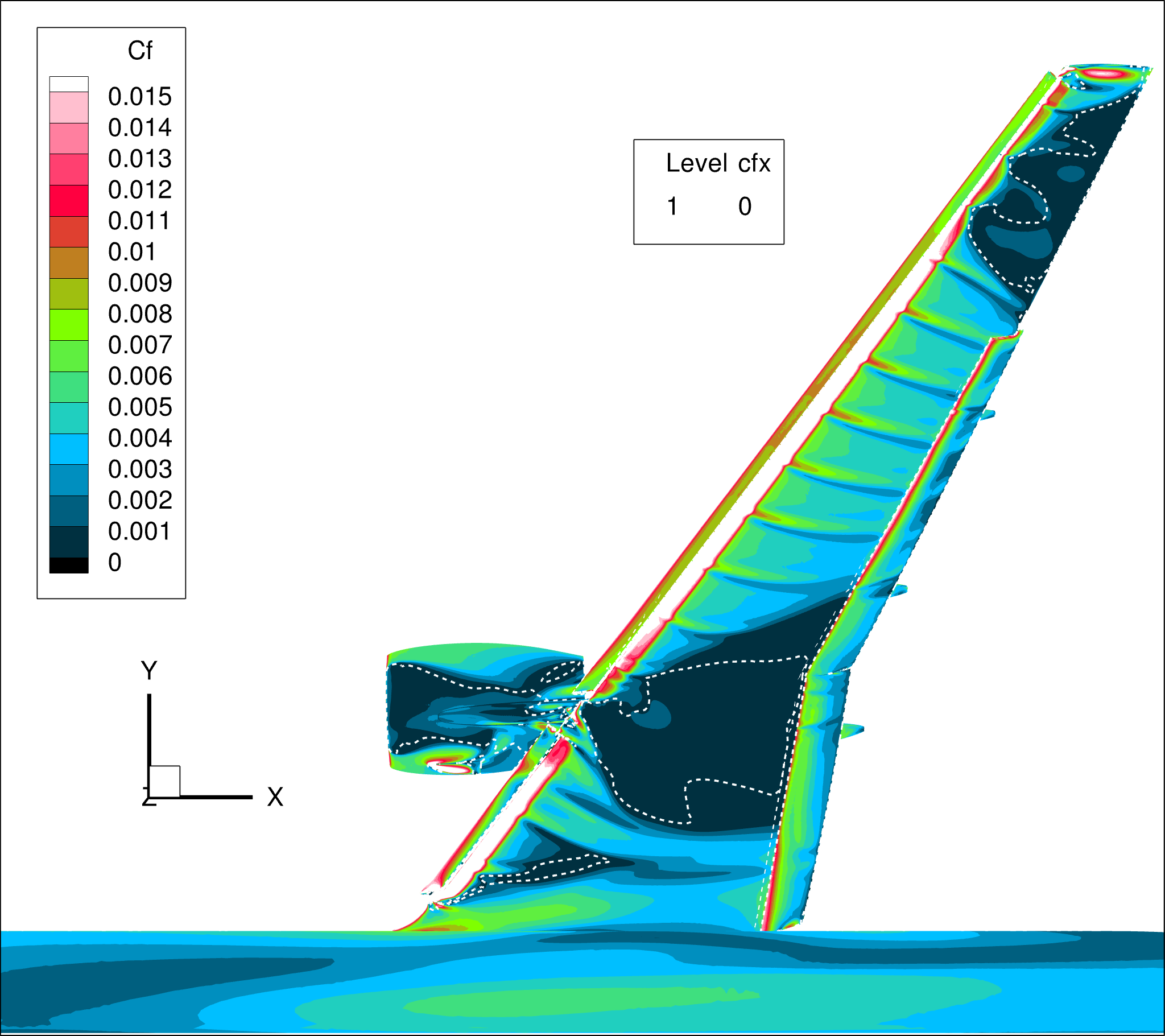}    
        %\caption{Subplot 3}
    \end{subfigure}\\
    \vspace{0.1cm} % adjust vertical spacing between rows of subplots
%%%%%%%%%%%%%%%%%%%%%%%%%%%%%%%%%%%%%%%%%%%%%%%%%%%%%%%%%%%%%%%%%%%%%%%%%%%%%%%%%%%%%%%%%%%%%%%%%%  
    \begin{subfigure}[b]{0.33\textwidth}
        g) SST-2003 \\
        \includegraphics[width=\textwidth, trim=10 10 10 10, clip]{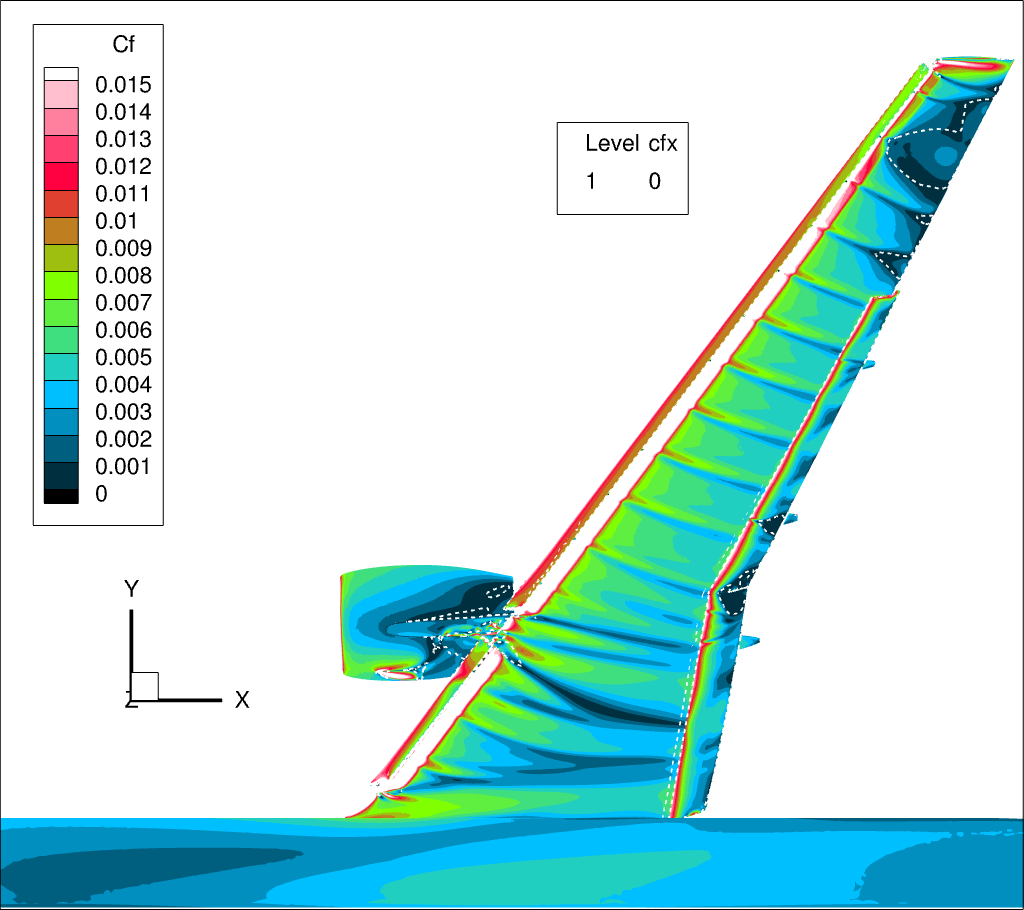}    
        %\caption{Subplot 1}
    \end{subfigure}
    \begin{subfigure}[b]{0.33\textwidth}
        h) \\
        \includegraphics[width=\textwidth, trim=10 10 10 10, clip]{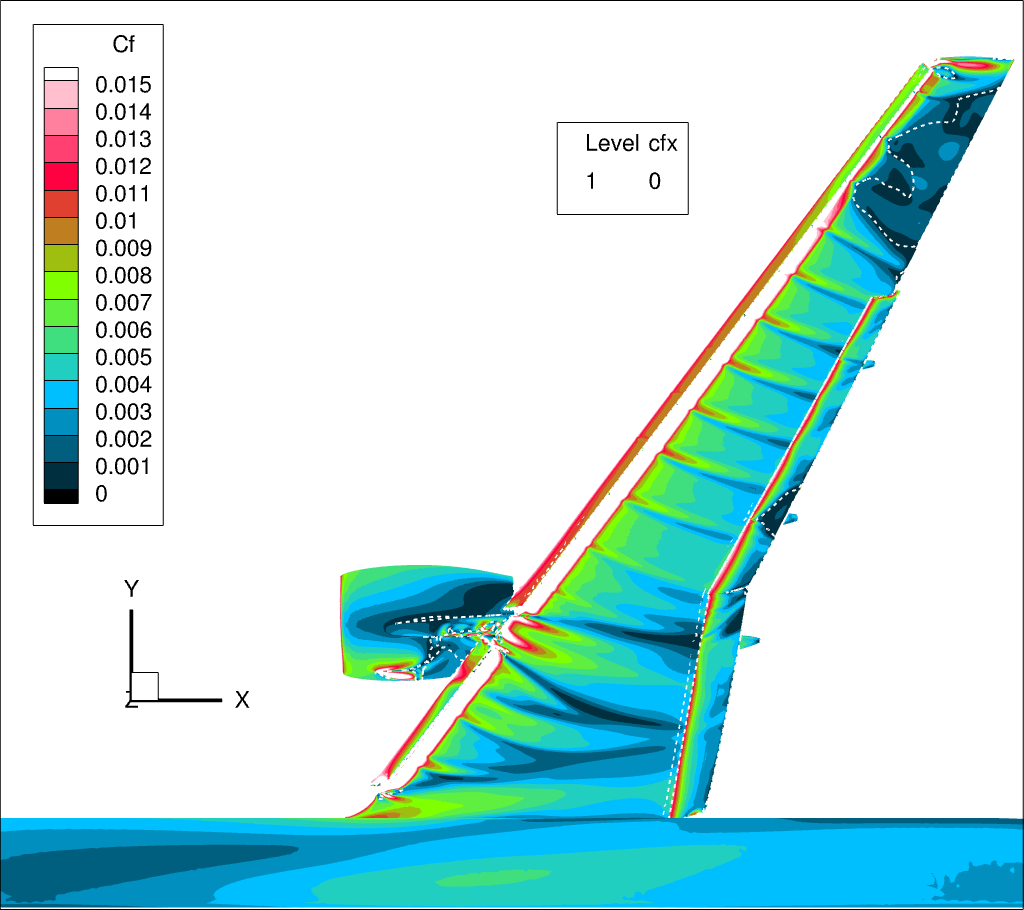}        
        %\caption{Subplot 2}
    \end{subfigure}
    \begin{subfigure}[b]{0.33\textwidth}
        i) \\
        \includegraphics[width=\textwidth, trim=10 10 10 10, clip]{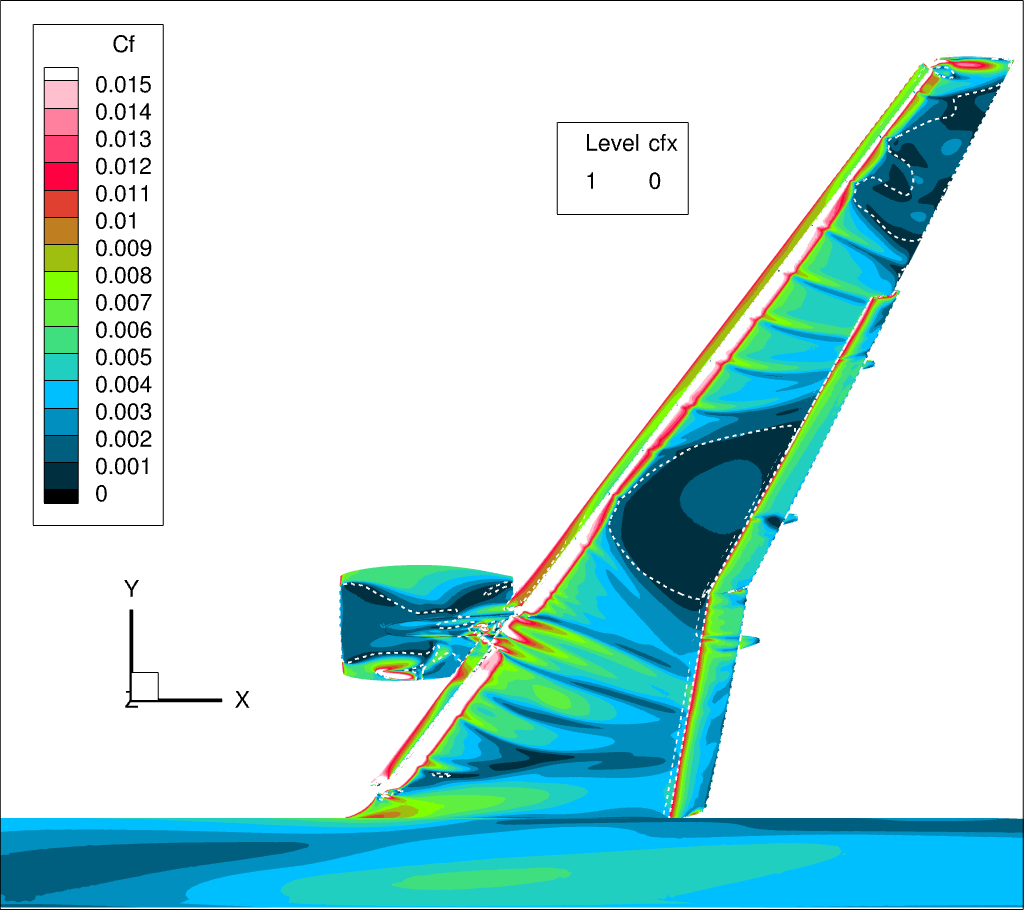}
        %\caption{Subplot 3}
    \end{subfigure}\\
    % \vspace{0.1cm} % adjust vertical spacing between rows of subplots
    \caption{
    % Contours showing skin-friction coefficient $C_f$ for warm-started RANS simulations using (a-c) SA-noft2, (d-f) SA-noft2-R-QCR, and (g-i) SST-2003 turbulence models. Columns correspond to angles of attack of $17.05^{\circ}$ (left-hand-side), $19.57^{\circ}$ (middle), and $21.47^{\circ}$ (right-hand-side). White-dashed iso-curves denote zero stream-wise skin-friction components.
    $C_f$-contours for warm-started RANS using (a-c) SA-noft2, (d-f) SA-noft2-R-QCR, and (g-i) SST-2003 turbulence models. Columns (left to right) correspond to $\alpha=17.05^{\circ}$, $19.57^{\circ}$, and $21.47^{\circ}$. Iso-curves denote $C_{f,x}=0$.
    }\label{fig:FA_RANS_Cf_init}
\end{figure}
Fig~\ref{fig:FA_RANS_Cf_init} shows warm-started RANS simulation results for the selected turbulence models over a range of angles of attack of $17.05^{\circ}$ (left-hand-side plots), $19.57^{\circ}$ (center plots), and $21.47^{\circ}$ (right-hand-side plots). While skin-friction contour plots (a-c) display results using our baseline SA-noft2 turbulence model, plots (d-f) show results with rotation and QCR corrections applied. Plots (g-i) correspond to the SST-2003 model. 
Firstly, it should be noted that below $17.05^{\circ}$, we do not observe significant differences between cold- and warm-started CFD solutions (see Fig~\ref{fig:FA_RANS_Coeff_Init}). Comparing the two SA turbulence models, we obtain good agreement up to $\alpha=19.57^{\circ}$. Although we observe large outboard flow separation appearing already at $\alpha=17.05^{\circ}$, we do not observe the massive secondary flow separation around mid-span (upstream of the out-board flaps), which is evident in cold-started simulation results in Fig~\ref{fig:FA_RANS_Cf}. For further increases of angle of attack, the SA-noft2 turbulence model shows similar flow topology to cold-started simulations. The SA-noft2-R-QCR turbulence model, on the other hand, shows massive non-physical flow separation behind the nacelle. It is interesting to note that the flow-separation on flaps is partly suppressed by these massively separated flow regions upstream. This is also the reason why the SA-noft2-R-QCR turbulence model (showing reduced flap separation), leads to increased values of pitch-up moment.
At $\alpha=17.05^{\circ}$ and $19.57^{\circ}$, RANS results obtained using the SST-2003 model are also similar to the SA results in Fig~\ref{fig:FA_RANS_Cf_init}, and those of present cold-started simulations.
At post-stall condition at $\alpha=21.47^{\circ}$, however, the region of non-physical flow separation around mid-span appears for SST results somewhere between those observed for the SA models shown in Fig~\ref{fig:FA_RANS_Cf_init}. This - potentially random - appearance of large bulks of separated flow near mid span may well be due to convergence problems at increased angles of attack (i.e. post-stall conditions). 
However, as we will show in the following sub-section, improved convergence (despite not machine-precision converged) has only a limited effect when compared to uncertainties associated with the choice of turbulence model.
While the simulations using QCR corrections reported in \cite{DHFMK2022} agree well with present results at $\alpha=19.57^{\circ}$, their baseline SA model and SST models deliver qualitatively different results to ours. Rather than a large bulk region of flow separation near the tip, they show more localised triangular-shaped patches of flow separation. The simulations of \cite{DHFMK2022} also show more sensitivity of the wing-root region to the turbulence models applied, whereas the present results are more consistent in that regard. For all turbulence models at $\alpha \le 19.57^{\circ}$, \cite{DHFMK2022} shows large flow separation on the nacelle similar to Fig~\ref{fig:FA_RANS_Cf_init}(f), which also differs from current observations. 
At $\alpha=21.47^{\circ}$, simulations using QCR corrections in \cite{DHFMK2022} also show massive flow separation near the wing root, which is not observed in present simulation results.
Based on this comparison of simulation results using grids of similar size (Type C grids), there seems to be significant sensitivity of turbulence models to the grid topology and local details of resolution. 

%%%%%%%%%%%%%%%%%%%%%%%%%%%%%%%%%%%%%%%%%%%%%%%%%%%%%%%%%%%%%%%%%%%%%%%%%%%%%%%%%%%%
\subsubsection{Assessment of Iterative Convergence}
Up until this point, we have only looked at the convergence parameter $CPA$, which is based on fluctuations of the aerodynamic coefficients. At increased angles of attack we have found some cases that had a less than optimal $CPA>10^{-4}$, which raises questions about the trustworthiness of the results in that regime. 
% Convergence
\begin{figure}
    \centering
    \begin{subfigure}[b]{0.45\textwidth}
        a) $\alpha=7.05^{\circ}$\\
        \includegraphics[width=\textwidth, trim=10 10 10 50, clip]{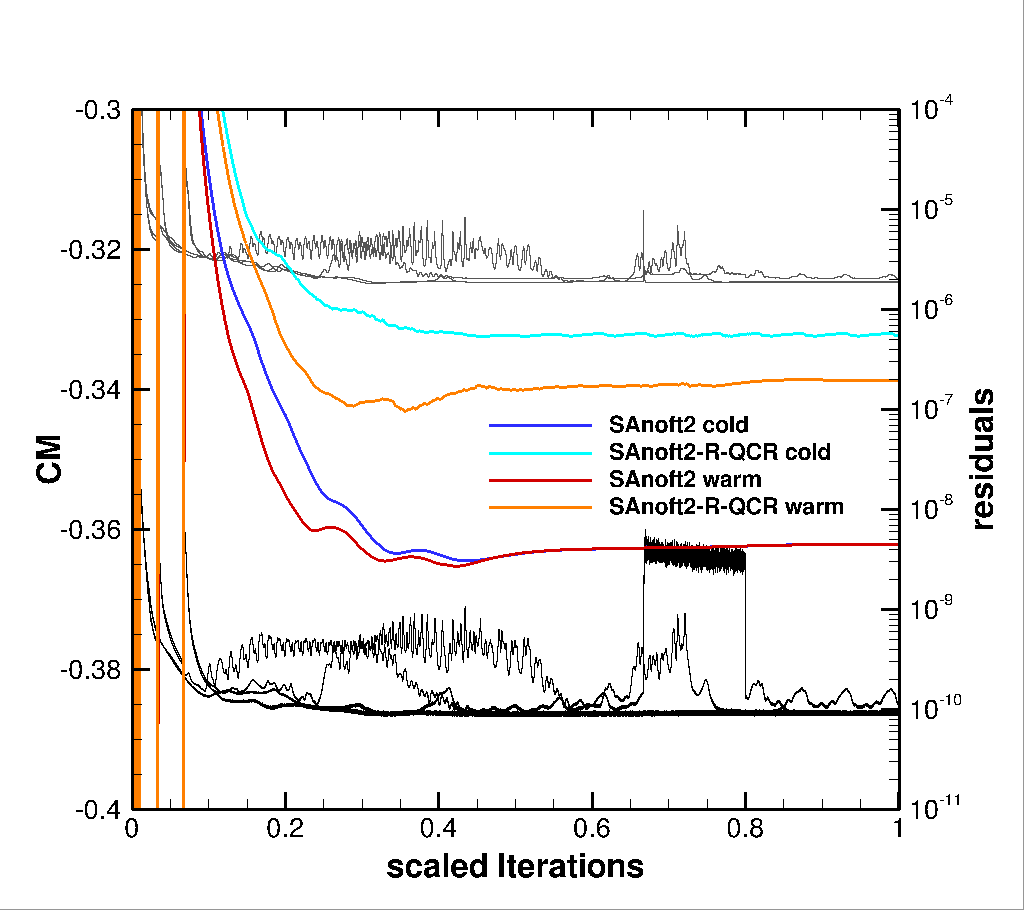}
        %\caption{Subplot 1}
    \end{subfigure}
    \begin{subfigure}[b]{0.45\textwidth}
        b)  $\alpha=17.05^{\circ}$\\
        \includegraphics[width=\textwidth, trim=10 10 10 50, clip]{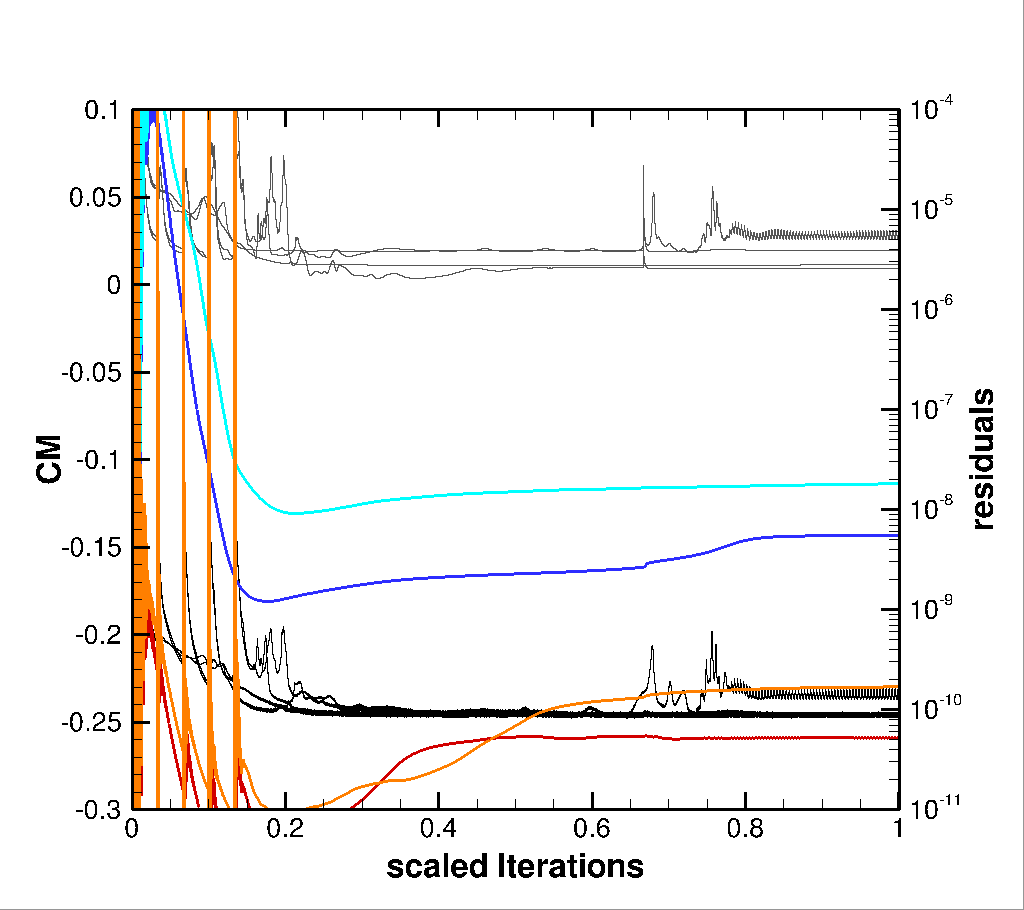}
        %\caption{Subplot 2}
    \end{subfigure}\\
    \vspace{0.5cm} % adjust vertical spacing between rows of subplots
    \begin{subfigure}[b]{0.45\textwidth}
        c)  $\alpha=19.57^{\circ}$\\
        \includegraphics[width=\textwidth, trim=10 10 10 50, clip]{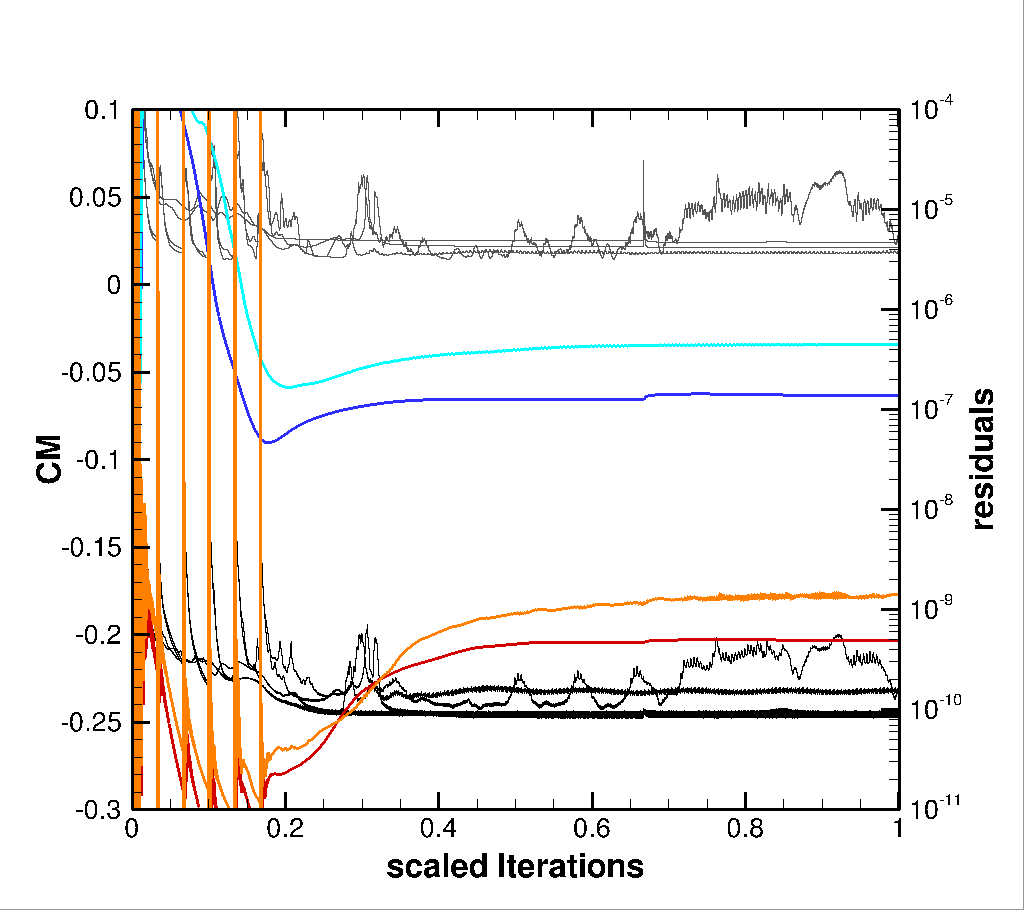}
        %\caption{Subplot 3}
    \end{subfigure}
    \begin{subfigure}[b]{0.45\textwidth}
        d)  $\alpha=21.47^{\circ}$\\
        \includegraphics[width=\textwidth, trim=10 10 10 50, clip]{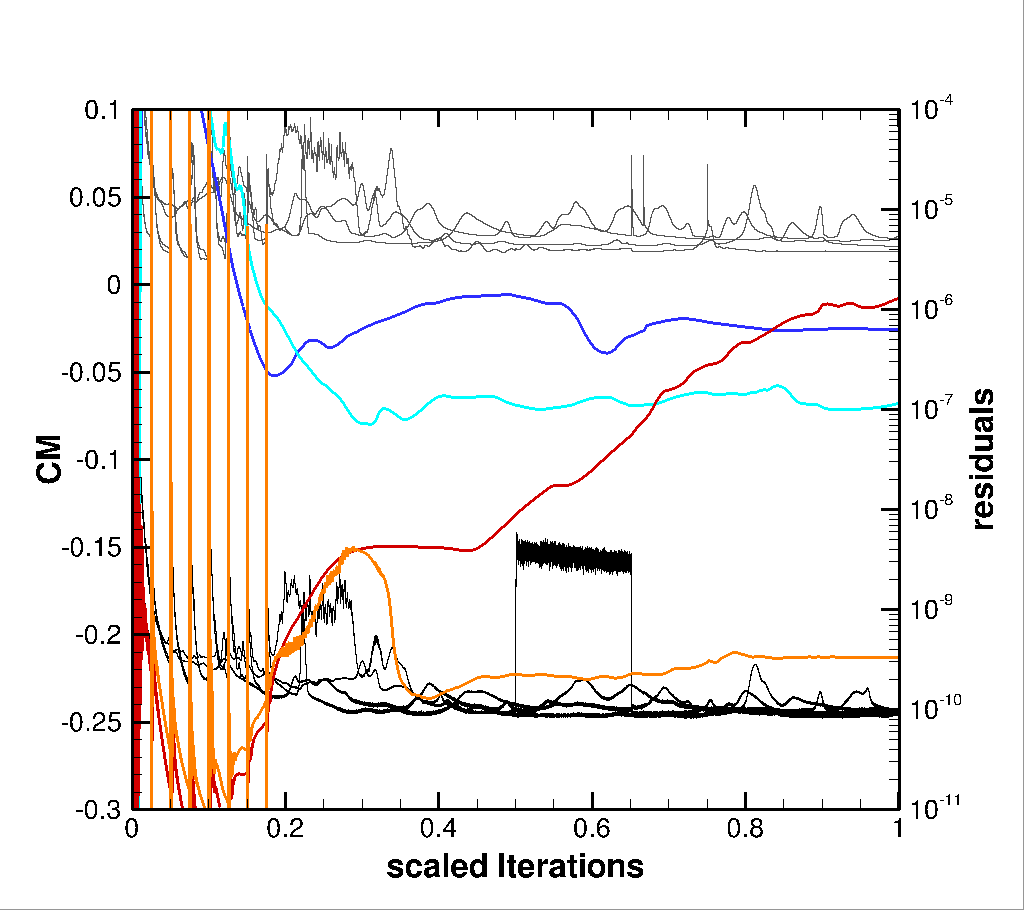}
        %\caption{Subplot 3}
    \end{subfigure}
    \caption{
    % Convergence history for cold- and warm-started RANS results using SA turbulence model for (a) $\alpha=7.05^{\circ}$, (b) $17.05^{\circ}$, (c) $19.57^{\circ}$, and (d) $21.47^{\circ}$. Coloured lines correspond to the left-hand-side scale showing the moment coefficient as a function of scaled iterations. Dark and light grey curves correspond to the right-hand-side scale showing residuals of the u-velocity component and turbulent terms, respectively.
    Coloured lines show $C_M$ as a function of scaled iterations. Dark and light grey curves showing residuals of $u$-velocity and turbulent terms, respectively.
    }\label{fig:FA_RANS_Conv}
\end{figure}
Coloured curves in Fig~\ref{fig:FA_RANS_Conv} correspond to the left-hand-side $y$-axis and show the moment coefficient as a function of iterations normalised by the total number of iterations of each corresponding RANS simulation at different angles of attack. Dark and light grey curves correspond to the right-hand-side scale showing residuals of the u-velocity component (representative of convective terms) and turbulent terms, respectively. For brevity, we show only results obtained with SA turbulence models. While we observe significant differences in the $C_M$ curves and CPA values (corresponding values are summarised in the tables \ref{tab:Sum_SA} and \ref{tab:Sum_SST} of the appendix), residuals reach a similar order of magnitude for all simulations. An attempt to increase the CFL number lead to the sudden increase in residuals of the u-velocity component in Fig~\ref{fig:FA_RANS_Conv}(a) at scaled iteration of approximately $0.65$. After resuming with $CFL=10$, the residuals dropped to their original level after scaled iterations of approximately $0.80$.
For $CPA>10^{-4}$, oscillations of the $C_M$ curves in Fig~\ref{fig:FA_RANS_Conv} are clearly evident. However, despite these oscillations, the mean values remain largely the same. Therefore, it is questionable whether improved convergence would correct the non-physical phenomena present in our results. 

In order to address this question directly by improving the convergence characteristics, we applied a Selective Frequency Damping (SFD) technique \cite{Richez2016,Plante2018,Liguori2021} to the steady RANS simulations using the SA-noft2 turbulence model. Results are summarised in table \ref{tab:Conv_SFD} showing that even though the oscillations of aerodynamic coefficients decreased significantly for results using SFD, their mean values changed only marginally. It should also be noted that the effectiveness of the present SFD approach depends on several parameters and may not always be successfully in improving the convergence characteristics. Additionally, while the oscillation amplitudes of aerodynamic coefficients and residuals are relatively low, SFD solutions do not reach machine-precision convergence. Therefore, we cannot conclude entirely whether RANS results would (theoretically) change when reaching machine precision, but we have demonstrated that they are only marginally affected by the better convergence levels provided by SFD. Further detailed analysis of the effect SFD has on steady RANS simulations is out of scope of the present contribution and will not be considered any further.
\begin{table}
  \begin{center}
\def~{\hphantom{0}}
  \begin{tabular}{r|ccccc|ccccc}
     & \multicolumn{5}{c|}{SA-noft2} & \multicolumn{5}{c}{SA-noft2 with SFD} \\
    AoA & $\log({CPA})$ & $\log(R_{turb})$ & $\overline{C_L}$ & $\overline{C_D}$ & $\overline{C_M}$ & $\log({CPA})$ & $\log(R_{turb})$ & $\overline{C_L}$ & $\overline{C_D}$ & $\overline{C_M}$ \\ 
	\hline
	07.05 & -5.69 & -5.72 & 1.79 & 0.188 & -0.362 & -5.89 & -5.72 & 1.79 & 0.188 & -0.363 \\
	17.05 & -3.91 & -5.41 & 2.31 & 0.369 & -0.143 & -5.87 & -5.39 & 2.31 & 0.368 & -0.146 \\
	19.57 & -4.32 & -5.33 & 2.30 & 0.420 & -0.063 & -5.68 & -5.31 & 2.30 & 0.419 & -0.066 \\
	21.47 & -2.95 & -5.30 & 2.14 & 0.453 & -0.026 & -5.61 & -5.28 & 2.20 & 0.444 & -0.057 \\
  \end{tabular}
  \caption{Summary of cold-started RANS results using SA-noft2 turbulence model with and without Selective Frequency Damping (SFD).}\label{tab:Conv_SFD}
  \label{tab:SFD}
  \end{center}
\end{table}

\subsubsection{Grid-convergence Study \label{sec:grid_study}}
% Grid study
Before concluding this sub-section, it is also important to briefly address sensitivity to the grid resolution for the cold-started RANS simulations. Using the SA-noft2 turbulence model, additional simulations were carried out on refined grids at $\alpha=7.05^{\circ}$ and $17.05^{\circ}$. 
According to the HLPW-4 guidelines, our `Standard RANS grid' is classified as a $C$-level grid, while a `Refined RANS grid' is denoted as a $D$-level RANS grid. Even though the `DDES grid' (provided by ANSA for the HLPW-4) is classified as a $C$-level grid, it is significantly larger compared to the RANS grids and will be also used later for unsteady hybrid RANS simulations in \S\ref{sec:results_freeair_ddes}.

\begin{table}
  \begin{center}
\def~{\hphantom{0}}
  %\begin{tabular}{lcc|ccccc}
  \begin{tabular}{lccccccc}
     Grid & Grid nodes & AoA & $\log({CPA})$ & $\log(R_{turb})$ & $\overline{C_L}$ & $\overline{C_D}$ & $\overline{C_M}$ \\ 
     \hline
     Standard RANS grid & $85,\!969,\!733$ & $7.05^{\circ}$ & -5.69 & -5.72 & 1.79 & 0.188 & -0.362 \\
     Refined RANS grid & $181,\!692,\!364$ & $7.05^{\circ}$ & -4.73 & -5.44 & 1.79 & 0.188 & -0.361 \\
     DDES grid & $218,\!077,\!184$         & $7.05^{\circ}$ & -4.26 & -5.67 & 1.78 & 0.187 & -0.361\\ % SA10
     \hline
     Standard RANS grid & $85,\!969,\!733$ & $17.05^{\circ}$ & -3.91 & -5.41 & 2.31 & 0.369 & -0.143 \\
     Refined RANS grid & $181,\!692,\!364$ & $17.05^{\circ}$ & -4.70 & -5.55 & 2.31 & 0.376 & -0.140 \\
     %DDES grid & $218,\!077,\!184$ & $17.05^{\circ}$ & -2.95 & -4.86 & 2.48 & 0.36 & -0.25\\ % SA11
     DDES grid & $218,\!077,\!184$ & $17.05^{\circ}$ & -3.97 & -4.92 & 2.47 & 0.357 & -0.246 \\ % SA10
     \hline
  \end{tabular}
  \caption{Summary of cold-started RANS results using SA-noft2 turbulence model for standard and refined grids.}\label{tab:GridStudy_FA_RANS}
  \end{center}
\end{table}
Aerodynamic coefficients and convergence characteristics are summarised for all three grids in table \ref{tab:GridStudy_FA_RANS}, considering two angles of attack. 
%The `Refined RANS grid' consists of more than twice the number of grid points compared to the `Standard RANS grid'. 
Using JAXA's in-house tool MEGG3D\cite{Ito2002, Ito2013b}, the `Standard RANS grid' was uniformly refined in each direction by about $33\%$. Despite the `Refined RANS grid' containing twice as many grid nodes compared to the standard RANS grid, we do not observe significant differences in the solution for both angles of attack.

When considering steady RANS results using the DDES grid, however, we observe significant differences in the results at $\alpha = 17.05^{\circ}$, leading to aerodynamic coefficients close to experimental measurements. 
\begin{figure}
    \centering
    \begin{subfigure}[b]{0.48\textwidth}
        a) \\
        \includegraphics[width=\textwidth, trim=10 10 10 300, clip]{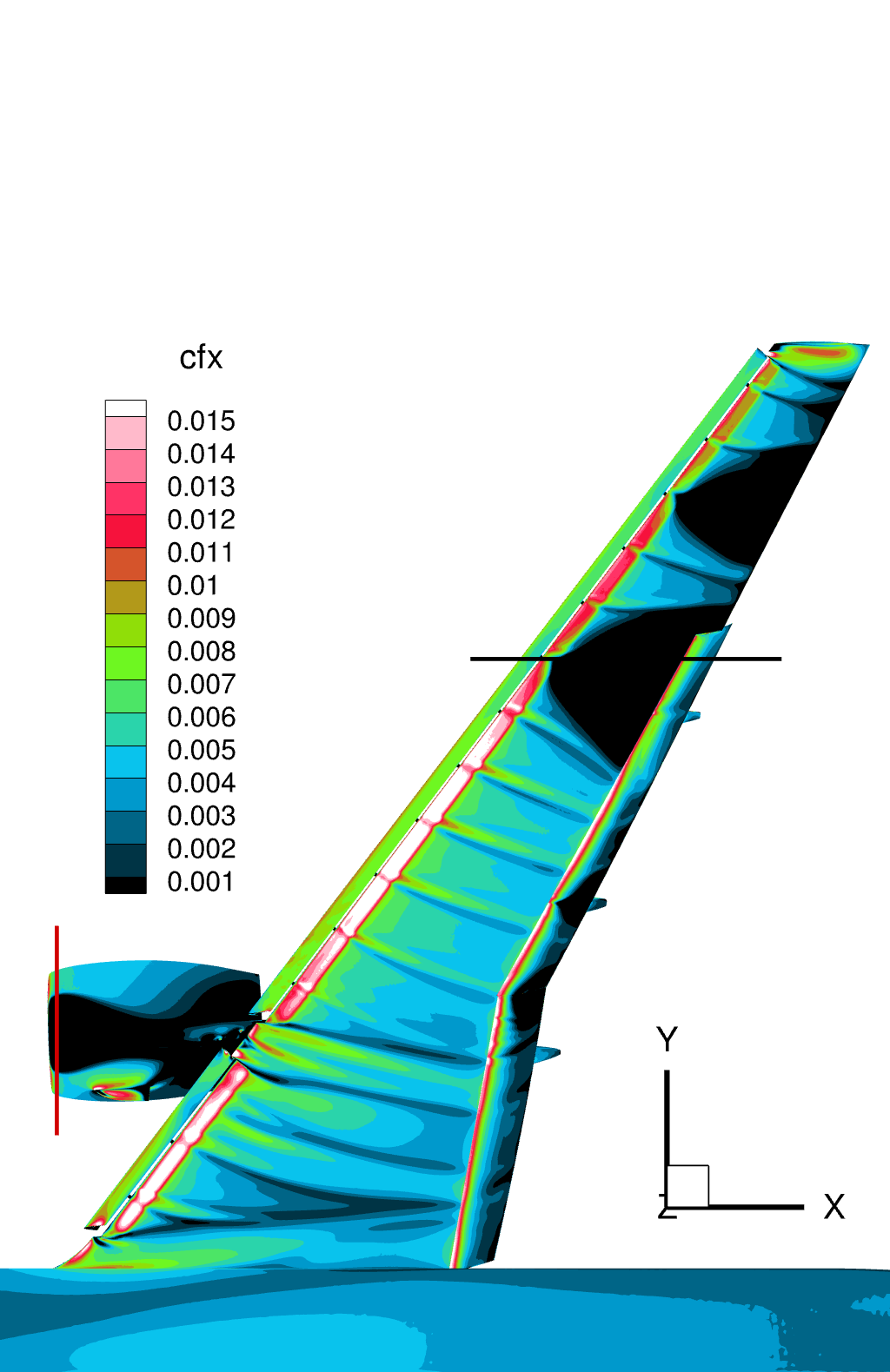}
        %\caption{Subplot 1}
    \end{subfigure}
    \begin{subfigure}[b]{0.48\textwidth}
        b) \\
        \includegraphics[width=\textwidth, trim=10 10 10 300, clip]{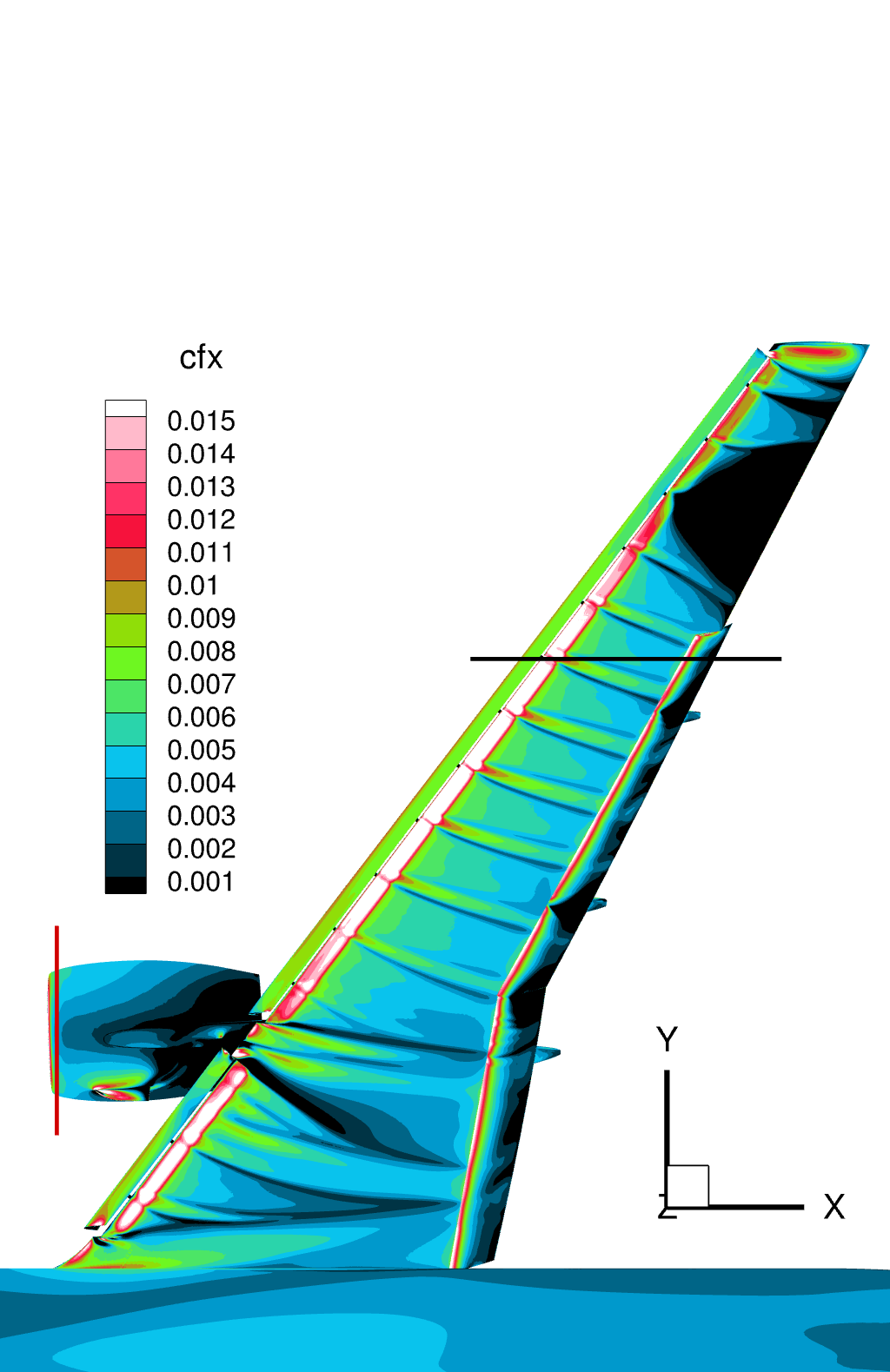}
        %\caption{Subplot 2}
    \end{subfigure}
    \caption{Contours showing skin-friction coefficient $C_f$ for cold-started RANS on the (a) refined RANS grid and (b) a DDES grid at $\alpha = 17.05^{\circ}$, using SA-noft2. Black and red lines indicate locations of cross-sections plotted in figures \ref{fig:FA_RANS_GridStudy_CU1} and \ref{fig:FA_RANS_GridStudy_CU1} of the appendix \S \ref{sec:AppGridRef}}\label{fig:FA_RANS_GridStudy}
\end{figure}
Fig~\ref{fig:FA_RANS_GridStudy} shows the cold-started RANS results at $\alpha = 17.05^{\circ}$ for the (a) `Refined RANS grid' and (b) `DDES grid'. While we observe for both grids large non-physical flow separation near the wing tips, the flow is notably different near the nacelle and near the end of the out-board flaps, where we do not observe the second large region of out-board separation. 
For brevity, figures showing details of the grids at cross-sections corresponding to black (Slat region) and red (upper crest of nacelle) lines are added to the appendix in \S \ref{sec:AppGridRef}. Generally speaking, despite different types of grid topologies being applied, the differences in grid resolutions between the two grids are minor. However, the leading edge of the slat does have higher resolution in the streamwise direction for the DDES grid compared to the refined RANS grid. This affects the evolution of the boundary layer and, subsequently, may lead to the observed differences in separation phenomena on the main wing (where grid resolutions are similar). For the nacelle, finer mesh resolution is present for both the standard and refined RANS grids compared to the DDES grid, which may also lead to the observed differences in $C_f$ contours. The large flow separation on the nacelle in Fig~\ref{fig:FA_RANS_GridStudy}(a) is reminiscent of the results of \cite{DHFMK2022} using overset grids.
It is remarkable how well results for the DDES grid compare with warm-started RANS results on the standard RANS grid shown before in Fig~\ref{fig:FA_RANS_Cf_init}. 
% Based on the present grid-refinement study, we cannot make any strong conclusions. However, 
Given the relatively minor sensitivity of RANS solutions to initialization strategies using over-set grids in \cite{DHFMK2022}, there is a high chance of the sensitivity to initial solutions being grid dependent. 
To strengthen present conclusions, future studies will apply adaptive mesh refinement methods to build upon the work in this section in the future.
%Further investigation on grid sensitivities for RANS solutions is out of scope for the present contribution, but will be subject of futures studies employing adaptive mesh refinement methods. 

%%%%%%%%%%%%%%%%%%%%%%%%%%%%%%%%%%%%%%%%%%%%%%%%%%%%%%%%%%%%%%%%%%%%%%%%%%%%%%%%%%%%
%\subsubsection{Intermediate conclusions}
% % Summary
% In the present subsection, we carried out an extensive study for steady RANS simulations testing different Spalart-Allmaras (SA) and Shear-Stress Transport (SST) turbulence models as well as initialization strategies. Furthermore, we tried to quantify the sensitivities to limited iterative convergence as well as grid resolution.

% For all considered approaches, we observed the appearance of non-physical flow features. Despite fair agreement with experimental force measurements, present steady RANS simulations could not predict flow conditions near $C_{L,max}$ sufficiently accurate. While non-physical flow separation near half-span can be suppressed in warm-started simulations near $C_{L,max}$, all considered turbulence models lead to excessive flow separation near wing tips. At $\alpha=21.47^{\circ}$, all simulations show massive non-physical regions of flow separation either around half span or behind the nacelle. Considering reference literature (e.g. \cite{DHFMK2022}), we can confirm rather arbitrary appearance of spurious flow separated, which appears to be highly sensitive to grid resolution and topology.

\subsubsection{Quantification of Sensitivities}
In this section we quantify the assessed sensitivities in terms of variations of absolute values of aerodynamic coefficients. Details of all present simulations are provided in the appendix \ref{App:FA_RANS} in tables \ref{tab:Sum_SA} and \ref{tab:Sum_SST}. 
Table \ref{tab:sensitivities_RANS_FA} lists differences in $C_L$, $C_D$, and $C_M$ for selected angles of attack of $\alpha=7.05^{\circ}$, $17.05^{\circ}$, $19.57^{\circ}$, and $21.47^{\circ}$, where $C_{L,max}$ is expected near $\alpha=19.57^{\circ}$.
The first segments of $\Delta$-values shows the spread across different turbulence models. %The values in brackets denote the spread considering Spalart-Allmaras turbulence models only. %We should emphasise at this point that here we are only comparing the results from the present study. 
% A reduced spread in results allows no conclusion about the physical correctness of results. 
Considering present results using Spalart-Allmaras turbulence model (values in brackets in table \ref{tab:sensitivities_RANS_FA}), we can see even at $\alpha=17.05^{\circ}$ a moderate spread of results.
However, we have seen in previous sub-sections that looking only at popular integral quantities such as aerodynamic coefficients can be misleading, when concluding on the physical correctness of results.
This becomes particularly evident at $\alpha=21.47^{\circ}$, where simulation results using SA turbulence models agree well with each other, but we have shown that they are far from accurate.
Near $C_{L,max}$, we observe a similar spread with respect to turbulence models and initialization strategies of $\Delta C_L \approx 0.25$ corresponding to a variability of $10\%$ with respect to experimental measurements. In the worst case that means an error of approximately $20\%$, assuming grid convergence and sufficient iterative convergence. At reduced angles of attack, however, steady RANS solutions appear fairly consistent in terms of the choice of turbulence models, while variations remain high for the choice of initialization. At $\alpha=7.05^{\circ}$, steady RANS predictions show good convergence characteristics and agree very well among each other as well as with experimental measurements.
\begin{table}
  \begin{center}
\def~{\hphantom{0}}
  \begin{tabular}{r|ccc|ccc|ccc}
  		& \multicolumn{3}{c|}{Turbulence Model} & \multicolumn{3}{c|}{Initialisation} & \multicolumn{3}{c}{Convergence}\\
    AoA & $\Delta \overline{C_L}$ & $\Delta \overline{C_D}$ & $\Delta \overline{C_M}$ 
    	& $\Delta \overline{C_L}$ & $\Delta \overline{C_D}$ & $\Delta \overline{C_M}$ 
    	& $\Delta \overline{C_L}$ & $\Delta \overline{C_D}$ & $\Delta \overline{C_M}$ \\ 
	\hline
	07.05 &  0.08 (0.07) & 0.01 (0.01) & 0.04 (0.04) & 0.03 & 0.00 & 0.01 & 0.00 & 0.00 & 0.00 \\
	% SA min 1.73 0.18 -0.37 SST min 1.72 0.18 -0.33 ColdWarm SA10 0 0 0
	% SA max 1.80 0.19 -0.33 SST max 1.73 0.18 -0.33 ColdWarm SA12 0.03 0 0.01
	17.05 &  0.10 (0.08) & 0.01 (0.00) & 0.10 (0.03) & 0.22 & 0.02 & 0.12 & 0.00 & 0.00 & 0.01\\
	% SA min 2.22 0.37 -0.14 SST min 2.20 0.36 -0.12 ColdWarm SA10 0.17 0.02 0.12
	% SA max 2.30 0.37 -0.11 SST max 2.26 0.37 -0.04 ColdWarm SA12 0.22 0.02 0.12
	19.57 &  0.27 (0.19) & 0.02 (0.01) & 0.10 (0.04) & 0.24 & 0.04 & 0.15 & 0.00 & 0.00 & 0.01\\
	% SA min 2.13 0.41 -0.07 SST min 2.05 0.41 -0.04 ColdWarm SA10 0.24 0.03 0.14
	% SA max 2.32 0.42 -0.03 SST max 2.22 0.43  0.03 ColdWarm SA12 0.22 0.04 0.15
	21.47 &  0.38 (0.05) & 0.06 (0.03) & 0.16 (0.05) & 0.18 & 0.01 & 0.14 & 0.06 & 0.01 & 0.03\\
    % SA min 2.09 0.43 -0.08 SST min 1.76 0.41 -0.08 ColdWarm SA10 0.17 0.01 0.02
	% SA max 2.14 0.46 -0.03 SST max 2.07 0.47  0.08 ColdWarm SA12 0.18 0.00 0.14
  \end{tabular}
  \caption{Sensitivities of RANS results to the choice of turbulence model, initialisation strategy, and level of iterative convergence.}%\label{tab:FA_RANS_Sens}
  \label{tab:sensitivities_RANS_FA}
  \end{center}
\end{table}

%Regarding to the grid sensitivity, we need to be careful drawing strong conclusions. 
% Considering \S \ref{sec:grid_study} and reference literature, we could show that grid refinement can reduce sensitivities to initial solutions.
At reduced angles of attack, we observe remarkably good agreement across results using different grids in terms of size as well as topology. 
%Despite careful manual grid refinement, however, present results as well as those reported in literature do not manage to overcome the appearance of non-physical flow phenomena near $C_{L,max}$. 
Considering \S \ref{sec:grid_study} and reference literature, the present work shows that even careful manual grid refinement struggles to overcome the appearance of non-physical flow phenomena near $C_{L,max}$. However, given the present results, it seems that targeted grid refinement can reduce sensitivities to initial solutions.
In the light of \cite{Alauzet2022} on the present test case, adaptive mesh refinement may reduce sensitivities to initialization strategies and grid resolution in the future, but uncertainties associated with turbulence models for physically accurate $C_{L,max}$ prediction remain. Results obtained by adaptive mesh refinement across different participants of the HLPW-4 appear fairly consistent, but always lead to questionable `pizza slice' patterns of flow separation near the wing tip.
Further increasing the grid size and the number of refinement cycles may help to improve the solutions, but compromise the cost-effectiveness of RANS.

\subsection{Sensitivities of Unsteady RANS Methods \label{sec:results_freeair_urans}}
In this sub-section, we want to determine if unsteady RANS simulations yield improved results, considering comparable computational resources to steady RANS simulations. We are particularly interested in the sensitivities of URANS simulations to initialization strategies and SA turbulence model choices. However, before we will assess sensitivities to the numerical URANS setup.

\subsubsection{Sensitivity Study to Time-integration Parameters}
The baseline parameter set-up for production runs of the present URANS study consists of a CFL number of $10$, a time step of $\Delta \tau = 0.0002$ CTUs, and $n_{i}=5$ inner iterations.
\begin{table}
  \begin{center}
\def~{\hphantom{0}}
  %\begin{tabular}{ccc|cccc}
  \begin{tabular}{ccccccc}
    CFL & Inner iterations & Time step & $\overline{C_L}$ & $C'_{L,rms}$ & $\overline{C_D}$ & $\overline{C_M}$\\ 
    \hline
    1     & 5     & 0.0002   & 2.54 & 1.03e-3 & 0.389 & -0.207  \\
    10    & 5     & 0.0002   & 2.54 & 4.64e-3 & 0.387 & -0.215  \\
    50    & 5     & 0.0002   & 2.54 & 4.08e-3 & 0.388 & -0.210  \\
    \hline
    50    & 5     & 0.0002   & 2.54 & 4.08e-3 & 0.388 & -0.210  \\
    50    & 5     & 0.0004   & 2.54 & 3.31e-3 & 0.388 & -0.205  \\   
    \hline
    50    & 5     & 0.0002   & 2.54 & 4.08e-3 & 0.388 & -0.210  \\
    50    & 10    & 0.0002   & 2.55 & 2.53e-3 & 0.386 & -0.218  \\
    50    & 20    & 0.0002   & 2.54 & 5.39e-3 & 0.385 & -0.211  \\
    50    & 20    & 0.0020   & 2.54 & 1.01e-3 & 0.389 & -0.205  \\
  \end{tabular}
  \caption{Summary of cold-started URANS results using the SA-noft2 turbulence model for different parameter settings at $\alpha=19.57^{\circ}$. Mean values are averaged over the final $30,\!000$ iterations.}\label{tab:FA_URANS_Sensitivity}
  \end{center}
\end{table}
Table \ref{tab:FA_URANS_Sensitivity} shows the results of the time-stepping sensitivity study, varying the CFL number, the maximum number of inner iterations $n_i$, as well as the time step $\Delta \tau$, using the baseline Spalart-Allmaras SA-noft2 model at $\alpha=19.57^{\circ}$. Given the trend of decaying unsteadiness (as shown later in Fig~\ref{fig:FA_URANS_CL}), simulations in this sensitivity study are only run for approximately $60,\!000$ iterations (corresponding to approximately $12-24$ CTUs, depending on the time step). Although oscillation amplitudes vary between the cases tested, mean values collected for the last $30,\!000$ iterations ($\approx 6-12$ CTUs) agree very well within a parameter space of $1<CFL<50$, $5<n_i<20$, and $0.0002<\Delta \tau<0.0004$. For the present standard RANS grid, $\Delta \tau > 0.0004$ or $CFL>50$ lead to numerical instabilities for $n_{i}=5$. Increasing the maximum number of inner iterations would allow for larger time steps to be taken, but even though mean results do not seem to be significantly affected, oscillation amplitudes decay even faster. For the example in table \ref{tab:FA_URANS_Sensitivity}, we observe for $n_i=20$ and $\Delta \tau=0.002$ significantly lower values for $C_{L,rms}'$ compared to our baseline setup using $n_i=5$ and $\Delta \tau=0.0002$.

Based on these sensitivity tests, a baseline set-up for production runs was selected according to the simulation with maximum unsteadiness (i.e. maximum $C'_{L,rms}$ values), using a CFL number of $10$, a time step of $\Delta \tau = 0.0002$ CTUs, and $n_{i}=5$ inner iterations.
% Given the increased cost of URANS without showing significant improvement compared to RANS, we did not extend this numerical experiment and limit our URANS study. %to sensitivities to initialization strategies and turbulence model choices.
Given the increased cost of URANS relative to RANS without showing significant improvement in accuracy, the URANS study is limited to the parameters shown here.

\subsubsection{Sensitivity of URANS to Initialization Strategies and Turbulence Models}
\begin{figure}
    \centering
    \begin{subfigure}[b]{0.33\textwidth}
        a) \\
        \includegraphics[width=\textwidth, trim=10 10 50 10, clip]{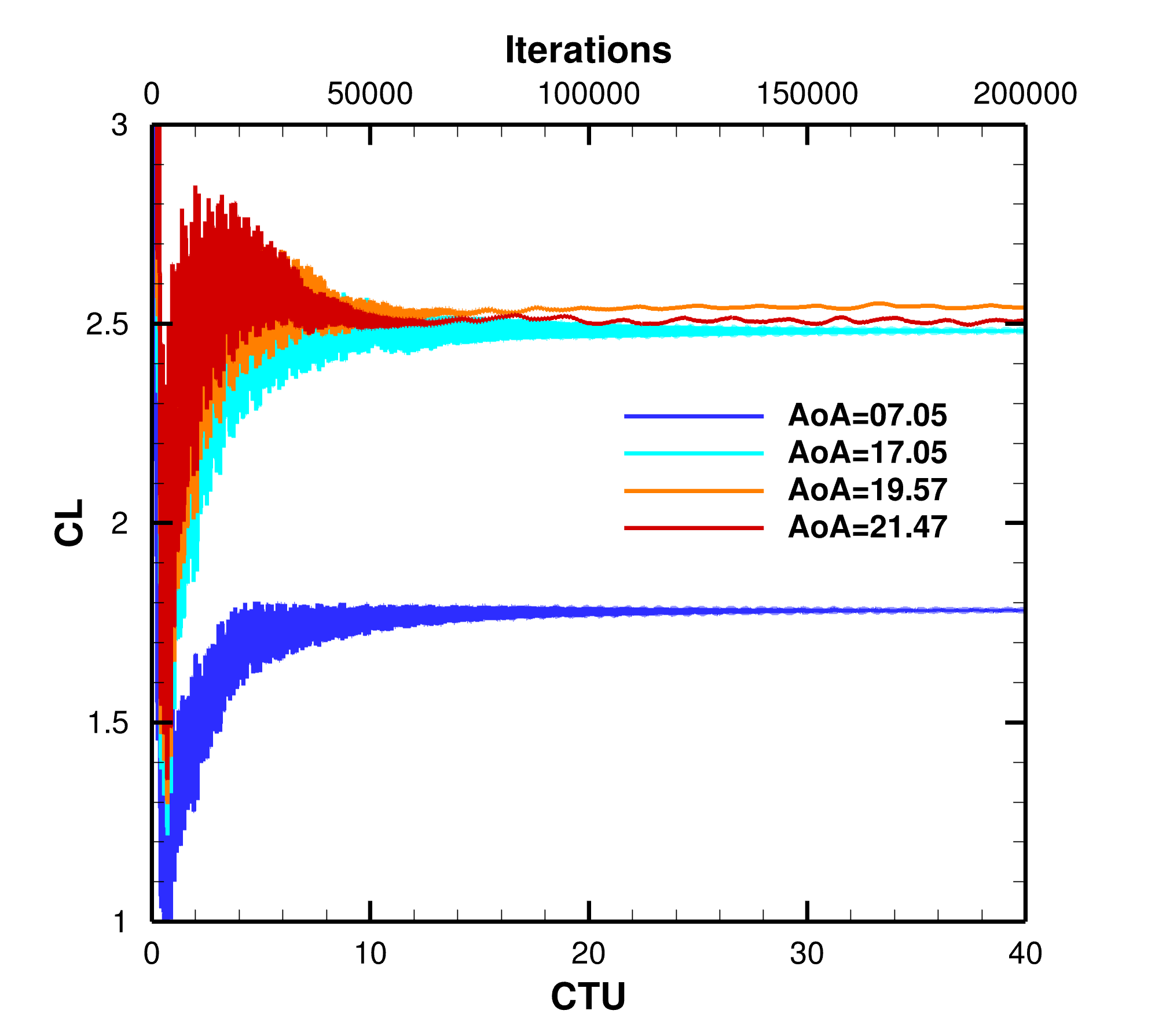}
        %\caption{Subplot 1}
    \end{subfigure}
    \begin{subfigure}[b]{0.33\textwidth}
        b) \\
        \includegraphics[width=\textwidth, trim=10 10 50 10, clip]{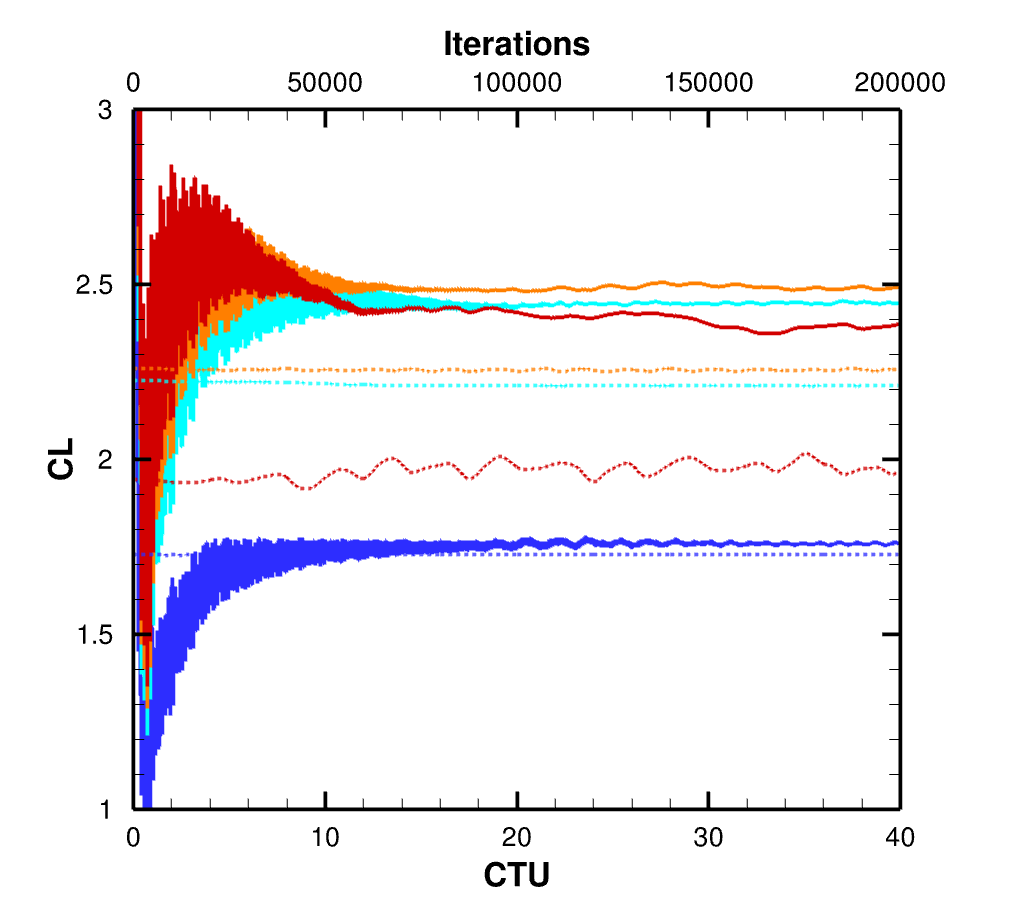}
        %\caption{Subplot 2}
    \end{subfigure}
    \begin{subfigure}[b]{0.33\textwidth}
        c) \\
        \includegraphics[width=\textwidth, trim=10 10 50 10, clip]{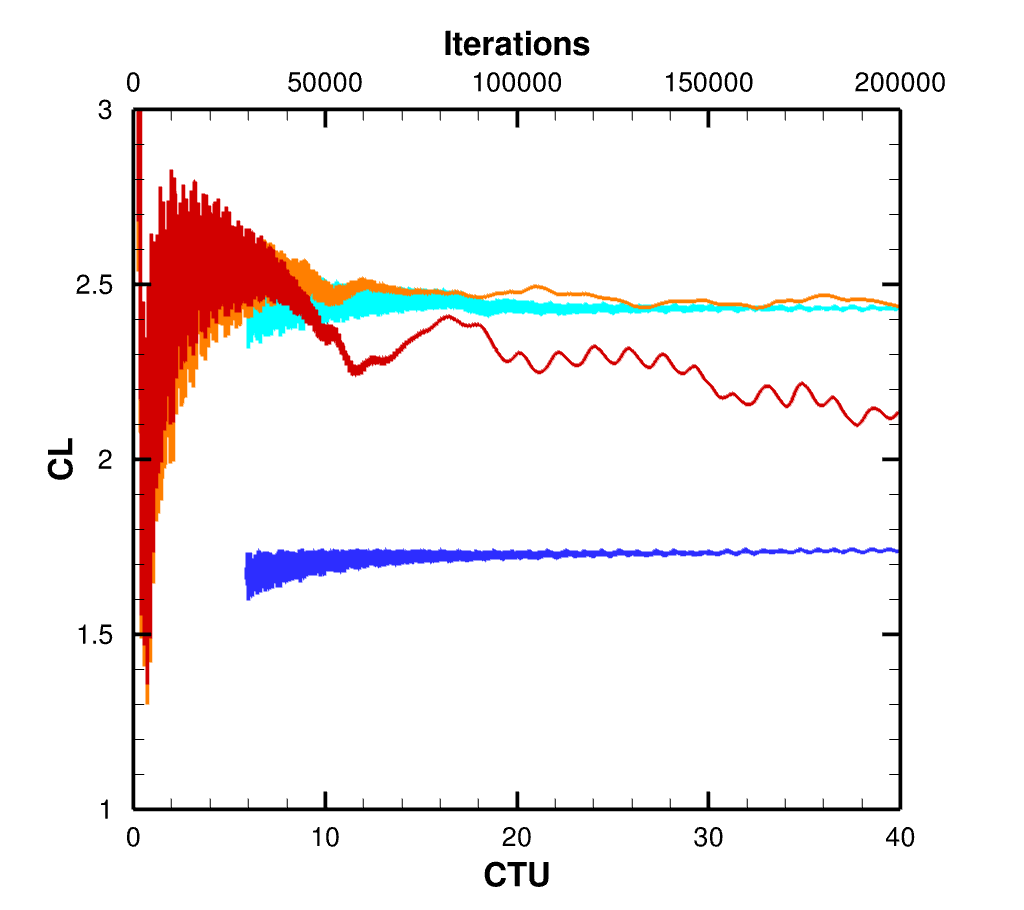}
        %\caption{Subplot 3}
    \end{subfigure}
    \caption{
    %Histories of lift coefficient for URANS simulations using (a) SA-noft2, (b) SA-noft2-R-QCR2000, and (c) SST turbulence models. Bottom and top horizontal axes correspond to convective time units (CTU) and iterations, respectively. Solid lines denote cold-started simulations, while dashed lines in (b) correspond to warm-started simulations. Mean values are summarised in table \ref{tab:URANS}.
    $C_L$ for URANS simulations using (a) SA-noft2, (b) SA-noft2-R-QCR2000, and (c) SST turbulence models. Solid and dashed lines denote cold-started and warm-started simulations (only in (b)), respectively.
    }\label{fig:FA_URANS_CL}
\end{figure}
Fig~\ref{fig:FA_URANS_CL} shows histories of lift coefficient for URANS simulations using (a) SA-noft2, (b) SA-noft2-R-QCR2000, and (c) SST turbulence models. The bottom and top horizontal axes correspond to convective time units and iterations respectively, to allow for cost comparison to RANS\footnote{Note that each (time) iteration requires 5 sub-iterations.}. Solid lines denote cold-started simulations, while dashed lines in (b) correspond to warm-started simulations. We obtain very good agreement between the cold-started test cases. Simulations which are warm-started from cold-started RANS solutions in Fig~\ref{fig:FA_URANS_CL}(b) (denoted by dotted curves) seem to be locked on the initial solution and show similar discrepancies when compared to the RANS study of the previous sub-section. Due to their poor performance, warm-started URANS simulations have not been carried out for the other turbulence models. For the remainder of this sub-section, we will focus on cold-started URANS simulations only. 
The baseline Spalart-Allmaras turbulence model \verb|SA-noft2| performs best at pre-stall conditions based on the cases shown in table \ref{tab:URANS}, whereas the R-QCR correction leads to increased deviations of $C_M$ compared to experiments near $C_{L,max}$.
However, for all present URANS results we shall see that we encounter similar problems as reported for our best-practice steady RANS.

\begin{table}
  \begin{center}
\def~{\hphantom{0}}
  %\begin{tabular}{lllr|cc|ccc}
  \begin{tabular}{lllrccccc}
    $\alpha$ & Type & Turbulence model & Initialization & $\overline{C_L}/C_{L,exp}$ & $\overline{C_M}/C_{M,exp}$ & $\overline{C_L}$ & $\overline{C_D}$ & $\overline{C_M}$\\ 
    \hline
    %$07.05^{\circ}$ & \multicolumn{3}{l|}{Experiment} & 1.778620005 & 0.186706007 & -0.370599985 \\
    $07.05^{\circ}$ & URANS & SA-noft2 & cold-started           & 1.00 & 0.97 & 1.78 & 0.187 & -0.357\\
    $07.05^{\circ}$ & URANS & SA-noft2-R-QCR & cold-started     & 0.99 & 0.92 & 1.76 & 0.184 & -0.339 \\
    $07.05^{\circ}$ & URANS & SST & cold-started                & 0.97 & 0.89 & 1.73 & 0.181 & -0.330\\
    $07.05^{\circ}$ & URANS & SA-noft2-R-QCR & warm-started     & 0.97 & 0.89 & 1.73 & 0.180 & -0.331\\
    \hline
    %$17.05^{\circ}$ & \multicolumn{3}{l|}{Experiment} & 2.495490074 & 0.332423002 & -0.287313014
    $17.05^{\circ}$ & URANS & SA-noft2 & cold-started           & 0.99 & 0.90 & 2.48 & 0.355 & -0.259\\
    $17.05^{\circ}$ & URANS & SA-noft2-R-QCR & cold-started     & 0.98 & 0.83 & 2.45 & 0.350 & -0.240\\
    $17.05^{\circ}$ & URANS & SST & cold-started                & 0.97 & 0.80 & 2.43 & 0.345 & -0.234\\
    $17.05^{\circ}$ & URANS & SA-noft2-R-QCR & warm-started     & 0.88 & 0.31 & 2.21 & 0.373 & -0.095\\   
    \hline
    %$19.57^{\circ}$ & \multicolumn{3}{l|}{Experiment} &  2.514909983 & 0.361550003 & -0.244862005
    $19.57^{\circ}$ & URANS & SA-noft2 & cold-started           & 1.01 & 0.86 & 2.54 & 0.386 & -0.207\\
    $19.57^{\circ}$ & URANS & SA-noft2-R-QCR & cold-started     & 0.99 & 0.74 & 2.49 & 0.382 & -0.176\\
    $19.57^{\circ}$ & URANS & SST & cold-started                & 0.98 & 0.61 & 2.46 & 0.379 & -0.144\\
    $19.57^{\circ}$ & URANS & SA-noft2-R-QCR & warm-started     & 0.90 & 0.12 & 2.26 & 0.421 & -0.029\\    
    \hline
    %$21.47^{\circ}$  & \multicolumn{3}{l|}{Experiment} &  2.318239927 & 0.436477989 & -0.303606004
    $21.47^{\circ}$ & URANS & SA-noft2 & cold-started           & 1.08 & 0.49 & 2.51 & 0.407 & -0.150\\
    $21.47^{\circ}$ & URANS & SA-noft2-R-QCR & cold-started     & 1.03 & 0.36 & 2.39 & 0.413 & -0.100\\
    $21.47^{\circ}$ & URANS & SST & cold-started                & 0.94 & 0.03 & 2.18 & 0.437 & -0.012\\
    $21.47^{\circ}$ & URANS & SA-noft2-R-QCR & warm-started     & 0.85 & 0.23 & 1.98 & 0.442 & -0.066\\
    \end{tabular}
  \caption{Summary of URANS results for selected turbulence models. 
  %In addition to the aerodynamic coefficients, $C_L$ and $C_M$ ratios of simulation results over experimental values are chosen as performance indicators. Mean values are averaged over the last $100,\!000$ iterations corresponding to 20 CTUs.
  Mean values are averaged over the last 20 CTUs.}\label{tab:URANS}
  \end{center}
\end{table}
% Table \ref{tab:URANS} shows for present URANS simulations the ratio between mean lift- and moment-coefficient over corresponding experimental measurements in addition to their mean values. All cold-started URANS simulation show lift coefficients reasonably close to experimental measurements and warm-started steady RANS results. The moment coefficient is also well predicted for low angles of attack, but the errors increase with $\alpha$. Even though our baseline Spalart-Allmaras turbulence model \verb|SA-noft2| performs best at pre-stall conditions compared to the cases presented in table \ref{tab:URANS}, we observe similar problems as reported for steady RANS.

%\subsubsection{Sensitivities to initialisation strategies}
\begin{figure}
    \centering
    \begin{subfigure}[b]{0.48\textwidth}
        a) \\
        \includegraphics[width=\textwidth, trim=10 10 10 10, clip]{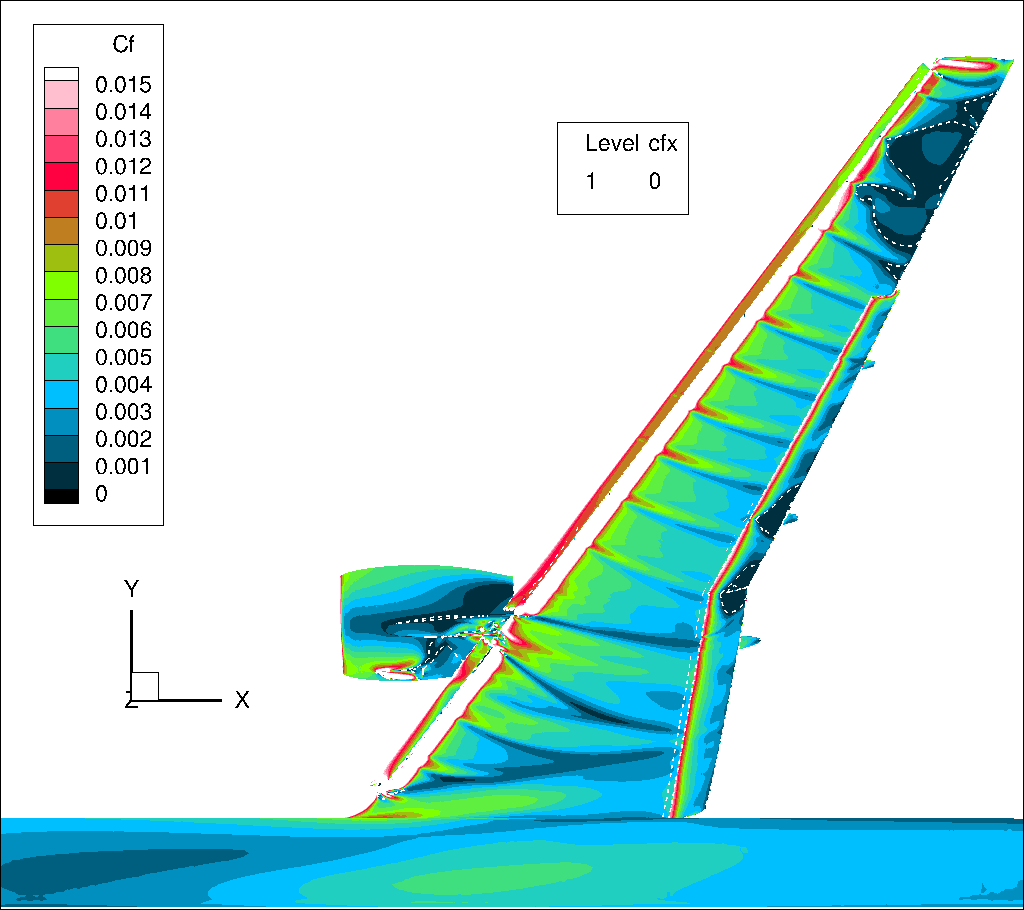}
        %\caption{Subplot 1}
    \end{subfigure}
    \begin{subfigure}[b]{0.48\textwidth}
        b) \\
        \includegraphics[width=\textwidth, trim=10 10 10 10, clip]{Figures_FA_RANS/Cf/SA10_warm_19p57.png}
        %\caption{Subplot 2}
    \end{subfigure}
    \caption{Contours showing skin-friction coefficient $C_f$ for (a) current best-practice unsteady (cold-started) and (b) steady (warm-started) RANS simulations at $\alpha = 19.57^{\circ}$, using SA-noft2.}\label{fig:FA_RANS_vs_URANS_CL}
\end{figure}
Fig~\ref{fig:FA_RANS_vs_URANS_CL} shows a comparison of skin-friction contours for (a) cold-started URANS (instantaneous snapshot\footnote{No statistical data has been collected for present URANS simulations. Even though oscillations are expected in the wake, due to the mild temporal variations it is sufficient to show a representative instantaneous snapshot of the wing surface.}) and (b) warm-started RANS simulations (using our best-practices for each of the respective methods), with the \verb|SA-noft2| turbulence model at $\alpha = 19.57^{\circ}$. Both methods exhibit non-physical large separation regions at the out-board region near stall conditions. While URANS is significantly more expensive compared to the steady RANS, it does not show improvement in resolving the issue of nonphysical separation. %The qualitative similarity between both contour plots near the nacelle and wing root is remarkable.
It is interesting to note that steady RANS delivers similar results (especially near the wing root and nacelle) as the present URANS approach, for significantly lower costs and complexity.

We have analysed URANS simulations to understand the sensitivity they exhibit to choice of numerical parameters (see table \ref{tab:FA_URANS_Sensitivity}). Present results agree well with the study of \cite{DHFMK2022}, who similarly only reported mild temporal variations of aerodynamic loads over time. Furthermore, in a similar manner to the present contribution, the authors of \cite{DHFMK2022} also highlighted the close similarities between URANS and steady RANS results. Based on these findings, we recommend cold-started URANS using the SA-noft2 turbulence model as the 
%best practice for the present study using the numerical set-up mentioned above.
most reliable and accurate method, and should be used as a best practice for future studies on similar configurations to those simulated here.
% Even though the R-QCR correction also delivers reasonable results, the baseline SA-noft2 model matches aerodynamic coefficients of experiments slightly better near $C_{L,max}$.
%However, 
With regards to comparison between the choice of using either RANS or URANS, unsteady RANS do not lead to significant improvements compared to steady RANS.
While some of the URANS cases in this study show mild variations in time, detailed assessment of unsteady flow phenomena will instead be examined in the next section in the context of hybrid RANS/LES methods where they are more prominent.

%The parameter-set of present URANS simulations has been carefully analysed and the time step was limited to relatively small values due to numerical problems for the present grid. 
% Furthermore, these URANS simulations have not been run for long enough to draw any conclusions on unsteady phenomena and should be considered as a numerical experiment. However, present results agree well with \cite{DHFMK2022} reporting virtually indiscernible temporal variations of aerodynamic loads over time, which are similar to steady RANS results.

% Based on these results, we recommend cold-started URANS without SA correction as best practice for the present study using the numerical set-up mentioned above. However, unsteady RANS do not lead to significant improvement compared to steady RANS and become more expensive than DDES. Therefore, we will move on to hybrid RANS/LES simulations in the following sub-section, where we will also assess also unsteady flow phenomena.  

\subsection{Sensitivities of Hybrid RANS/LES (DDES) Methods \label{sec:results_freeair_ddes}}

% Sensitivities to initialization strategy
Having demonstrated the significant limitations of pure RANS-based methods for simulating the CRM-HL configuration near stall conditions ($C_{L,max}$), we now extend the sensitivity study to scale-resolving unsteady methods using Delayed Detached-Eddy Simulations (DDES), which is a hybrid RANS/LES approach. Unless stated otherwise, all simulations in this section are performed with the SA-noft2-R turbulence model and a C-level grid provided by ANSA for HLPW4 (103-Case2a\footnote{\url{https://hiliftpw.larc.nasa.gov/Workshop4/grids_downloads.html}}) consisting of 218 million nodes (276 million cells).
As a first step, we will investigate unsteady characteristics observed during cold-started DDES. For a representative test case at an angle of attack of $\alpha = 17.05^{\circ}$, we will compare between cold- and warm-started simulations and comment on the statistical convergence. We will then focus on sensitivities of the mean-flow characteristics to initial conditions, and compare results to those presented in the previous sub-sections and experiments. 
Eventually, we will investigate differences between DDES results using SA- or SST-based turbulence models.

\begin{table}
  \begin{center}
\def~{\hphantom{0}}
    %\begin{tabular}{cccc|cc|cccc}
  \begin{tabular}{cccccccccc}
%    \multicolumn{4}{c}{Settings} & \multicolumn{2}{c}{Performance} & \multicolumn{4}{c}{Statistics of Aerodyn. Coeff.}\\
    $\alpha$ & Init. & Turb. & Sampling Interval & $\overline{C_L}/C_{L,exp}$ & $\overline{C_M}/C_{M,exp}$ & $\overline{C_L}$ & $C'_{L,rms}$ & $\overline{C_D}$ & $\overline{C_M}$\\ 
    \hline
    7.05    & cold & SA & $10.9<t<47.1$ & 0.92 & 0.81 & 1.64   &  2.97e-3 & 0.167 & -0.299  \\
    7.05    & warm & SA & $10.9<t<47.1$ & 0.92 & 0.81 & 1.64   &  4.87e-3 & 0.168 & -0.300  \\ 
    \hline
    17.05   & cold & SA & $36.3<t<72.5$    & 0.96 & 0.88 & 2.39  & 1.02e-2   & 0.339  & -0.253 \\
    17.05   & warm & SA & $36.3<t<72.5$    & 0.97 & 0.87 & 2.42  & 7.03e-3   & 0.336  & -0.250 \\
    17.05   & warm & SA & $116.0<t<170.4$  & 0.97 & 0.87 & 2.419  & 6.47e-3   & 0.336  & -0.250 \\
    \hline
    19.57   & cold & SA & $36.3<t<72.5$    & 0.93 & 1.00 & 2.35  & 1.69e-2   & 0.371  & -0.245 \\
    19.57   & warm & SA & $36.3<t<72.5$    & 0.94 & 1.01 & 2.36  & 2.49e-2   & 0.373  & -0.248 \\
    % 19.57   & warm from scratch       &      &      & 2.31  & 1.18e-2   & 0.36  & -0.24 \\
    19.57   & cold & SST    & $36.3<t<72.5$    & 0.91 & 0.97 & 2.30  & 2.47e-2   & 0.373  & -0.238 \\
    \hline  
    21.47   & cold & SA & $36.3<t<72.5$      & 0.94 & 1.17 & 2.14  & 3.28e-2   & 0.453  & -0.355 \\
    21.47   & warm & SA & $36.3<t<72.5$      & 0.95 & 1.12 & 2.17  & 7.88e-2   & 0.454  & -0.340 \\
  \end{tabular}
  \caption{Summary of cold- and warm-started DDES results.}\label{tab:FA_DDES_Sensitivity}
  \end{center}
\end{table}
%As a standard procedure (denoted as `warm interpolated'), warm-started simulations used steady RANS solutions from \S \ref{sec:results_freeair_rans}, which were interpolated on the present Hybrid RANS/LES mesh. 
%Alternatively, we also tested warm starts from a RANS simulation, which was cold-started from scratch on the present Hybrid RANS/LES mesh (denoted as `warm from scratch').
%While we will focus in the present sub-section on mean-flow properties, we will discuss unsteady flow characteristics of free-air simulations together with in-tunnel configurations later in sub-section \ref{sec:results_freeair_vs_intunnel}.
\begin{figure}
    \centering
    \begin{subfigure}[b]{0.48\textwidth}
        a) \\
        \includegraphics[width=\textwidth, trim=10 10 100 50, clip]{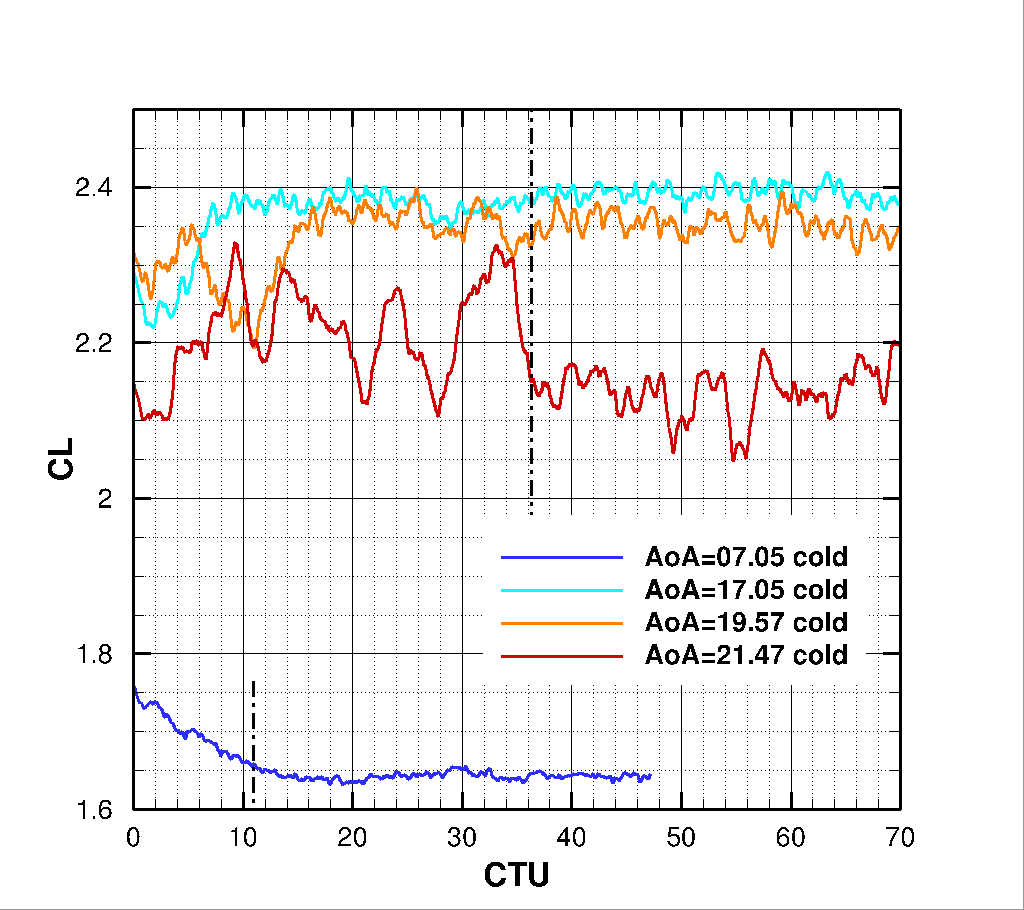}
        %\caption{Subplot 1}
    \end{subfigure}
    \begin{subfigure}[b]{0.48\textwidth}
        b) \\
        \includegraphics[width=\textwidth, trim=10 10 100 50, clip]{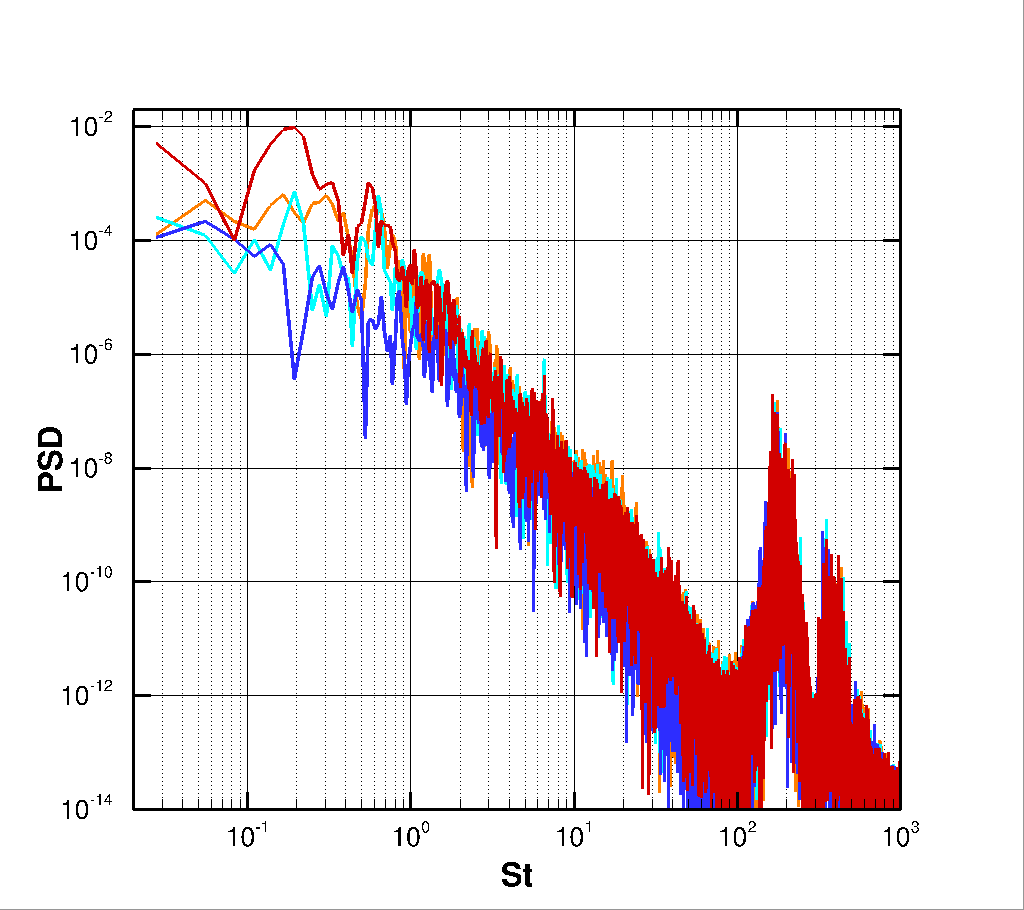}
        %\caption{Subplot 2}
    \end{subfigure}
    \caption{(a) $C_L$ histories of cold-started DDES and (b) corresponding $PSD$ for $0.1<St<1000$, considering a
    %A single Hanning window is applied to the 
    sampling interval starting at the vertical dash-dotted lines.}\label{fig:DDES_unsteady}
\end{figure}
Table \ref{tab:FA_DDES_Sensitivity} summarizes simulation details of all DDES carried out in the scope of this work. 
%As DDES results do not show great sensitivity to initialization, w
For clarity, we show in Fig~\ref{fig:DDES_unsteady}(a) only $C_L$ histories of cold-started DDES at different angles of attack. Comparison to warm-started histories will be shown later in Fig~\ref{fig:DDES_unsteady_2}(a) only near $C_{L,max}$ at $\alpha = 17.05^{\circ}$, where main differences occur.
For cold-started simulations in Fig~\ref{fig:DDES_unsteady}(a), we observe the duration of the initial transients changing with increasing angle of attack. 
While the initial transient at $\alpha = 7.05^{\circ}$ is relatively short ($\Delta t_0 \approx 10$ CTUs), it increases at higher $\alpha$ to $\Delta t_0 \approx 36$ CTUs. Due to the presence of large-scale unsteadiness associated with large time scales relative to the length of the present time signal, it is not clear whether the flow at $\alpha=21.47^{\circ}$ is already fully developed. It is interesting to note that simulations of \cite{Mavriplis2022} using a different grid show significantly weaker oscillation amplitudes of aerodynamic coefficients, acknowledging a limited runtime of $30$ CTUs though. 
%However, post-stall flow prediction is not scope of the present contribution, as we focus on the flow near $C_{L,max}$. 
However, as pre-stall conditions are the primary focus of this work, further discussion of post-stall predictions is omitted for conciseness.
%To confirm fully developed flow at $\alpha=17.05^{\circ}$, the warm-started DDES has been run for $170$ CTUs in total and we could confirm no significant change in the aerodynamic coefficients.

Fig~\ref{fig:DDES_unsteady}(b) shows Power Spectral Densities (PSD) of the lift histories shown in Fig~\ref{fig:DDES_unsteady}(a). For all cold-started simulations, the last $36$ CTUs of the signals are used for the PSD calculation (starting at the black dash-dotted lines in Fig~\ref{fig:DDES_unsteady}(a)), where statistics of aerodynamic coefficients are sufficiently well converged. After first subtracting the mean value, all spectra are obtained by Fast Fourier Transformation applying a single Hanning window to the raw data.
The high-frequency range (defined here as $St>1.0$) of the spectrum agrees well for all simulations, where we observe spectral peaks at $St \approx 370$, $170$, and $6.5$. Considering the acoustic analysis of \cite{Lockard2021}\footnote{For comparison, frequencies in Hertz were computed according to $f=St \cdot U^{*}_{\infty}/c^{*}_{mac} \approx St \cdot 66.3/0.7$.} for a similar configuration, frequencies at $St>100$ were attributed to the presence of slat noise. In the present study, we cannot determine the exact physical phenomena associated with frequencies at $St \approx 6.5$, but it is reasonable to expect them being related to vortex shedding.
The low-frequency part of the spectrum (defined here as $St<1$) shows more significant variations between the different angles of attack. At $\alpha > 7.05^{\circ}$, we observe spectral peaks around $St\approx0.4-0.6$, which are in the range of frequencies obtained by global stability analysis\cite{Sansica2024}. Such instabilities could be associated with unsteady flow-phenomena appearing within the hollow nacelle. 
While we observe for $\alpha = 17.05^{\circ}$ a sharp spectral peak near $St\approx0.2$, it becomes comparatively less pronounced at $\alpha = 19.57^{\circ}$, and eventually becomes a bump for $\alpha = 21.47^{\circ}$. However, at increased angles of attack we observe also fairly high PSD-levels near cut-off frequencies of $St<0.1$. As the signals of the cold-started simulations are limited ($\Delta t \approx 36$ CTUs) compared to the time scales of $\Delta t \ge 10$ associated with low-frequency peaks at $St \le 0.1$, it is worth analysing a warm-started DDES near $C_{L,max}$ with longer runtime.

\begin{figure}
    \centering
    \begin{subfigure}[b]{0.48\textwidth}
        a) \\
        \includegraphics[width=\textwidth, trim=10 10 100 50, clip]{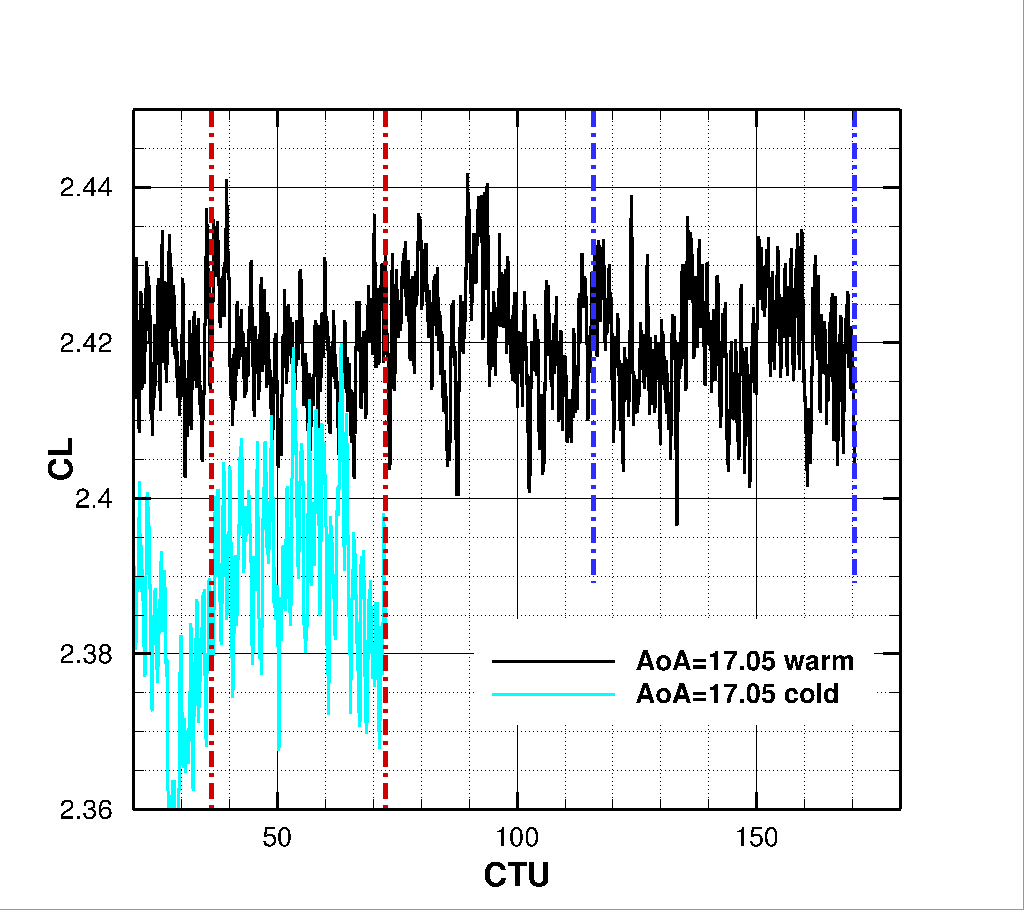}
        %\caption{Subplot 1}
    \end{subfigure}
    \begin{subfigure}[b]{0.48\textwidth}
        b) \\
        \includegraphics[width=\textwidth, trim=10 10 100 50, clip]{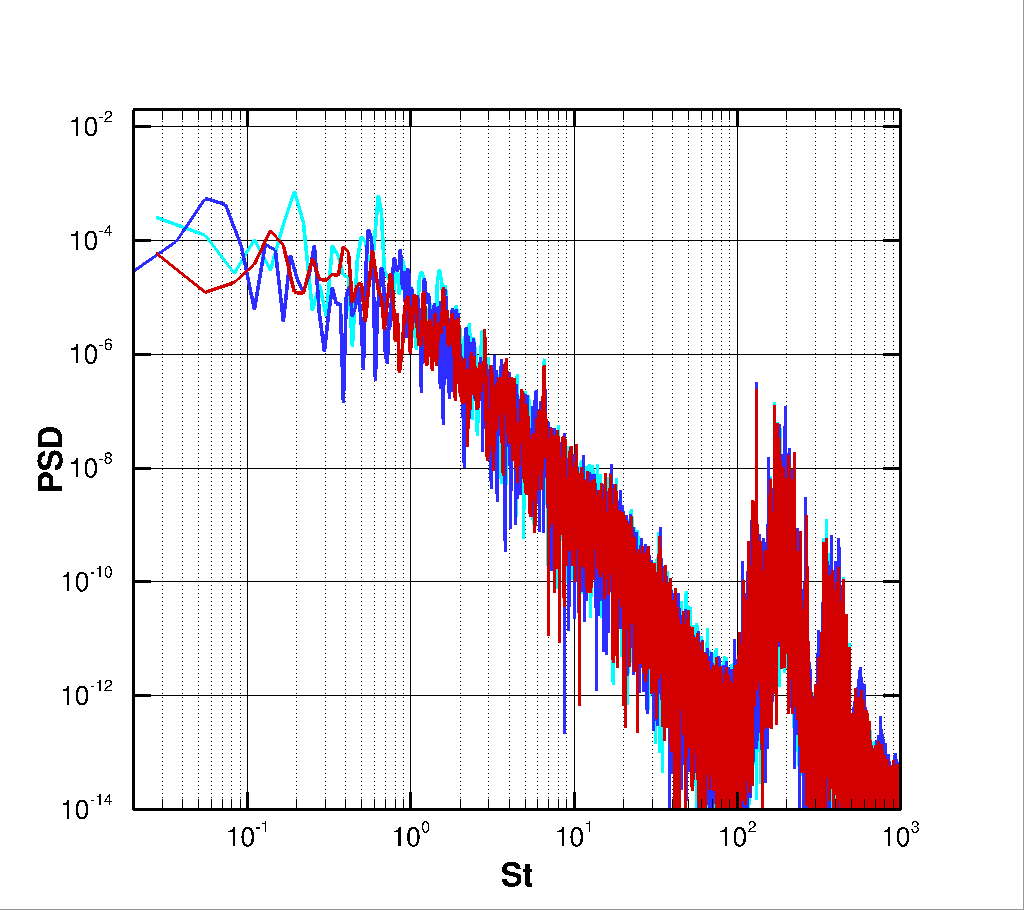}
        %\caption{Subplot 2}
    \end{subfigure}
    \caption{(a) $C_L$-histories of cold- (cyan curve) and warm-started DDES (black curve) at $\alpha = 17.05^{\circ}$. (b) $PSD$ for $0.1<St<1000$, %A single Hanning window is applied to the 
    considering sampling intervals between $36.3<t<72.5$ (red curves) and $116.0<t<170.4$ (blue curves).
    }\label{fig:DDES_unsteady_2}
\end{figure}
Fig~\ref{fig:DDES_unsteady_2}(a) shows histories of the lift coefficient $C_L$ of cold- (cyan curve) and warm-started DDES (black curve) at $\alpha = 17.05^{\circ}$. For a comparison of warm- and cold-started results, we consider a time interval between $36.3<t<72.5$, which is indicated by red vertical dash-dotted lines. 
While $\overline{C_L}$ at $\alpha = 17.05^{\circ}$ is about $\Delta \overline{C_L} \approx 0.03$ lower for cold-started simulations, oscillation amplitudes remain similar with $\Delta C_L\approx0.04$. We would like to emphasize that we observe the largest differences between warm- and cold-started DDES at $\alpha = 17.05^{\circ}$. To confirm the better agreement between cold- and warm-started DDES in terms of statistical quantities, numerical data is reported in table \ref{tab:FA_DDES_Sensitivity}.
Looking at spectra over the same sampling interval in Fig~\ref{fig:DDES_unsteady_2}(a), however, we note some differences in the low frequency range. Peaks at $St \approx 0.6$ and $St \approx 0.2$ for the cold-started DDES (cyan curve in Fig~\ref{fig:DDES_unsteady_2}(b)) are instead found in the spectra of the warm-started simulation histories (red curve in Fig~\ref{fig:DDES_unsteady_2}(b)) at the slightly lower frequencies of $St \approx 0.5$ and $St\approx 0.15$, respectively. In the warm-started simulation results these peaks are also less pronounced. Looking at the black curve in Fig~\ref{fig:DDES_unsteady_2}(a), we can visually identify a low-frequency phenomenon at a time scale of $\Delta t \approx 15$ being well pronounced after $100$ CTUs. Considering a sampling interval covering about $54$ CTUs indicated by the vertical blue dash-dotted lines, we obtain in Fig~\ref{fig:DDES_unsteady_2}(b) (blue curve) a spectral peak between $St \approx 0.05-0.06$.
% Whether this peak can be associated with low-speed buffet remains pure speculation at this point. However, we do see also at post-stall conditions (red curve in Fig~\ref{fig:DDES_unsteady}(a)) low-frequency content, even though these oscillations show intermittency and irregular behaviour in terms of amplitude and occurrence.
We can also observe at post-stall conditions (red curve in Fig~\ref{fig:DDES_unsteady}(a)) low-frequency content, even though these oscillations show intermittency and irregular behaviour in terms of amplitude and occurrence. 
% Near stall conditions and reduced $M$, the operational safe-flight envelope of such aircraft configurations is often limited by an aerodynamic instability at low frequencies called low-speed buffet. 
However, further analysis (e.g. modal decomposition) is required to confirm the physical phenomena associated with these spectral peaks, which will be the topic of future work. Here we want to emphasise the need for sufficiently long run times to (a) overcome the initial transients and (b) capture low-frequency oscillations near $C_{L,max}$. Even though present simulations are run for long enough to obtain well-converged mean values for aerodynamic coefficients (compare mean values for different sampling windows at $\alpha = 17.05$ in table \ref{tab:FA_DDES_Sensitivity}), low-frequency oscillations may still develop or become more pronounced for longer run times. However, present run times of approximately $70$ CTUs may still be sufficient for practical application to estimate the mean-flow properties near $C_{L,max}$. %, which we will the focus of the remainder of this sub-section.

% Initialisation
%\subsubsection{Initialisation}
\begin{figure}
    \centering
    \begin{subfigure}[b]{0.33\textwidth}
        a) \\
        \includegraphics[width=\textwidth, trim=10 10 50 50, clip]{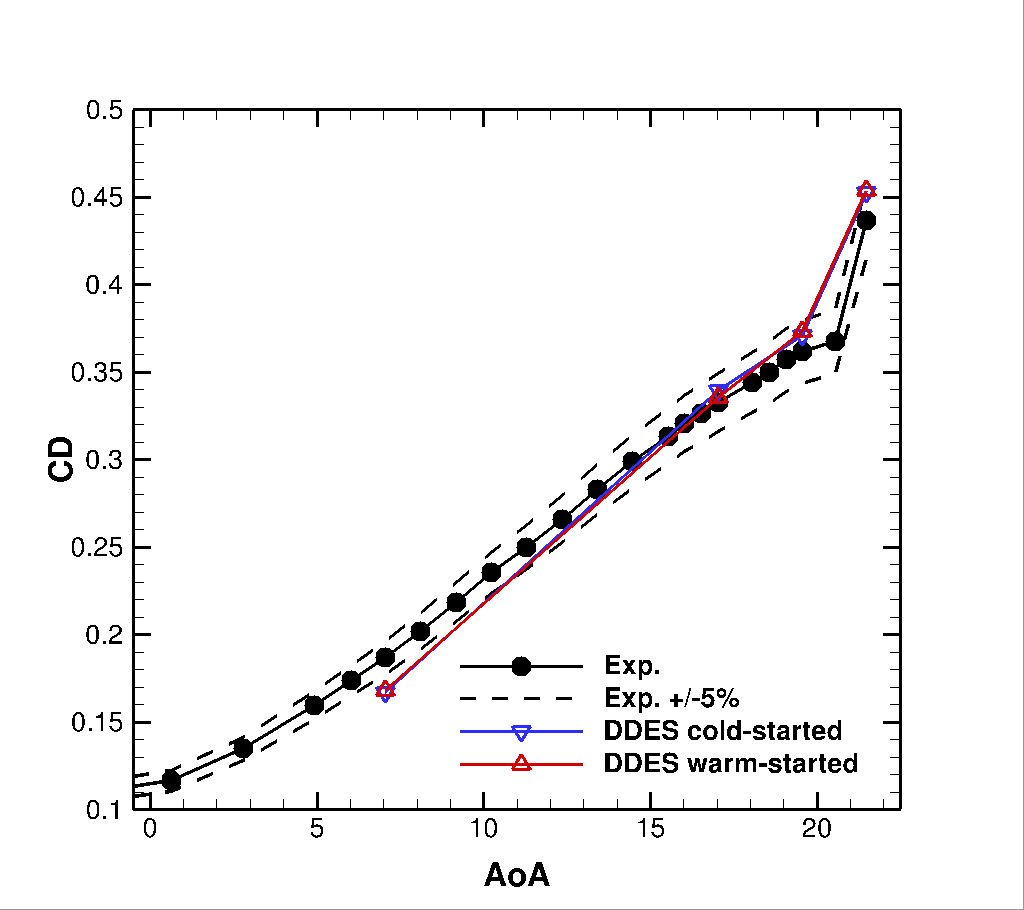}
        %\caption{Subplot 1}
    \end{subfigure}
    \begin{subfigure}[b]{0.33\textwidth}
        b) \\
        \includegraphics[width=\textwidth, trim=10 10 50 50, clip]{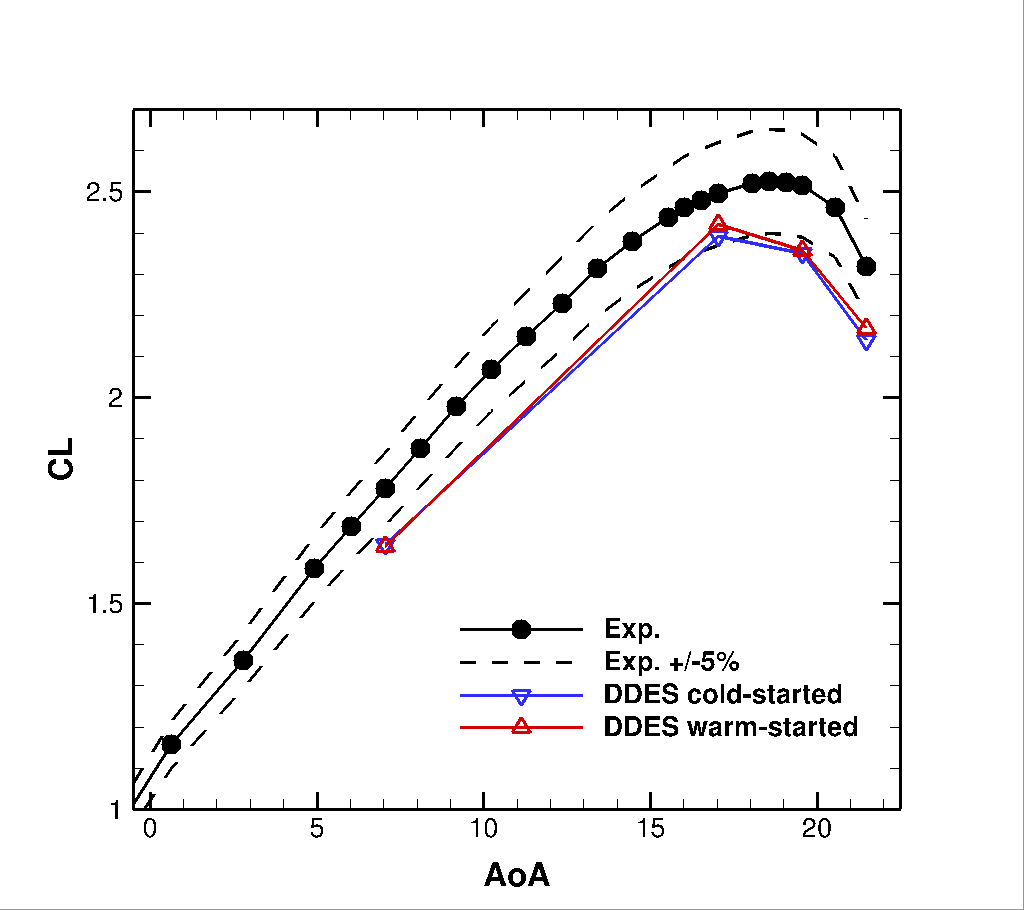}
        %\caption{Subplot 2}
    \end{subfigure}
    \begin{subfigure}[b]{0.33\textwidth}
        c) \\
        \includegraphics[width=\textwidth, trim=10 10 50 50, clip]{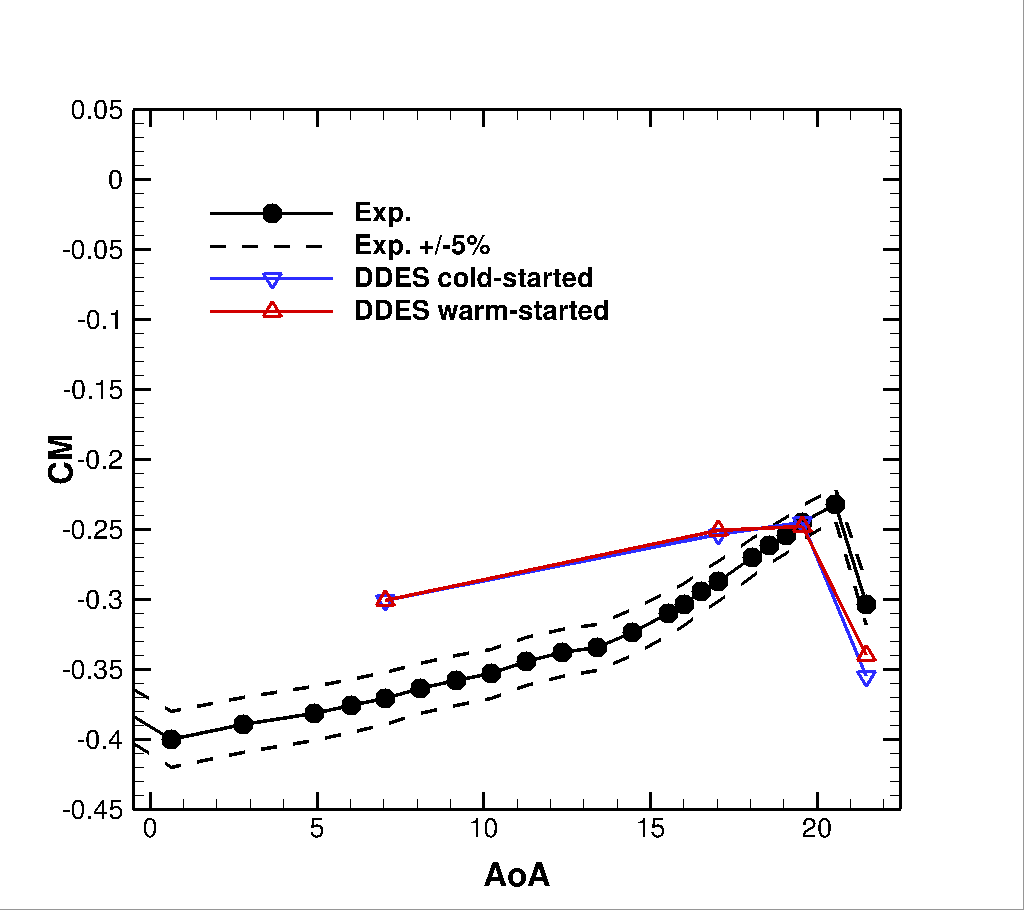}
        %\caption{Subplot 3}
    \end{subfigure}
    \caption{(a) Drag, (b) lift, and (c) moment coefficient for warm- and cold-started DDES as a function of the angle of attack. Black symbols denote experimental measurements and black dashed lines correspond to $\pm 5\%$ error margins.}\label{fig:FA_DDES_Coeff_Init}
\end{figure}
Fig~\ref{fig:FA_DDES_Coeff_Init} shows now time-averaged (a) drag, (b) lift, and (c) moment coefficient for warm- and cold-started DDES as a function of the angle of attack. Black symbols denote experimental measurements and black dashed lines correspond to $\pm 5\%$ error margins. 
At increased angles of attack, the relative difference between simulation and experimental results remain within $5\%$ even for the moment coefficient. At low angles of attack, however, RANS results on the same grid (see \S \ref{sec:grid_study}) appear in better agreement with experimental data compared to the DDES results. 
%This may be due to imperfections of the shielding function. 
It is important to note that flaps experience high aerodynamic loads at reduced angles of attack and are prone to separation, which also has significant effect on the moment coefficient, and leads to notable deviations from experimental results in Fig~\ref{fig:FA_DDES_Coeff_Init}.

%While lift and drag coefficients remain close to the error margin of experimental measurements for all considered angles of attack, the lift coefficient is under-predicted even at low angles of attack. It is important to note that the flaps experience high aerodynamic loads at reduced angles of attack and are prone to separation, which also has significant effect on the moment coefficient shown in Fig~\ref{fig:FA_DDES_Coeff_Init}(c). 
 % At increased angles of attack, the relative difference between simulation and experimental results remain within $5\%$ even for the moment coefficient. At low angles of attack, however, RANS results on the same grid (see \S \ref{sec:grid_study}) appear in better agreement with experimental data compared to the DDES results. This may be due to imperfections of the shielding function.

\begin{figure}
    \centering
    \begin{subfigure}[b]{0.48\textwidth}
        a) \\
        \includegraphics[width=\textwidth, trim=10 10 10 10, clip]{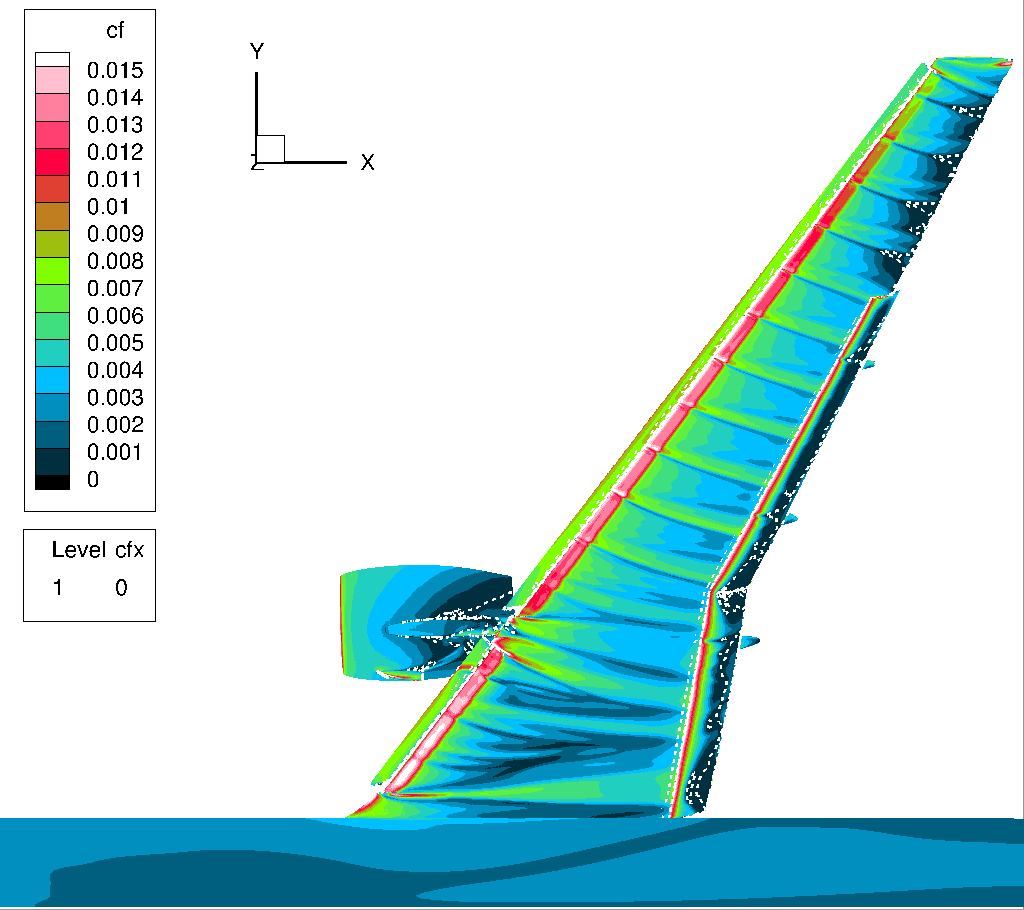}
        %\caption{Subplot 1}
    \end{subfigure}
    \begin{subfigure}[b]{0.48\textwidth}
        b) \\
        \includegraphics[width=\textwidth, trim=10 10 10 10, clip]{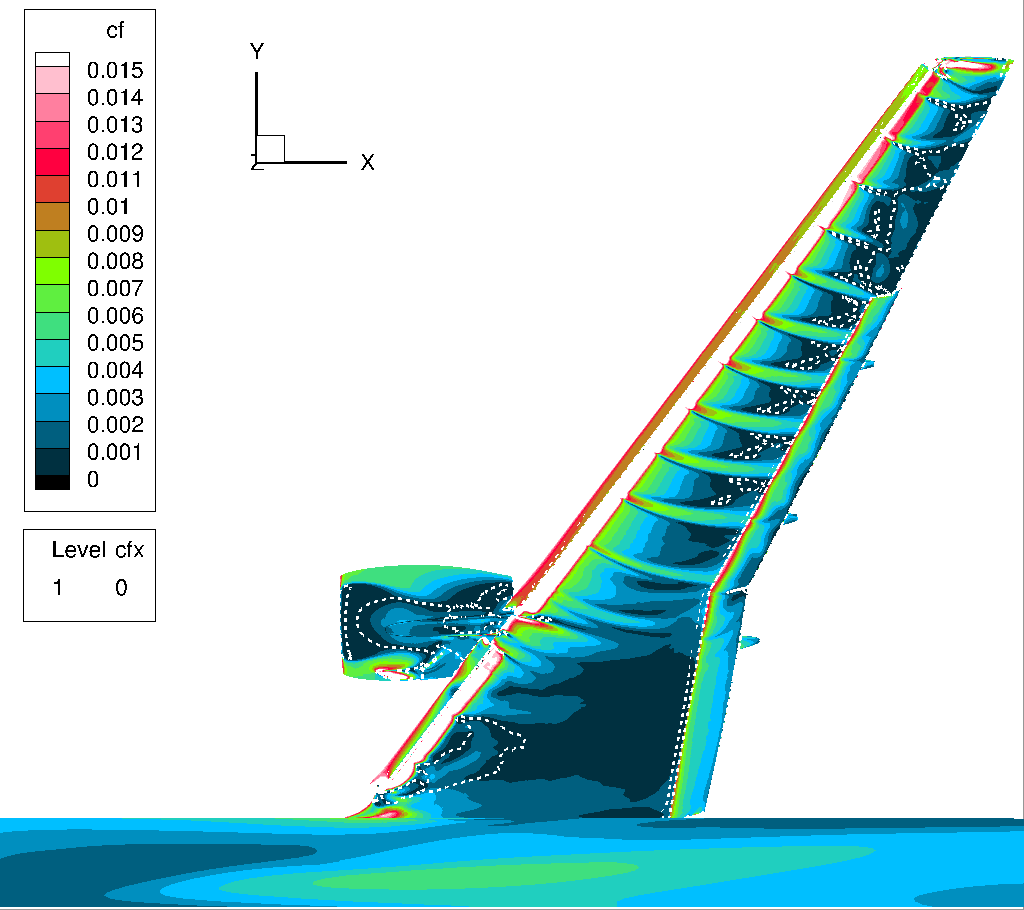}
        %\caption{Subplot 2}
    \end{subfigure}
    % \begin{subfigure}[b]{0.32\textwidth}
    %     c) \\
    %     \includegraphics[width=\textwidth, trim=10 10 10 10, clip]{Figures_FA_DDES/FA_DDES_Cf_19p57_cold.png}
    %     %\caption{Subplot 2}
    % \end{subfigure}
    \caption{Contours showing time-averaged skin-friction coefficient $C_f$ for cold-started DDES at (a) $\alpha = 7.05^{\circ}$ and (b) $\alpha = 19.57^{\circ}$. 
    %(c) DDES at $\alpha = 19.57^{\circ}$ warm-started simulation from a RANS, which was cold-started from scratch on the present DDES grid. 
    The last 36 CTUs are considered for time-averaging.}\label{fig:Cf_FA_DDES}
\end{figure}
Excessive flow separation along the flaps at $\alpha = 7.05^{\circ}$ can be confirmed in the skin-friction contours of Fig~\ref{fig:Cf_FA_DDES}(a). 
Adaptively refined Cartesian HRLES grids of \cite{Mavriplis2022} deliver higher lift at the same angle of attack. Their results show reduced flow separation along the flaps, where their grid is significantly refined.
At an increased angle of attack, our results near $C_{L,max}$ match well with experiments of \cite{ELSR2020}. Contrasting RANS solutions, we do not observe non-physically large bulks of flow separation near the wing tip in Fig~\ref{fig:Cf_FA_DDES}(b).
% Without showing additional contour plots of present DDES results for brevity, it is remarkable that we do not observe significant differences in skin-friction contour plots comparing present warm- and cold-started DDES solutions (even at $\alpha 17.05^{\circ}$), which confirms the similarities observed for time-averaged aerodynamic coefficients.
Comparison was made between skin-friction contours for both cold- and warm-started DDES, however, they are omitted here for brevity as we did not observe any significant differences.

\begin{figure}
    \centering
    \begin{subfigure}[b]{0.32\textwidth}
        a) \\
        \includegraphics[width=\textwidth, trim=20 10 100 50, clip]{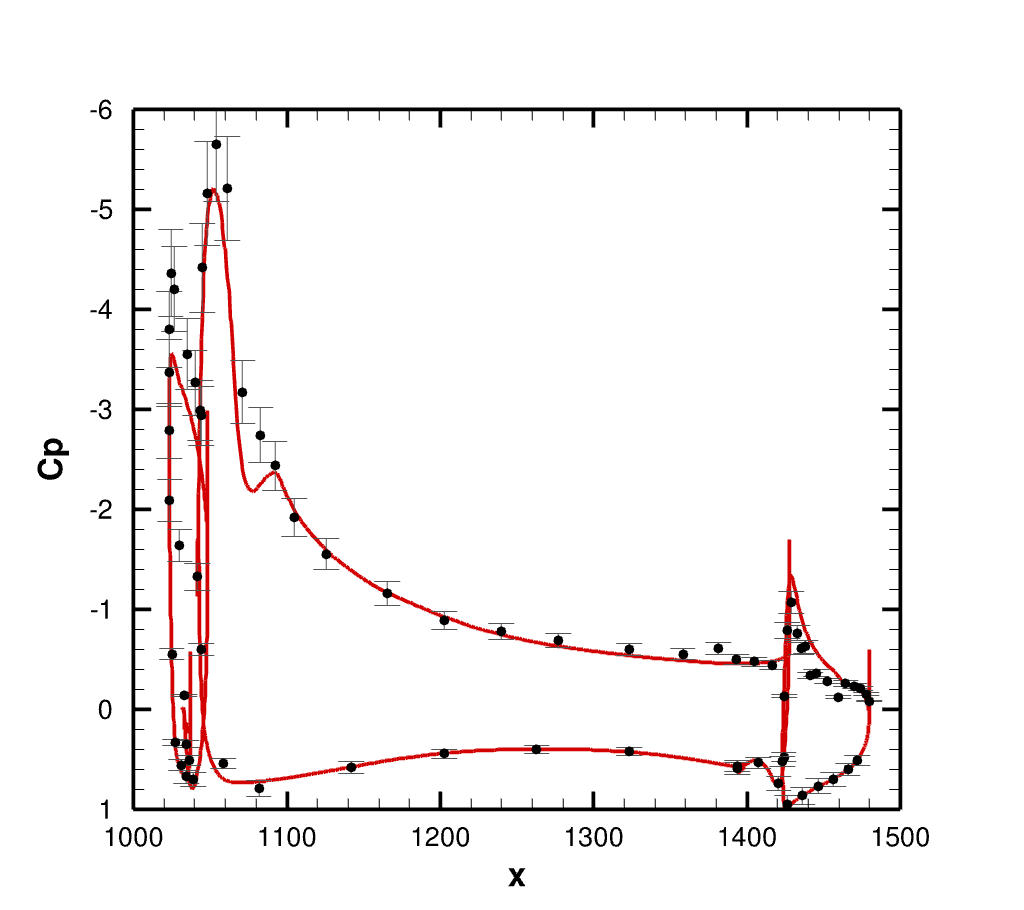}
        %\caption{Subplot 1}
    \end{subfigure}
    \begin{subfigure}[b]{0.32\textwidth}
        b) \\
        \includegraphics[width=\textwidth, trim=20 10 100 50, clip]{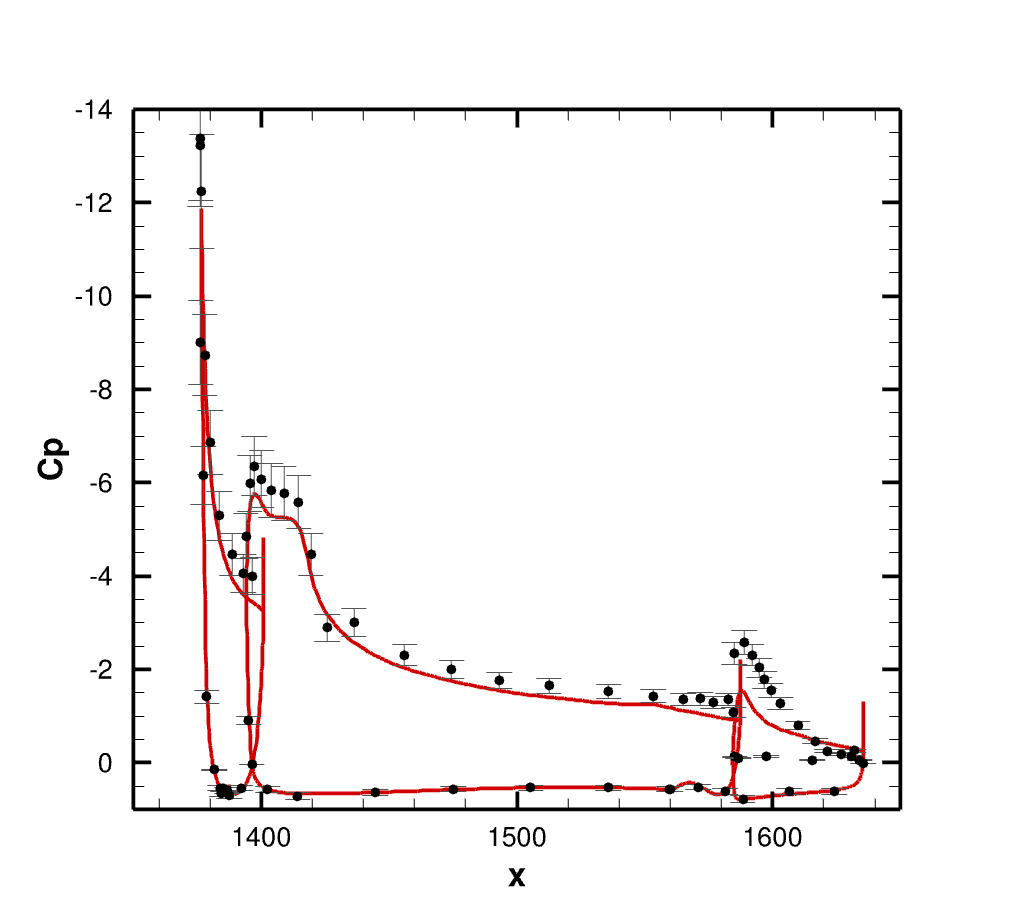}
        %\caption{Subplot 2}
    \end{subfigure}
    \begin{subfigure}[b]{0.32\textwidth}
        c) \\
        \includegraphics[width=\textwidth, trim=20 10 100 50, clip]{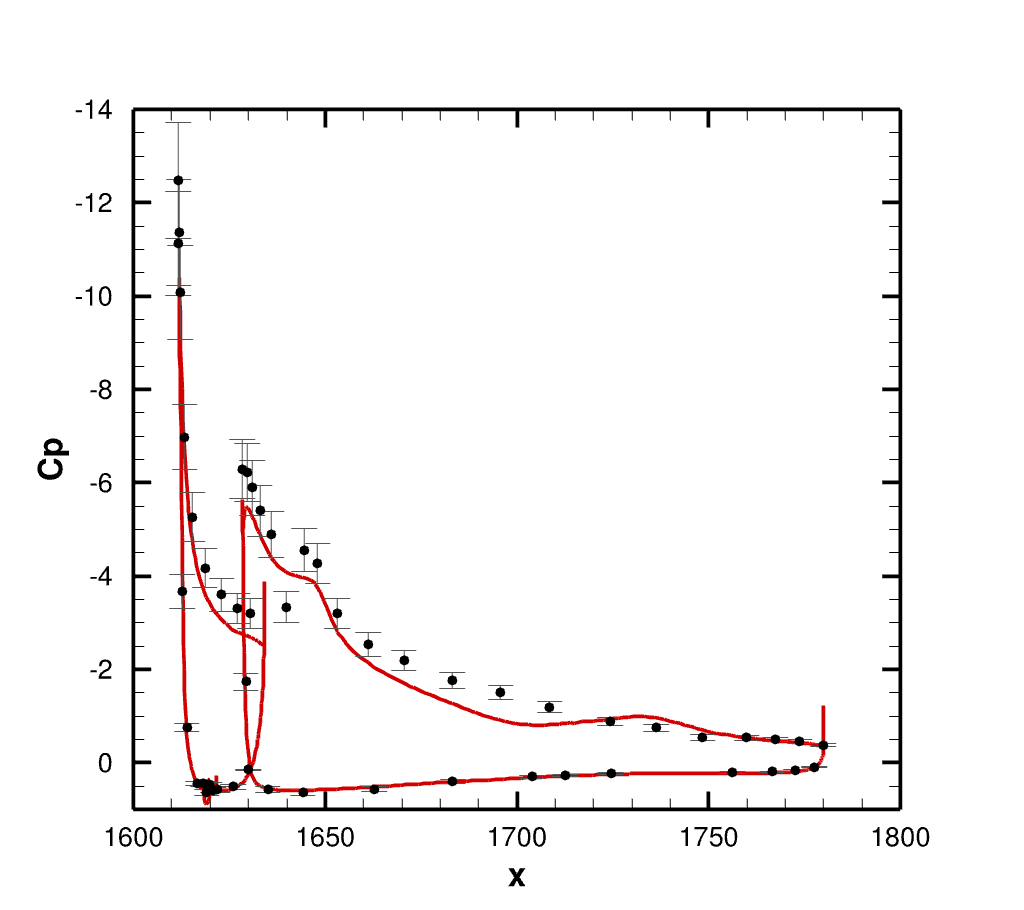}
        %\caption{Subplot 2}
    \end{subfigure}
    \caption{Red curves show wall-pressure $C_p$ distributions of cold-started DDES at $\alpha = 19.57^{\circ}$, taken at cross-sections (a) $A$ $\eta = 0.15$, (b) $E$ $\eta = 0.55$, and (c) $G$ $\eta = 0.82$ according to Fig~\ref{fig:Intro}(b). Black symbols denote experimental data and error bars indicating $10\%$ deviation from experimental measurements.}\label{fig:Cp_FA_DDES}
\end{figure}
Fig~\ref{fig:Cp_FA_DDES} shows wall-pressure coefficient as a function of $x$ for cold-started DDES at $\alpha = 19.57^{\circ}$, taken at spanwise locations (a) $\eta = 0.15$, (b) $\eta = 0.55$, and (c) $\eta = 0.82$. The locations of these cross-sections are respectively labelled in Fig~\ref{fig:Intro}(b) as $A$, $E$, and $G$. Black symbols denote experimental data and a $\pm 5\%$ error bar. While the pressure side agrees very well with experiments at all locations, we observe difference on the suctions side over the slats as well as front sections of main wing and flaps. Given the deficit in lift prediction, the flow acceleration over the suction side is under-predicted. Potential reasons for this could be excessive flow separation, over-predicted boundary-layer thickness, or even transition effects. There may also be sensitivities of the shielding function, which controls the switching between RANS and LES schemes based on a characteristic length scale computed for each cell.
In the hybrid RANS/LES study of \citet{Browne2022} using Zonal Detached Eddy Simulation (ZDES), significant grid-sensitivity was shown near the out-board region (i.e. section G), where present DDES results show better performance on a level C grid. Near the wing root (i.e. section A), however, their results do not show significant grid-dependency. In particular, they do not show a kink in the $C_p$ distributions that we observe in Fig~\ref{fig:Cp_FA_DDES}(a) around $x \approx 1100$. Looking at skin-friction contours of \ref{fig:Cf_FA_DDES}(b), we can observe a pocket of separated flow behind the slat at $\eta = 0.15$. Whether these differences are due to variations in grid resolution/topology or numerical approach (e.g. shielding function) is difficult to answer and requires further investigations.

% DDES using SST
Before summarizing the lessons learnt from this work and suggesting best practises for future studies, sensitivities of DDES results to the turbulence model are reported here at an angle of attack of $\alpha = 19.57^{\circ}$, representative of $C_{L,max}$ conditions. Table \ref{tab:FA_DDES_Sensitivity} shows that aerodynamic coefficients are very similar for cold-started DDES using SA- and SST-based turbulence models at $\alpha = 19.57^{\circ}$. 
\begin{figure}
    \centering
    \begin{subfigure}[b]{0.48\textwidth}
        a) \\
        \includegraphics[width=\textwidth, trim=10 10 100 50, clip]{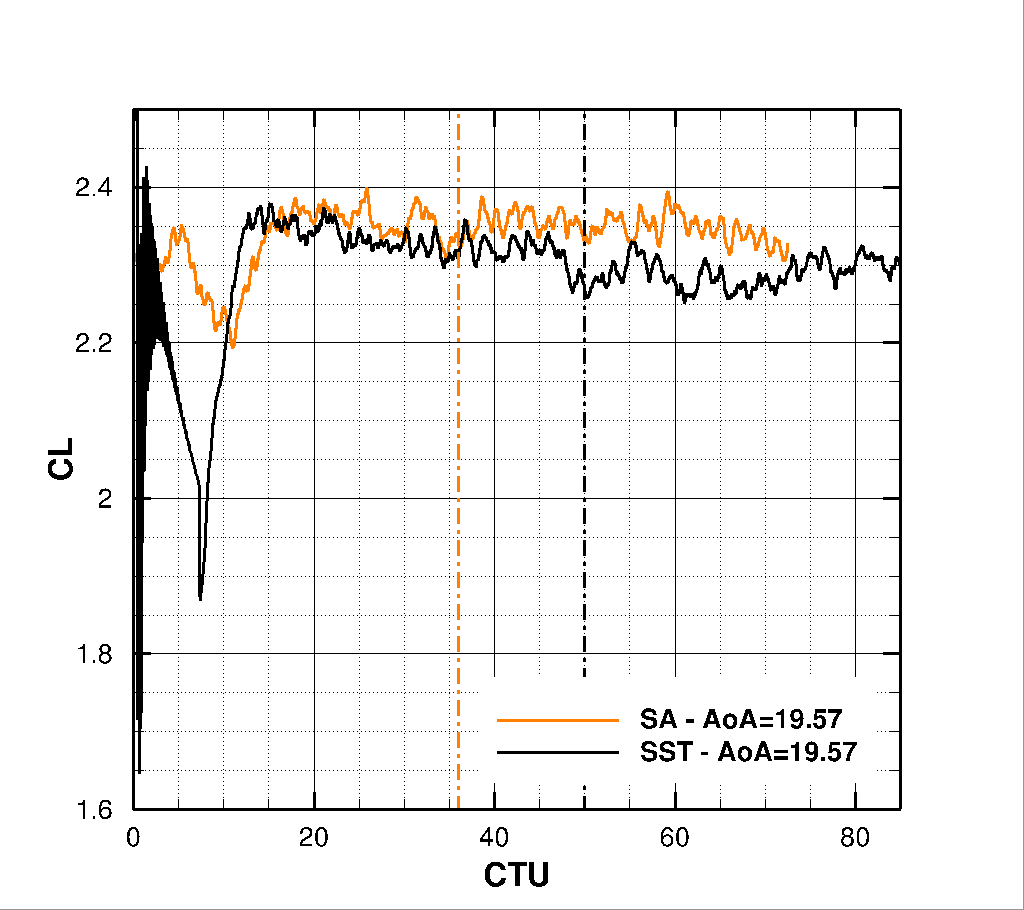}
        %{Figures_FA_DDES/Histories_SA_SST_19p57.png}
        %\caption{Subplot 1}
    \end{subfigure}
    \begin{subfigure}[b]{0.48\textwidth}
        b) \\
        \includegraphics[width=\textwidth, trim=10 10 100 50, clip]{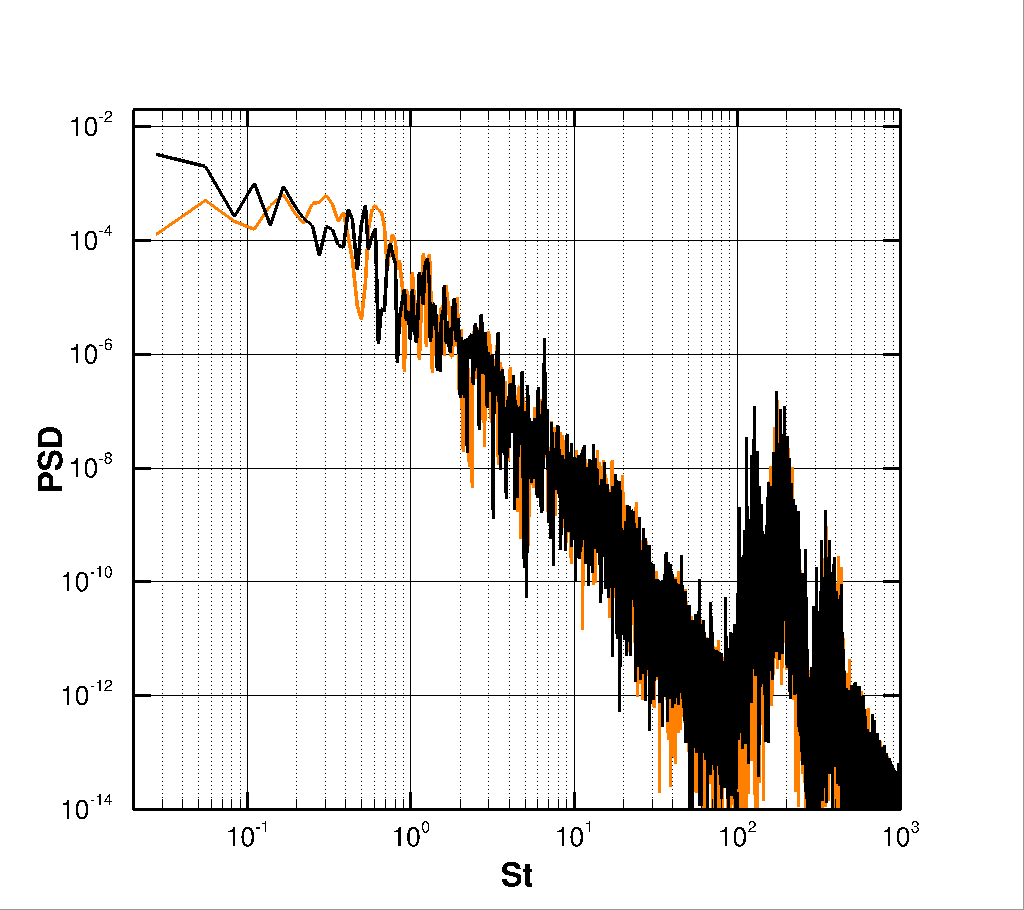}
        %{Figures_FA_DDES/PSD_Welch_1seg_SA_SST.png}
        %\caption{Subplot 2}
    \end{subfigure}
    \caption{(a) Histories of the lift coefficient $C_L$ of cold-started DDES and (b) corresponding Power-Spectral-Density $PSD$ for Strouhal numbers $0.1<St<1000$ using SA- (orange) and SST-based (black) turbulence models. A single Hanning window is applied to the sampling interval after the end of an initial transient indicated by vertical dash-dotted lines.}\label{fig:DDES_unsteady_SA_SST}
\end{figure}
Time histories showing $C_L$ for cold-started DDES in Fig~\ref{fig:DDES_unsteady_SA_SST}(a) confirm the good agreement between SA- (orange curve) and SST-based (black curve) DDES results. 
%We can see that after the initial transient indicated by the vertical dash-dotted line, 
The lift coefficient drops for the SST-based DDES after $\approx 45$ CTUs, leading to an extended initial transient, indicated by the vertical orange line. Therefore the overall run-time of the simulation has been extended to maintain a sampling interval of $\approx 36$ CTUs to compute spectra in Fig~\ref{fig:DDES_unsteady_SA_SST}(b).
For frequencies at $St>0.1$, we observe good agreement between DDES results of both turbulence models.
However, due to intermittent or irregular phenomena in the lift signals of Fig~\ref{fig:DDES_unsteady_SA_SST}(a) after the initial transients, we observe increased PSD at $St<0.1$. Given the limited runtime of the corresponding simulations, we cannot trust this part of the spectrum. However, we have to acknowledge the possibility that for significantly longer run-times, a low-frequency peak may develop at $St\approx0.03$. 
%This causes increased power-spectral density in the black curve of Fig~\ref{fig:DDES_unsteady_SA_SST}(b) at low frequencies around $St\approx0.1$. Elsewhere, the remaining parts of the spectra agree very well between both turbulence models.

\begin{figure}
    \centering
    \begin{subfigure}[b]{0.48\textwidth}
        a) \\
        \includegraphics[width=\textwidth, trim=10 10 10 10, clip]{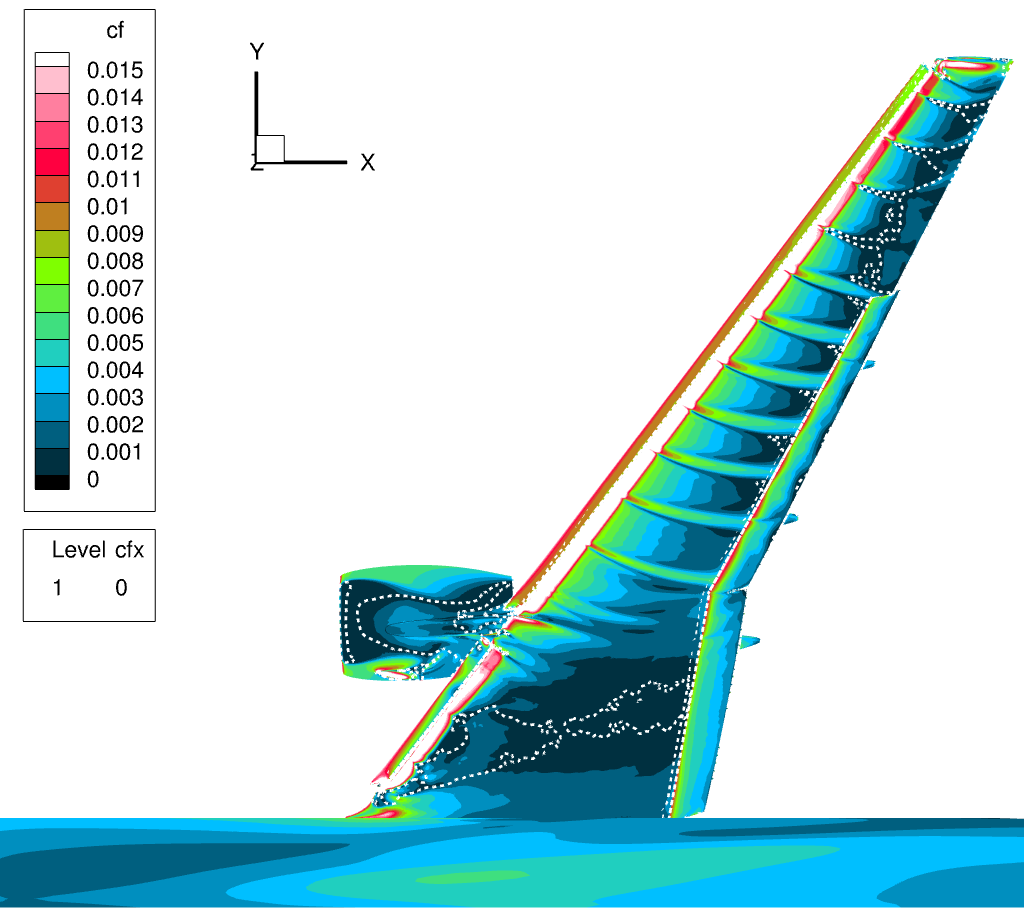}
        %\caption{Subplot 1}
    \end{subfigure}
    \begin{subfigure}[b]{0.48\textwidth}
        b) \\
        \includegraphics[width=\textwidth, trim=10 10 10 10, clip]{Figures_FA_DDES/FA_DDES_Cf_19p57_cold.png}
        %\caption{Subplot 2}
    \end{subfigure}
    \caption{Contours showing time-averaged skin-friction coefficient $C_f$ for cold-started DDES at $\alpha = 19.57^{\circ}$ using (a) SST and (b) SA turbulence models. 
    The last 36 CTUs are considered for time-averaging.}\label{fig:Cf_FA_DDES_SA_SST}
\end{figure}
Finally, Fig~\ref{fig:Cf_FA_DDES_SA_SST} shows time-averaged $C_f$ contours for cold-started DDES at $\alpha = 19.57^{\circ}$ using (a) SST and (b) SA turbulence models. Again we can see strong similarities between both results. We observe slight difference in the separation region near the wing root, which may lead to the slight decrease in $C_L$. In contrast to RANS and URANS results in previous sections, DDES results seem very consistent for the considered flow conditions, with far less sensitivity to initialization strategy and turbulence model.

\section{Recommended Practices and Lessons Learned\label{sec:results_bestpractices}}
Before presenting final conclusions, this section first discusses computational costs, before summarizing the main lessons learnt from this study. This includes detailing recommended practices for steady, unsteady, and hybrid scale-resolving RANS methods in the context of the CRM-HL configuration.
\begin{figure}
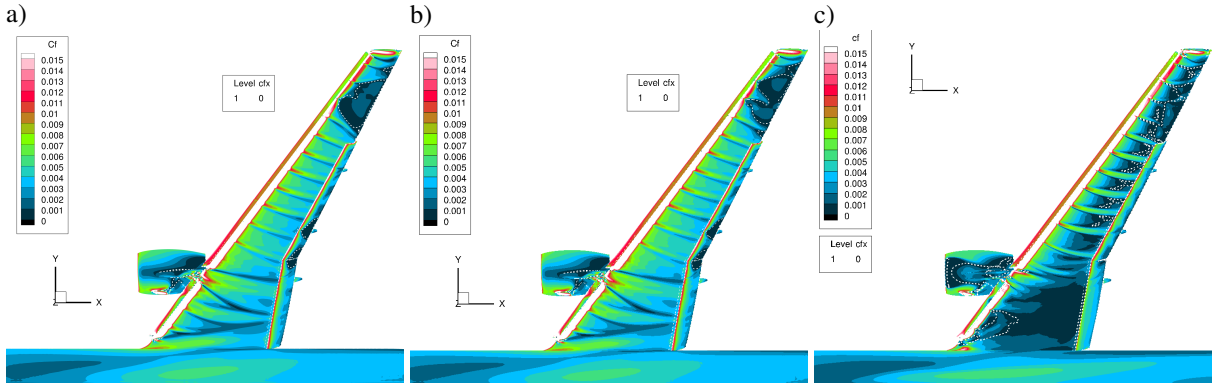

    \centering
    \begin{subfigure}[b]{0.32\textwidth}
        a) \\
        \includegraphics[width=\textwidth, trim=10 10 10 10, clip]{Figures_FA_RANS/Cf/SA10_warm_19p57.png}
    \end{subfigure}
    \begin{subfigure}[b]{0.32\textwidth}
        b) \\
        \includegraphics[width=\textwidth, trim=10 10 10 10, clip]{Figures_FA_URANS/Cf_URANS_SA10_coldstart_19p57_TS0p001_CFL10.png}
    \end{subfigure}
    \begin{subfigure}[b]{0.32\textwidth}
        c) \\
        \includegraphics[width=\textwidth, trim=10 10 10 10, clip]{Figures_FA_DDES/FA_DDES_Cf_19p57_cold.png}
    \end{subfigure}
    \caption{Contours showing skin-friction coefficient $C_f$ for best practice of (a) RANS (warm-started), (b) URANS (cold-started), and (c) DDES (cold-started) at $\alpha = 19.57^{\circ}$}\label{fig:Cf_FA_comparison}
\end{figure}
To emphasize that only scale-resolving hybrid RANS/LES simulations were capable to show significant improvement in terms of flow physics, Fig~\ref{fig:Cf_FA_comparison} shows skin-friction coefficient $C_f$ for best practice of (a) RANS, (b) URANS, and (c) DDES near $C_{L,max}$ at $\alpha = 19.57^{\circ}$, which can be compared with experimental results in \cite{ELSR2020}. After commenting on the computational costs, we will summarize best practises and lessons learned for each approach individually. 

\subsection{Computational Costs}
% Simulations in this work were computed on the JAXA Supercomputing System Generation 3 (JSS3) consisting of $5,\!760$ standard nodes (Fujitsu PRIME-HPC FX1000). Each node contains $48$ cores (Fujitsu A64FX) delivering a total peak performance of $19.4$ PFLOPS. 
Simulations in this work were computed on the JAXA Supercomputing System Generation 3 (JSS3) consisting of $5,\!760$ standard nodes (Fujitsu PRIME-HPC FX1000). Each node contains $48$ cores (Fujitsu A64FX) delivering a total peak performance of $19.4$ PFLOPS. Costs are measured in core hours (coreh), considering a typical runtime to generate simulation results according to recommended practises discussed in previous sections. We consider the entire simulation loop as well as I/O procedures, while pre- and post-processing have not been included. The aim is to provide an overview of the costs and time required to generate simulation results for given flow conditions using different grids and methods presented in this work.

\begin{table}
    \centering
    \begin{tabular}{ccccc}
        Method & Grid size & Turb. Model & Cores &  Core-hours per $2 \cdot 10^5$ iterations\\
        \hline
        RANS & $85,\!969,\!733$ & SA  & $576$ & $32,\!000$ \\ % 13.7875h for 49131 iterations
        RANS & $85,\!969,\!733$ & SST & $576$ & $37,\!000$ \\ % 16.033h for 49999 iterations     
    \end{tabular}
    \caption{Computational costs of present steady simulations.}
    \label{tab:cost_steady}
\end{table}
Table \ref{tab:cost_steady} summarizes simulation details and corresponding costs for the steady RANS cases. A typical RANS simulation required $200,\!000$ iterations to obtain sufficient convergence (see \S \ref{sec:results_freeair_rans}). 
%We report the costs in core-hours (coreh) per $200,\!000$ iterations, which corresponds to a typical run-time required to obtain sufficient convergence (see \S \ref{sec:results_freeair_rans}). 
Considering the standard RANS grid consisting of $85,\!969,\!733$ nodes and using $576$ cores ($12$ nodes) on JSS3 requires for typical RANS jobs about $55$ hours elapsed real time ($\approx 32,\!000$ coreh) using Spalart-Allmaras turbulence models, while present SST models are $\approx 16\%$ more expensive. As shown in section \S \ref{sec:results_freeair_rans}, iterative convergence can vary for different angles of attack, turbulence models, or grids and eventually leads to differences in computational costs, which are not detailed here. However, based on information provided in table \ref{tab:cost_steady}, one can scale projected costs accordingly.
\begin{table}
    \centering
    \begin{tabular}{cccccc}
        Method & Grid size & Turb. Model & Time Step & Cores &  Core-hours per $72.5$ CTU \\
        \hline
        URANS & $85,\!969,\!733$ & SA & 0.00020 & $576$ & $300,\!000$ \\ 
        % 23.12h for 16200 iterations (5 inner iterations, CFL 10, deltat 0.002) -> 164385coreh for 40CTUs -> 
        URANS & $85,\!969,\!733$ & SST & 0.00020 & $960$ & $360,\!000$  \\
        % 23.02h for 22099 iterations (5 inner iterations, CFL 10, deltat 0.0002)
        DDES & $218,\!077,\!184$ & SA & 0.00036 & $1920$ & $320,\!000$ \\ 
% 15.2827 for 10000 iterations (5 inner iterations, CFL 10, deltat 0.5) -> 3.6258 CTUs original fastar version
% 23.989 h for 28864 iterations (5 inner iterations, CFL 10, deltat 0.5) -> 10.46555 CTUs accelerated fastar version
        DDES & $218,\!077,\!184$ & SST & 0.00036 & $1920$ & $340,\!000$ \\ 
% 23.19h for 26387 iterations (5 inner iterations, CFL 10, deltat 0.5) -> 9.567 CTUs accelerated fastar version
% 17.2936 for 20000 iterations (5 inner iterations, CFL 10, deltat 0.5) -> 7.2516 CTUs accelerated fastar version
    \end{tabular}
    \caption{Computational costs of present unsteady simulations. 
    %using $\rm{CFL}=10$ and a non-dimensional time step of $\Delta \tau=0.00036$. Costs for URANS simulations are estimated based on \S \ref{sec:results_freeair_urans}.
    }
    \label{tab:cost_unsteady}
\end{table}
Table \ref{tab:cost_unsteady} summarises the numerical costs for the unsteady (URANS and DDES) simulations using FaSTAR. For the unsteady methods, we measure the costs in core-hours per $72.5$ convective time units, which is considered as a minimum run time to capture unsteady phenomena near $C_{L,max}$. 
Using a time steps of $\Delta \tau = 0.00036$ and $\Delta \tau = 0.0002$ for DDES and URANS, respectively, run-times corresponds to approximately $200,\!000$ and $360,\!000$ iterations (For URANS $5$ sub-iterations per time step are considered). %, which has been the cost measure for steady simulations. 
It was shown in \S \ref{sec:results_freeair_ddes} that even longer run times may be required to collect sufficient statistics associated with post-stall low-frequency phenomena at $St\approx0.1$. 
The estimates of costs for URANS simulations are based on the numerical experiment in \S \ref{sec:results_freeair_urans} assuming a time step of $\Delta \tau = 0.0002$. Based on the URANS sensitivity study we could increase the time step cutting down costs, but it has been shown that unsteadiness gets even more damped for increased $\Delta \tau$ and numerical instabilities may occur.
In any case, as shown in \S \ref{sec:results_freeair_urans}, URANS simulations did not show significantly improved results compared to RANS. Computational costs for the DDES are approximately one order of magnitude higher compared to steady RANS, but lead to significantly improved results due to the improved fidelity. Only marginal increase of costs is observed for SST-based DDES. As reported in Fig~\ref{fig:Cf_FA_DDES_SA_SST}, results between the two turbulence models were very consistent for the grid tested. %While DDES based on the SST turbulence model is slightly more expensive than those using a SA model, no significant differences in results were observed for the present grid.
It may be interesting to note that present DDES required only $0.6 \%$ of the computational resources of the JSS3 cluster. 
Using $5 \%$ of the cluster, DDES results for a given set of flow conditions can be produced within $24$ hours. Therefore, we want to emphasize the technical feasibility of complementing RANS simulations with scale-resolving simulations at conditions near stall on modern clusters.

\subsection{Steady RANS}
Based on present steady RANS results for a given level C grid, it is hard to conclude on the best choice of turbulence model. No correction of the Spalart-Allmaras turbulence model lead to significant improvement of solutions. SST models also showed similar deficiencies as SA models, but were more expensive and sometimes even led to more difficulties in terms of convergence. As a result, we consider the baseline Spalart-Allmaras turbulence model without $f_{t2}$ term for our current best practice recommendations.
For all considered turbulence models, we could already see at $\alpha=17.05$ the appearance of spurious patches of outboard separation.
At increased angles of attack, we observe additional non-physically large separation zones, which show high sensitivity to the initial solutions. These simulations were performed on the standard RANS grid, which has been hand-crafted according to the HLPW-4 guidelines and best practices. 
For warm-started simulations, non-physical flow separation could be delayed to higher angles of attack. 
Even though larger grids may lead to reduced sensitivity to initial conditions, due to their increased costs, such grids would not be practical for real-life application. Despite increased computational costs, cold-started steady RANS solutions of significantly refined grids seem to suffer from the same non-physical flow separation near wing tips as those of warm-started solutions computed on the standard grid. Therefore, we consider warm-started steady RANS simulations performed on a $C$-level RANS grid as a recommended practice.
However, we reiterate that all our efforts for steady RANS computations lead to the appearance of non-physical flow phenomena near $C_{L,max}$, despite fair agreement with experiments in terms of aerodynamic coefficients. Therefore, we cannot trust the state-of-the-art RANS methods for $C_{L,max}$ predictions over such a complex aircraft configuration.
In the future, better turbulence models or adaptive mesh refinement techniques may provide solutions to the present deficiencies observed for steady RANS simulations.

Considering present state-of-the-art methods and turbulence models, RANS can still be used for high-lift configurations at reduced angles of attack (i.e. the linear part of the $C_L$-$\alpha$ curve). The challenge is to assess (quantify) the trustworthiness (uncertainty) of results at a given angle of attack. In other words, the challenge is to identify a certain threshold angle of attack, beyond which higher-fidelity methods are needed to maintain sufficient accuracy of results. % While proper (statistical) uncertainty quantification is still prohibitive using RANS, 
As an additional indicator of trustworthiness, one could test two turbulence models (e.g. standard SA with and without R-QCR corrections) for their consistency. 
As shown in \S \ref{sec:results_freeair_rans}, results using different turbulence models agree well (e.g. $\Delta \overline{C_M}<0.05$\footnote{While $C_L$ and $C_D$ are less sensitive due to error-compensation, $C_M$ appears more sensitive according to table \ref{tab:sensitivities_RANS_FA}.}), when errors to experimental measurements are low (i.e. less than $\pm 5 \%$). At increased angles of attack, where non-physical phenomena appear in RANS solutions, variations between different turbulence models become significant (e.g. $\Delta \overline{C_L}>0.05$). 

\subsection{Unsteady RANS}
%Present URANS simulations lead to similar problems encountered for steady RANS simulations, while computational costs would notably increase. 
%However, it is interesting to note that cold-started URANS simulations lead to similar results compared to warm-started steady RANS simulations.
Cold-started URANS simulations lead to similar results compared to warm-started steady RANS simulations. Warm-started URANS simulations, on the other hand, appeared to be "locked" on their initial (cold-started) RANS solution. 
In any case, present URANS simulations lead to similar problems encountered for steady RANS simulations, while computational costs would notably increase. 
While cold-started initialization for URANS is preferred, we recommend the use of steady RANS methods. 
%For future work, one could attempt cold-started URANS simulations for the initialisation and then switch to steady RANS to converge the solution in order to (a) circumvent the sensitivity to initial solutions of (pure) steady RANS results and (b) avoid increased computational costs.

\subsection{Hybrid RANS/LES}
The use of higher-fidelity DDES methods provide a potential solution to the shortcomings of RANS methods discussed in this work.
We observed reduced sensitivities to the initial solution and turbulence model for present simulations compared to RANS methods. While aerodynamic coefficients remain within a $\pm 5\%$ error range with respect to experimental results, regions of flow separation have been more localised and agree better with experimental observations. Even though computational costs are a factor of $10$ higher compared to steady RANS results\footnote{Computational costs of present methods are summarised in table \ref{tab:cost_unsteady}.}, we still recommend scale resolving methods for future best practise to predict $C_{L,max}$ more accurately. As SA-based DDES are cheaper compared to SST-based approaches and deliver results of similar accuracy, we recommend the SA-noft2-R turbulence model for current best practise.
Despite the promising results for Hybrid RANS/LES methods, there is still additional work required to assess the sensitivity to grid resolution and the choice of shielding function leading to excessive flap separation at reduced angles of attack. 
%Furthermore, we need to study whether wind-tunnel effects can answer remaining discrepancies to our simulations.

% SA11 or SA11-based DDES \\
% best initialization for each method \\
% comparisons (full-polar)
%CPU time RANS vs DDES

\section{Conclusions \label{sec:conclusions}}
An extensive sensitivity study has been performed using steady and unsteady RANS as well as hybrid RANS/LES methods to predict stall for NASA's high-lift configuration of the Common Research Model (CRM-HL) at free-air conditions. 
The present contribution provides quantitative results for a large number of simulations testing widely-used turbulence models and initialisation strategies to the community, to further establish the current state-of-the-art of RANS-based simulations in terms of accuracy and computational costs.
The present study extends the work of \citet{DHFMK2022} and \citet{Browne2022},  who considered RANS, URANS and Hybrid RANS/LES for structured over-set to unstructured grids. 
While \cite{Browne2022} limited their ZDES study to SA-based turbulence models, \cite{DHFMK2022} considered one SST model for their RANS analysis. The present work tests a significantly wider range of turbulence models including variations of Menter's Shear-Stress Transport model for steady and unsteady RANS as well as DDES.
In terms of unsteady flow characteristics, present simulation times are extended up to $>160$ CTUs, which is more than twice as long as those in \cite{Browne2022}.
Eventually, the present study and learned learned allow us to define recommended practises for high-lift predictions of aircraft models.

Near stall ($C_{L,max}\approx2.5$) at an angle of attack of $\alpha=19.57$, a large spread of RANS results was observed for different turbulence models ($\Delta C_L=0.27$) and initialization strategies ($\Delta C_L=0.24$). Iterative convergence appeared less crucial for present simulations, but the use of Selective Frequency Damping (SFD) did not deliver machine-precision convergence of RANS solutions.
Even though results agree well with experimental measurements at low angles of attack, steady as well as unsteady RANS simulations were unable to predict the correct flow physics at increased $\alpha$. 
Non-physical flow separation was observed for all results near the wing tips even for relatively fine grids. However, we found strong indications that sensitivities to initialization strategies are grid dependent. 
Further grid refinement or adaptive mesh refinement may reduce sensitivity to initial solutions near $C_{L,max}$, but may not solve issues related to turbulence models without compromising the cost-effectiveness of RANS\citep{Alauzet2022}.

Delayed Detached Eddy Simulations (DDES), showed good accuracy with wind-tunnel measurements of approximately $\pm 5\%$ near $C_{L,max}$. At reduced angles of attack, however, present DDES results showed excessive separation along the flaps leading to discrepancies with experimental measurements ($\overline{C_M}/C_{M,exp}\approx0.8$ at $\alpha=7.05^{\circ}$). Even though \cite{Browne2022} showed that $C_L$ approaches experimental values using D level grids at low angles of attack, they still observe significant discrepancies in $C_M$. Considering SST-based DDES instead of an SA-based approach does not improve results, but leads to increased costs. Considering different initialization strategies and turbulence models, it is remarkable how consistent DDES results are for a given grid, compared to purely RANS-based methods.

At lower angles of attack, warm-started RANS simulations using SA models are recommended. Due to the limitations discussed in \ref{sec:results_freeair_rans}, we suggest to consider scale-resolving methods - such as hybrid RANS/LES - near stall. Even though the increased fidelity leads to ten times higher costs compared to RANS, considering modern CPU clusters like JSS3, such simulations can still be performed within $24$ hours using $5\%$ of the available computational resources.
% While selecting warm-started RANS simulations using SA models for present best-practice at lower angles of attack, we suggest to consider scale-resolving methods - such as hybrid RANS/LES - near stall, even though costs increase for DDES by a factor of $10$ compared to RANS. 
%Besides potential sensitivities to the shielding function of hybrid RANS/LES, it is also important to study wind-tunnel confinement effects in the future.
Future studies will focus on wind-tunnel confinement effects and the effect of shielding function formulation for hybrid RANS/LES.

\section*{Appendix}
\subsection{Steady RANS}\label{App:FA_RANS}
\begin{figure}
    \centering
    \begin{subfigure}[b]{0.33\textwidth}
    	a)SA-noft2\\
        \includegraphics[width=\textwidth, trim=10 10 10 10, clip]{Figures_FA_RANS/Cf/SA10_cold_07p05.png}
        %\caption{Subplot 1}
    \end{subfigure}
    \begin{subfigure}[b]{0.33\textwidth}
    	b)\\
        \includegraphics[width=\textwidth, trim=10 10 10 10, clip]{Figures_FA_RANS/Cf/SA10_cold_17p05.png}
        %\caption{Subplot 2}
    \end{subfigure}
    \begin{subfigure}[b]{0.33\textwidth}
    	c)\\
        \includegraphics[width=\textwidth, trim=10 10 10 10, clip]{Figures_FA_RANS/Cf/SA10_cold_19p57.png}
        %\caption{Subplot 3}
    \end{subfigure}\\
    \vspace{0.1cm} % adjust vertical spacing between rows of subplots
%%%%%%%%%%%%%%%%%%%%%%%%%%%%%%%%%%%%%%%%%%%%%%%%%%%%%%%%%%%%%%%%%%%%%%%%%%%%%%%%%%%%%%%%%%%%%%%%%%
    \begin{subfigure}[b]{0.33\textwidth}
    	d)SA-noft2-R\\
        \includegraphics[width=\textwidth, trim=10 10 10 10, clip]{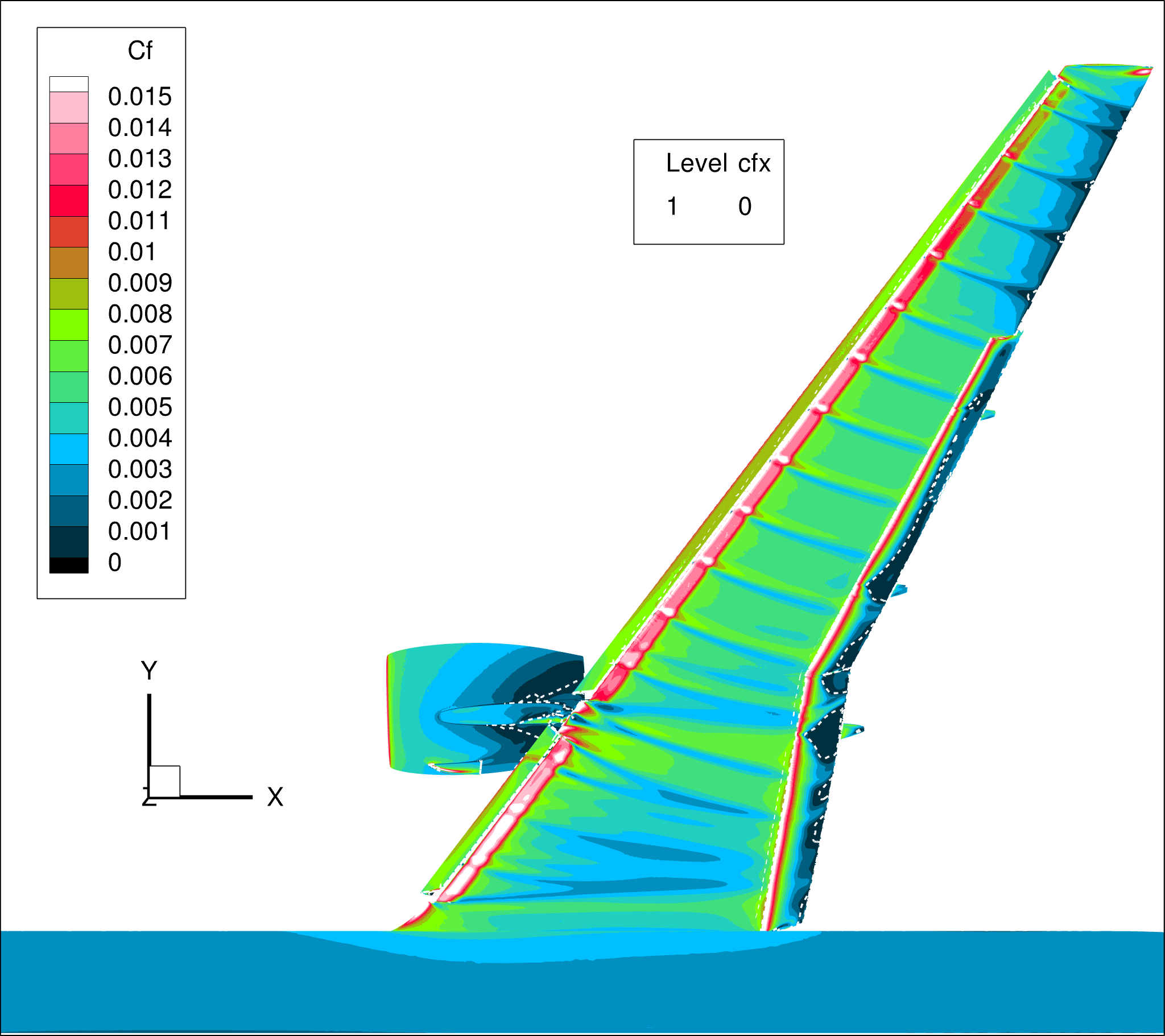}
        %\caption{Subplot 1}
    \end{subfigure}
    \begin{subfigure}[b]{0.33\textwidth}
    	e)\\
        \includegraphics[width=\textwidth, trim=10 10 10 10, clip]{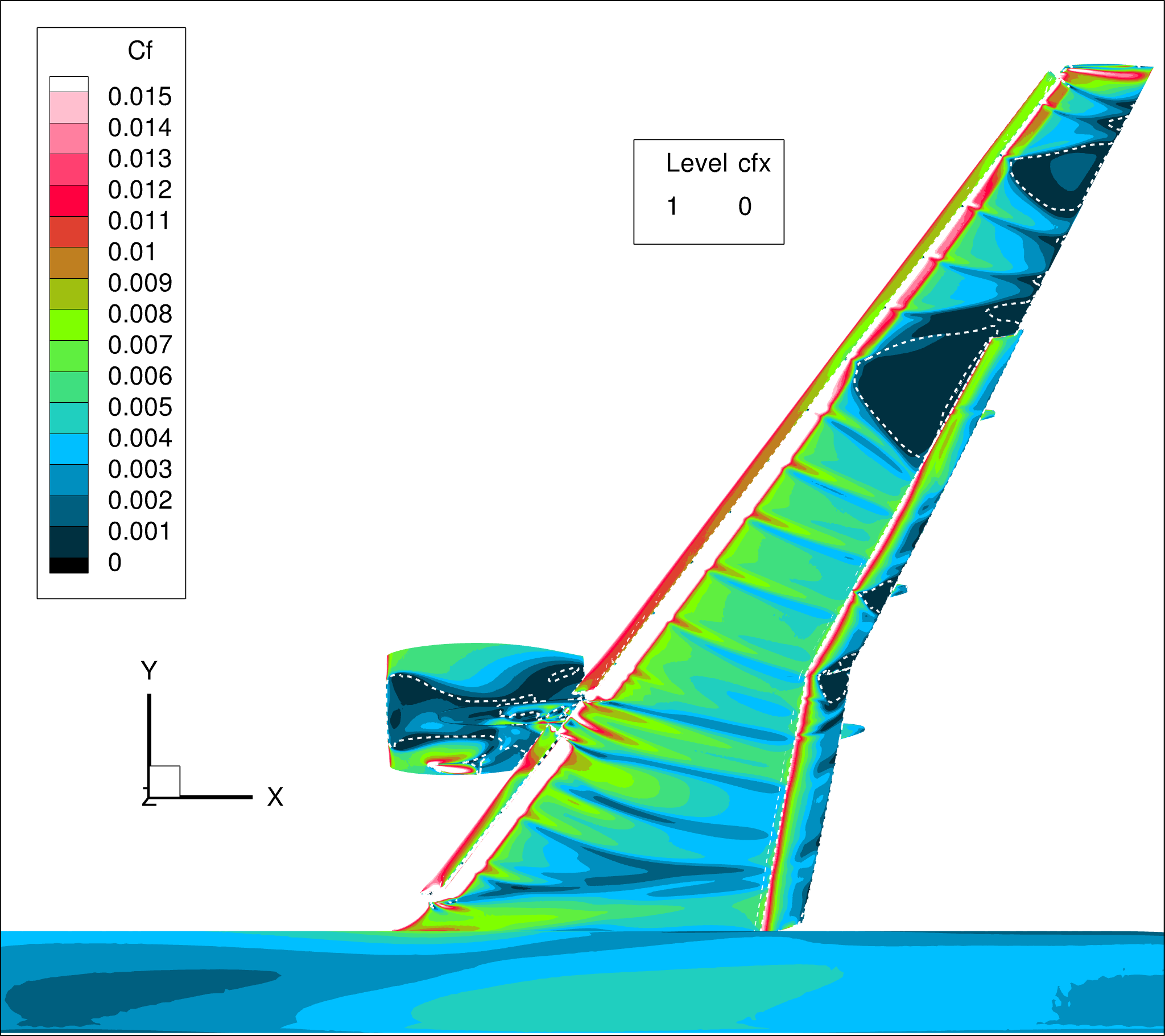}
        %\caption{Subplot 2}
    \end{subfigure}
    \begin{subfigure}[b]{0.33\textwidth}
    	f)\\
        \includegraphics[width=\textwidth, trim=10 10 10 10, clip]{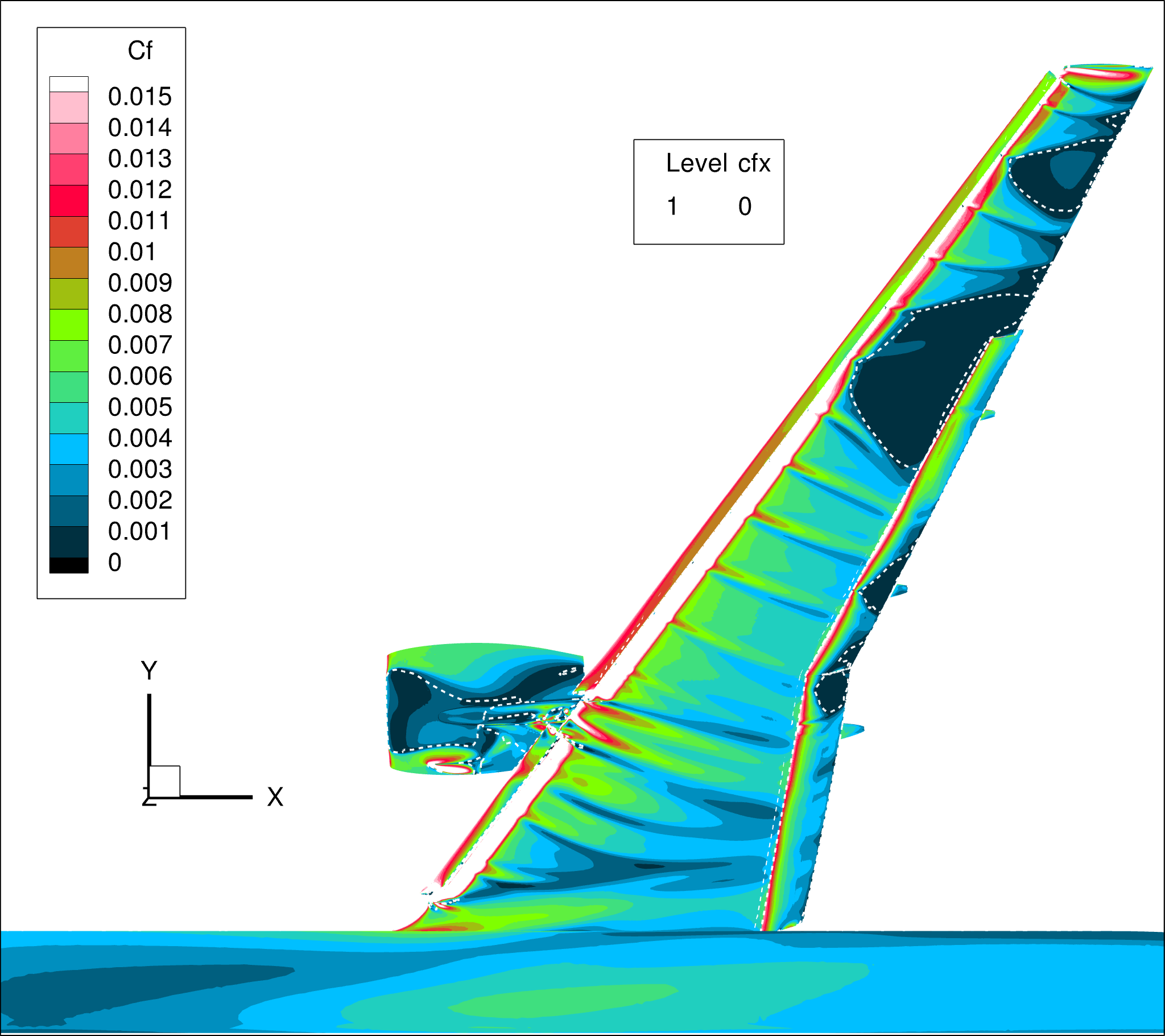}
        %\caption{Subplot 3}
    \end{subfigure}\\
    \vspace{0.1cm} % adjust vertical spacing between rows of subplots
%%%%%%%%%%%%%%%%%%%%%%%%%%%%%%%%%%%%%%%%%%%%%%%%%%%%%%%%%%%%%%%%%%%%%%%%%%%%%%%%%%%%%%%%%%%%%%%%%%
    \begin{subfigure}[b]{0.33\textwidth}
    	g)SA-noft2-R-QCR2000\\
        \includegraphics[width=\textwidth, trim=10 10 10 10, clip]{Figures_FA_RANS/Cf/SA12_cold_07p05.png}
        %\caption{Subplot 1}
    \end{subfigure}
    \begin{subfigure}[b]{0.33\textwidth}
    	h)\\
        \includegraphics[width=\textwidth, trim=10 10 10 10, clip]{Figures_FA_RANS/Cf/SA12_cold_17p05.png}
        %\caption{Subplot 2}
    \end{subfigure}
    \begin{subfigure}[b]{0.33\textwidth}
    	i)\\
        \includegraphics[width=\textwidth, trim=10 10 10 10, clip]{Figures_FA_RANS/Cf/SA12_cold_19p57.png}
        %\caption{Subplot 3}
    \end{subfigure}\\
    \vspace{0.1cm} % adjust vertical spacing between rows of subplots
%%%%%%%%%%%%%%%%%%%%%%%%%%%%%%%%%%%%%%%%%%%%%%%%%%%%%%%%%%%%%%%%%%%%%%%%%%%%%%%%%%%%%%%%%%%%%%%%%%
    \begin{subfigure}[b]{0.33\textwidth}
    	j)SA-noft2-RC\\
        \includegraphics[width=\textwidth, trim=10 10 10 10, clip]{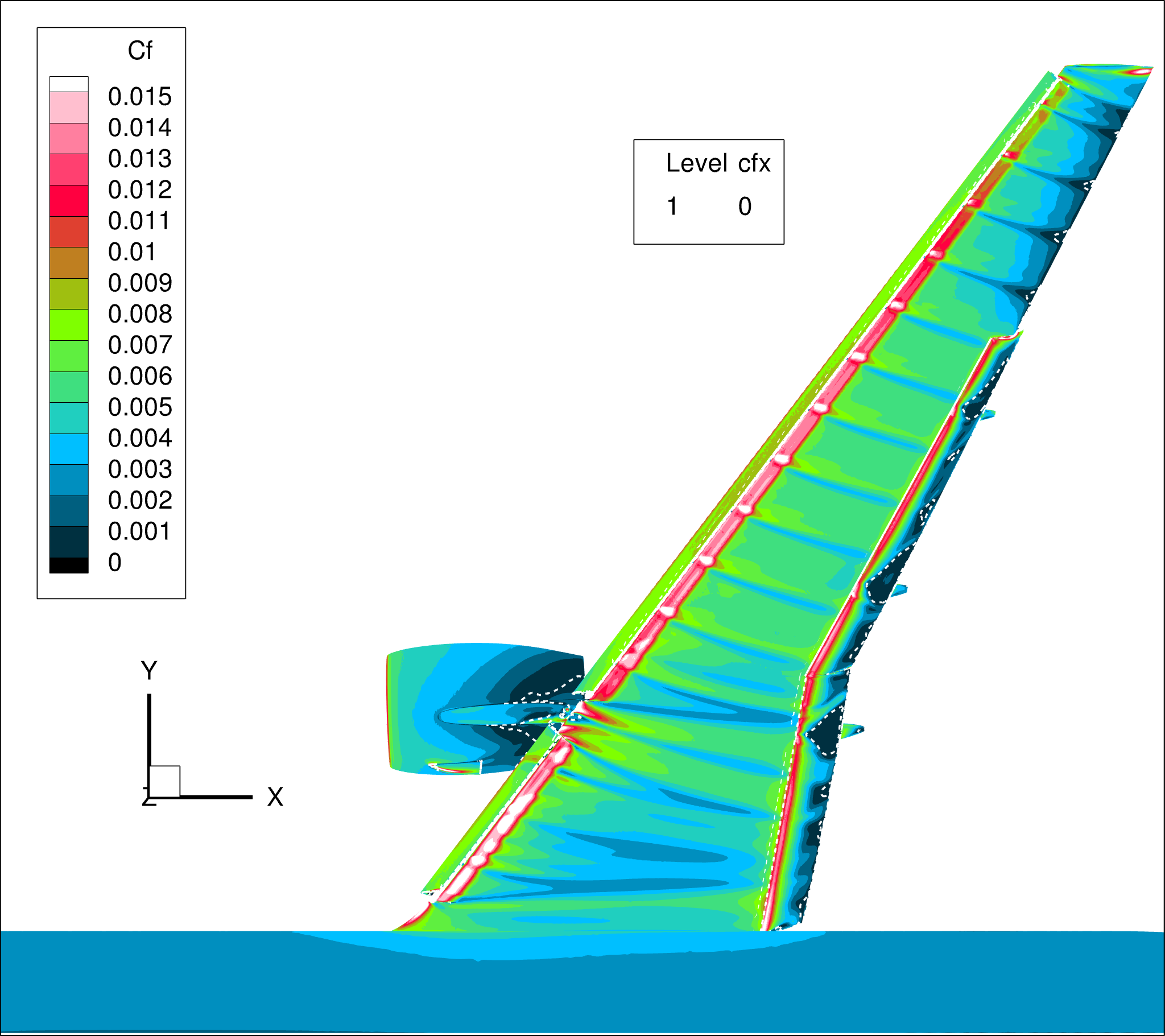}
        %\caption{Subplot 1}
    \end{subfigure}
    \begin{subfigure}[b]{0.33\textwidth}
    	k)\\
        \includegraphics[width=\textwidth, trim=10 10 10 10, clip]{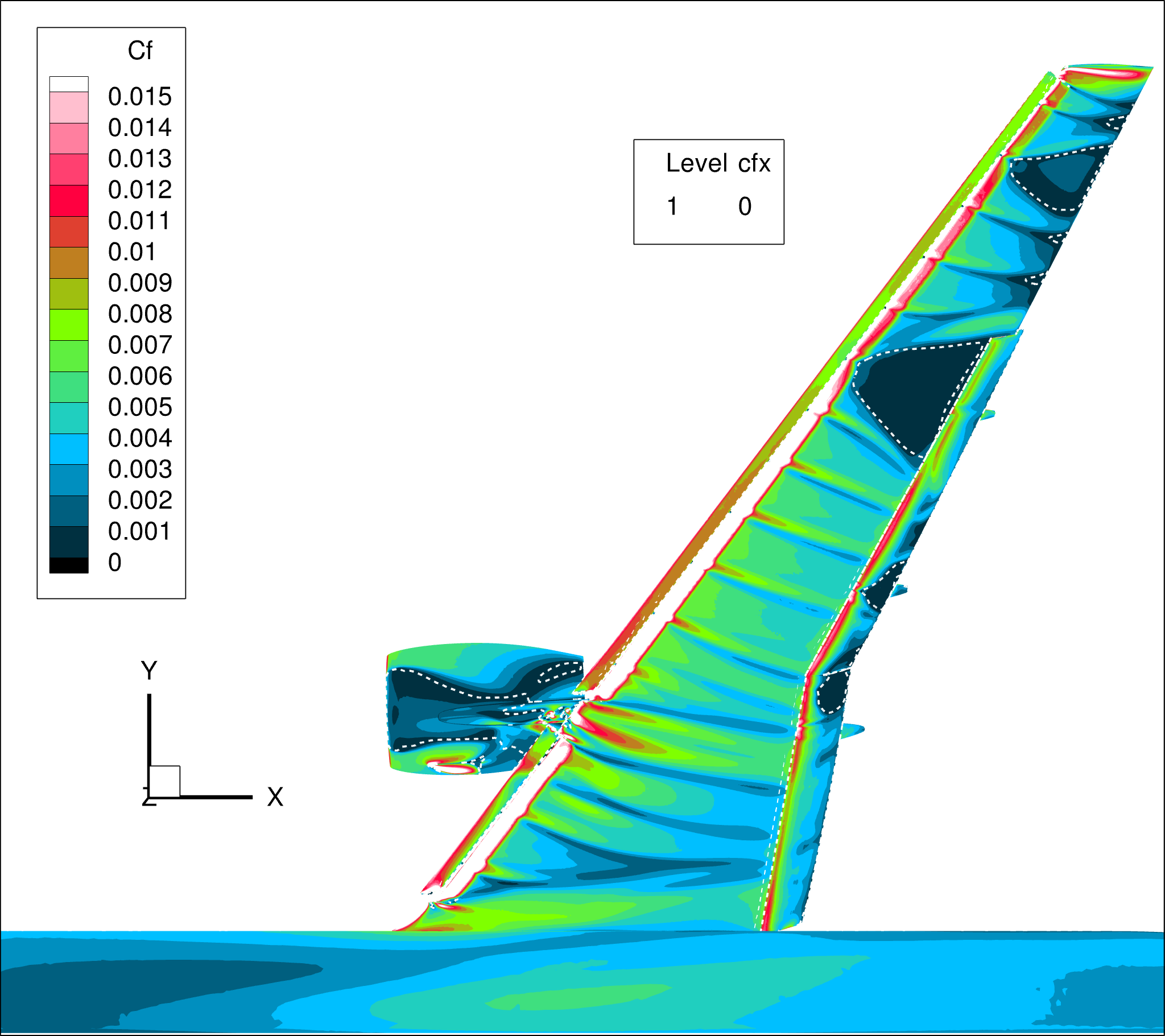}
        %\caption{Subplot 2}
    \end{subfigure}
    \begin{subfigure}[b]{0.33\textwidth}
    	l)\\
        \includegraphics[width=\textwidth, trim=10 10 10 10, clip]{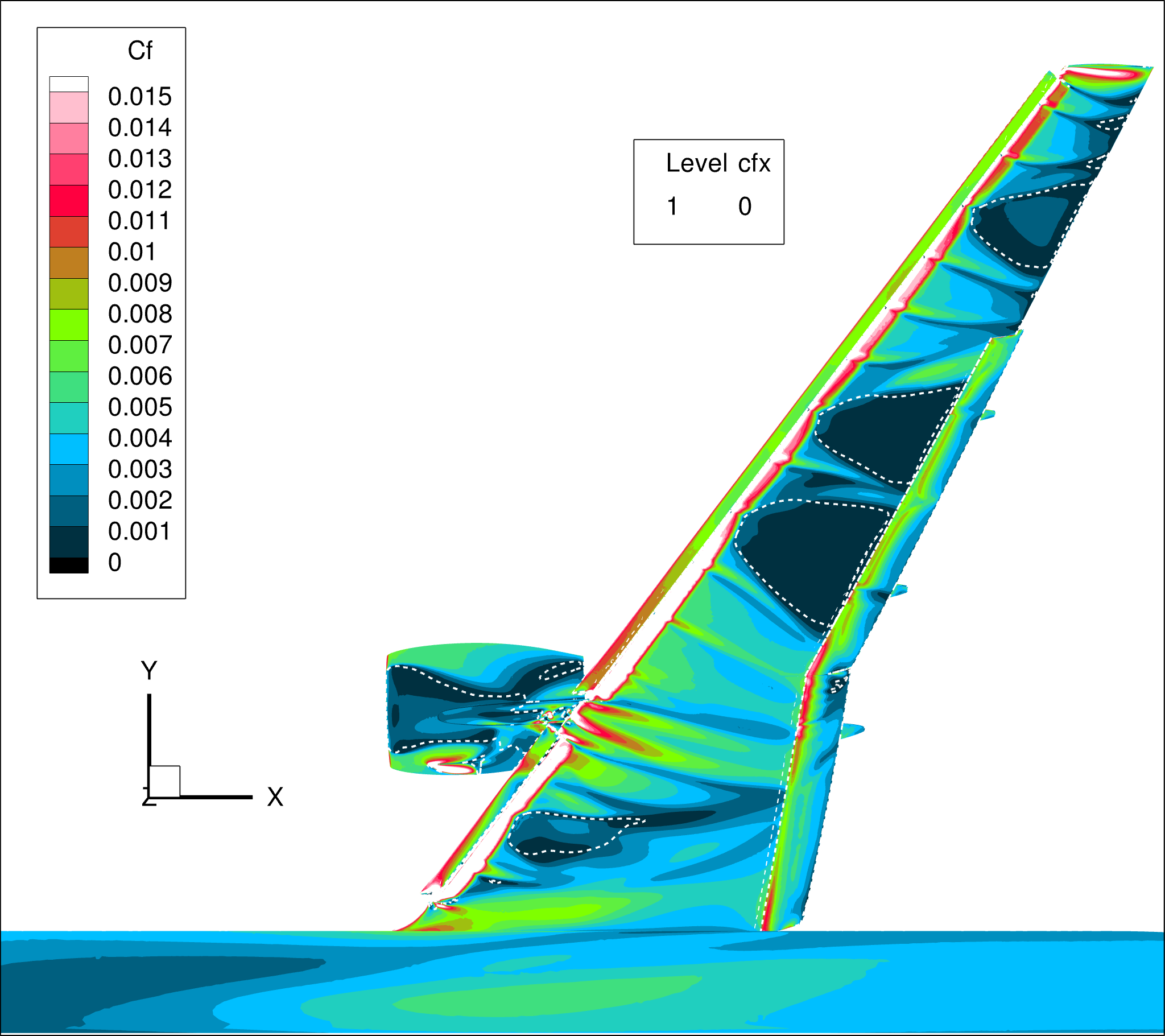}
        %\caption{Subplot 3}
    \end{subfigure}\\
%    \vspace{0.1cm} % adjust vertical spacing between rows of subplots
%%%%%%%%%%%%%%%%%%%%%%%%%%%%%%%%%%%%%%%%%%%%%%%%%%%%%%%%%%%%%%%%%%%%%%%%%%%%%%%%%%%%%%%%%%%%%%%%%%
    \caption{Skin-friction coefficient contour plots for SA turbulence models}
    \label{fig:APP_SA_Cf}
\end{figure}

Tables \ref{tab:Sum_SA} and \ref{tab:Sum_SST} summarize aerodynamic coefficients and performance parameters for all steady RANS simulations using Spalart-Allmaras and Shear-Stress-Transport models, respectively.
The convergence parameter CPA together with the $C_L$ and $C_M$ ratios of simulation results over experimental values are chosen as performance indicators. In addition, the aerodynamic coefficients as well as the L2-norm of residuals of density and model conservative variable are listed as well. Mean values are averaged over the last $10,\!000$ iterations.
Figures \ref{fig:APP_SA_Cf}, \ref{fig:APP_SST_Cf}, and \ref{fig:APP_SST_Cf_cont} show skin-friction contour plots for all cold-started simulations using Spalart-Allmaras as well as Shear-Stress Transport turbulence models. These figures complement \S \ref{sec:results_freeair_rans} and confirm that no turbulence model is capable of overcoming the appearance of non-physical flow separation at increased angles of attack.
\begin{table}
  \begin{center}
\def~{\hphantom{0}}
  \begin{tabular}{rcc|ccc|ccc|cc}
    \multicolumn{3}{c|}{Settings} & \multicolumn{3}{c|}{Performance indicators} & \multicolumn{3}{c|}{Aerodyn. coefficients} & \multicolumn{2}{c}{Residuals} \\
    AoA & SA corr. & Initial & $\log({CPA})$ & $\overline{C_L}/C_{L,exp}$ & $\overline{C_M}/C_{M,exp}$ & $\overline{C_L}$ & $\overline{C_D}$ & $\overline{C_M}$ & $R_{\rho}$ & $R_{turb}$ \\ 
	\hline
	7.05 & \textit{-} & cold & -5.69 & 1.005 & 0.98 &1.79 & 0.19 & -0.36 & 7.7e-10 & 1.9e-6 \\
	7.05 & \textit{-} & warm & -5.49 & 1.005 & 0.98 &1.79 & 0.19 & -0.36 & 7.7e-10 & 1.9e-6 \\
	7.05 & \textit{R-QCR} 	& cold & -3.26 & 0.97 & 0.90 & 1.73 & 0.18 & -0.33 & 7.3e-11 & 2.6e-6 \\
	7.05 & \textit{R-QCR} 	& warm 	& -5.31	& 0.99 	& 0.91 	& 1.76 	& 0.18 	& -0.34	& 7.0e-11	& 1.9e-6 \\
	7.05 & \textit{R} 		& cold	& -6.50 & 0.99	& 0.94 	& 1.76	& 0.18	& -0.35 & 7.1e-11	& 4.0e-1 \\
	7.05 & \textit{RC} 		& cold	& -3.74 & 1.01	& 1.00	& 1.80	& 0.19	& -0.37 & 1.2e-10	& 2.7e-6 \\
	\hline
	17.05 & \textit{-} & cold & -3.91 & 0.92 & 0.50 & 2.31 & 0.37 & -0.14 & 7.4e-11 & 3.9e-6 \\
	17.05 & \textit{-} & warm & -3.07 & 0.99 & 0.90 & 2.48 & 0.35 & -0.26 & 6.7e-11 & 5.5e-6 \\
	17.05 & \textit{R-QCR} & cold & -3.24 & 0.89 & 0.38 & 2.22 & 0.37 & -0.11 & 7.5e-11 & 2.7e-6 \\
	17.05 & \textit{R-QCR} & warm & -3.98 & 0.98 & 0.80 & 2.44 & 0.35 & -0.23 & 6.7e-11 & 2.6e-6 \\
	17.05 & \textit{R} 		& cold	&  -3.61 & 0.92 & 0.45 & 2.30 & 0.37 & -0.13 & 7.3e-11 & 2.8e-6 \\	
	17.05 & \textit{R} 		& warm	&  -3.63 & 0.99 & 0.88 & 2.47 & 0.35 & -0.25 & 6.1e-11 & 2.5e-6\\		
    17.05 & \textit{RC} 	& cold	&  -2.39 & 0.91 & 0.49 & 2.28 & 0.37 & -0.14 & 9.0e-10 & 1.3e-4\\
	\hline
	19.57 & \textit{-} & cold & -4.32 & 0.91 & 0.26 & 2.30 & 0.42 & -0.06 & 8.3e-11 & 4.7e-6 \\
	19.57 & \textit{-} & warm & -5.16 & 1.01 & 0.83 & 2.54 & 0.39 & -0.20 & 7.1e-11 & 4.1e-6 \\
	19.57 & \textit{R-QCR} & cold & -2.35 & 0.90 & 0.12 & 2.26 & 0.42 & -0.03 & 8.9e-11 & 3.9e-6 \\
	19.57 & \textit{R-QCR} & warm & -2.37 & 0.98 & 0.73 & 2.48 & 0.38 & -0.18 & 9.3e-11 & 8.1e-6 \\
	19.57 & \textit{R} 		& cold	&  -4.27 & 0.92 & 0.29 & 2.32 & 0.41 & -0.07 & 7.2e-11 & 3.0e-6\\
	19.57 & \textit{RC} 	& cold	&  -1.62 & 0.85 & 0.20 & 2.13 & 0.42 & -0.05 & 1.1e-9 & 1.4e-4 \\
	\hline
	21.47 & \textit{-} & cold & -2.95 & 0.92 & 0.09 & 2.14 & 0.45 & -0.03 & 6.7e-11 & 5.0e-6 \\
	21.47 & \textit{-} & warm & -0.90 & 0.99 & 0.76 & 2.31 & 0.44 & -0.01 & 7.9e-11 & 5.2e-5 \\
	21.47 & \textit{R-QCR} & cold & -1.95 & 0.84 & 0.22 & 1.94 & 0.43 & -0.07 & 7.2e-11 & 3.9e-6\\
	21.47 & \textit{R-QCR} & warm & -3.35 & 0.91 & 0.70 & 2.12 & 0.43 & -0.21 & 8.0e-11 & 3.8e-6 \\
	21.47 & \textit{R} 		& cold	& -1.76 & 0.90 & 0.20 & 2.09 & 0.45 & -0.06 & 8.0e-11 & 5.0e-6\\
	21.47 & \textit{RC} 	& cold	& -2.36 & 0.92 & 0.26 & 2.13 & 0.46 & -0.08 & 8.3e-10 & 1.3e-4\\
  \end{tabular}
  \caption{
  Summary of selected cold- and warm-started results for baseline SA-noft2 and SA-noft2-R-QCR2000 model.
  %Summary of selected cold- and warm-started results for baseline SA-noft2 and SA-noft2-R-QCR2000 models, supplemented by additional cold-started results using SA corrections. The convergence parameter CPA together with the $C_L$ and $C_M$ ratios of simulation results over experimental values are chosen as performance indicators. In addition, the aerodynamic coefficients as well as the L2-norm of residuals of density and model conservative variable are listed as well. Mean values are averaged over the last $10,\!000$ iterations.
  }
  \label{tab:Sum_SA}
  \end{center}
\end{table}
\begin{table}
  \begin{center}
\def~{\hphantom{0}}
  \begin{tabular}{rcc|ccc|ccc|cc}
    \multicolumn{3}{c|}{Settings} & \multicolumn{3}{c|}{Performance indicators} & \multicolumn{3}{c|}{Aerodyn. coefficients} & \multicolumn{2}{c}{Residuals} \\
    AoA & SST corr. & Initial & $\log({CPA})$ & $\overline{C_L}/C_{L,exp}$ & $\overline{C_M}/C_{M,exp}$ & $\overline{C_L}$ & $\overline{C_D}$ & $\overline{C_M}$ & $R_{\rho}$ & $R_{turb}$ \\ 
	\hline
	07.05 & \textit{-} & cold 			& -3.41 & 0.97 & 0.89 & 1.72 & 0.18 & -0.33 & 6.9e-11 & 1.6e-9 \\
	07.05 & \textit{2003} & cold 		& -4.39 & 0.97 & 0.89 & 1.72 & 0.18 & -0.33 & 7.8e-11 & 1.6e-9 \\ % 22
    07.05 & \textit{V} & cold 			& -3.19 & 0.97 & 0.89 & 1.72 & 0.18 & -0.33 & 1.2e-10 & 1.6e-9 \\ %21
	07.05 & \textit{sust} & cold 		& -3.41 & 0.97 & 0.89 & 1.73 & 0.18 & -0.33 & 5.9e-11 & 1.6e-9 \\ %23
	07.05 & \textit{V-sust} & cold 		& -4.74 & 0.97 & 0.89 & 1.73 & 0.18 & -0.33 & 8.3e-11 & 1.6e-9 \\ %24
	07.05 & \textit{2003-sust} & cold 	& -5.17 & 0.97 & 0.89 & 1.72 & 0.18 & -0.33 & 7.7e-11 & 1.6e-9 \\ %25	
	\hline
	17.05 & \textit{-} & cold 			& -2.33 & 0.91 & 0.42 & 2.25 & 0.36 & -0.12 & 6.1e-10 & 1.6e-9\\
	17.05 & \textit{2003} & cold 		& -3.52 & 0.89 & 0.35 & 2.20 & 0.37 & -0.10 & 6.6e-11 & 1.6e-9 \\ % 22
    17.05 & \textit{2003} & warm 		& -3.65 & 0.98 & 0.85 & 2.44 & 0.34 & -0.25 & 6.8e-11 & 1.6e-9 \\ % 22
    17.05 & \textit{V} & cold 			& -3.18 & 0.91 & 0.42 & 2.26 & 0.36 & -0.12 & 1.8e-10 & 1.6e-9\\ %21
	17.05 & \textit{sust} & cold 		& -2.36 & 0.91 & 0.38 & 2.25 & 0.37 & -0.11 & 6.2e-10 & 1.6e-9 \\ %23
	17.05 & \textit{V-sust} & cold 		& -3.03 & 0.91 & 0.38 & 2.25 & 0.37 & -0.11 & 1.6e-10 & 1.6e-9  \\ %24
	17.05 & \textit{2003-sust} & cold 	& -4.04 & 0.90 & 0.14 & 2.24 & 0.37 & -0.04 & 7.2e-11 & 1.6e-9 \\ %25	
	\hline
	19.57 & \textit{-} & cold 			& -1.75 & 0.88 & 0.12  & 2.22 & 0.41 & -0.03 & 7.6e-10 & 1.6e-9\\
	19.57 & \textit{2003} & cold 		& -2.10 & 0.85 & -0.04 & 2.13 & 0.43 & 0.01 & 8.2e-11 & 1.5e-9 \\ % 22
    19.57 & \textit{2003} & warm 		& -2.35 & 0.98 & 0.70  & 2.48 & 0.38 & -0.17 & 2.3e-10 & 1.5e-9\\ % 22
	19.57 & \textit{V} & cold 			& -1.64 & 0.83 & 0.00  & 2.09 & 0.43 & 0.00 & 2.0e-10 & 1.5e-9 \\ %21
	19.57 & \textit{sust} & cold 		& -1.81 & 0.91 & 0.16  & 2.28 & 0.41 & -0.04 & 7.0e-10 & 1.6e-9\\ %23
	19.57 & \textit{V-sust} & cold 		& -2.56 & 0.82 & -0.08 & 2.06 & 0.43 & 0.02 & 1.9e-10 & 1.5e-9 \\ %24
	19.57 & \textit{2003-sust} & cold	& -1.48 & 0.82 & -0.12 & 2.05 & 0.43 & 0.03 & 3.1e-10 & 1.5e-9 \\ %25	
	\hline
	21.47 & \textit{-} & cold 			& -2.44 & 0.79 & 0.23 & 1.84 & 0.413 & -0.07 & 5.4e-10 & 1.6e-9\\
	21.47 & \textit{2003} & cold 		& -2.32 & 0.80 & 0.26 & 1.86 & 0.42 & -0.08 & 8.1e-11 & 1.6e-9 \\ % 22
    21.47 & \textit{2003} & warm 		& -1.08 & 0.91 & 0.16 & 2.13 & 0.45 & -0.06 & 2.3e-10 & 1.5e-9\\ % 22
	21.47 & \textit{V} & cold 			& -1.03 & 0.86 &-0.07 & 2.00 & 0.44 & 0.02 & 2.2e-10 & 1.5e-9 \\ %21
	21.47 & \textit{sust} & cold 		& -0.90 & 0.76 & 0.56 & 1.76 & 0.44 & -0.17 & 5.0e-10 & 1.6e-9 \\ %23
	21.47 & \textit{V-sust} & cold 		& -2.95 & 0.88 &-0.23 & 2.04 & 0.46 & 0.07 & 2.2e-10 & 1.5e-9 \\ %24
	21.47 & \textit{2003-sust} & cold 	& -2.50 & 0.89 &-0.26 & 2.07 & 0.47 & 0.08 & 8.4e-11 & 1.5e-9 \\ %25	
  \end{tabular}
  \caption{Summary of selected cold-started results using SST models. 
  %The convergence parameter CPA together with the $C_L$ and $C_M$ ratios of simulation results over experimental values are chosen as performance indicators. In addition, the aerodynamic coefficients as well as the L2-norm of residuals of density and model conservative variable are listed as well. Mean values are averaged over the last $10,\!000$ iterations.
  }
  \label{tab:Sum_SST}
  \end{center}
\end{table}

\begin{figure}
    \centering
    \begin{subfigure}[b]{0.33\textwidth}
    	a) Menter SST \\
        \includegraphics[width=\textwidth, trim=10 10 10 10, clip]{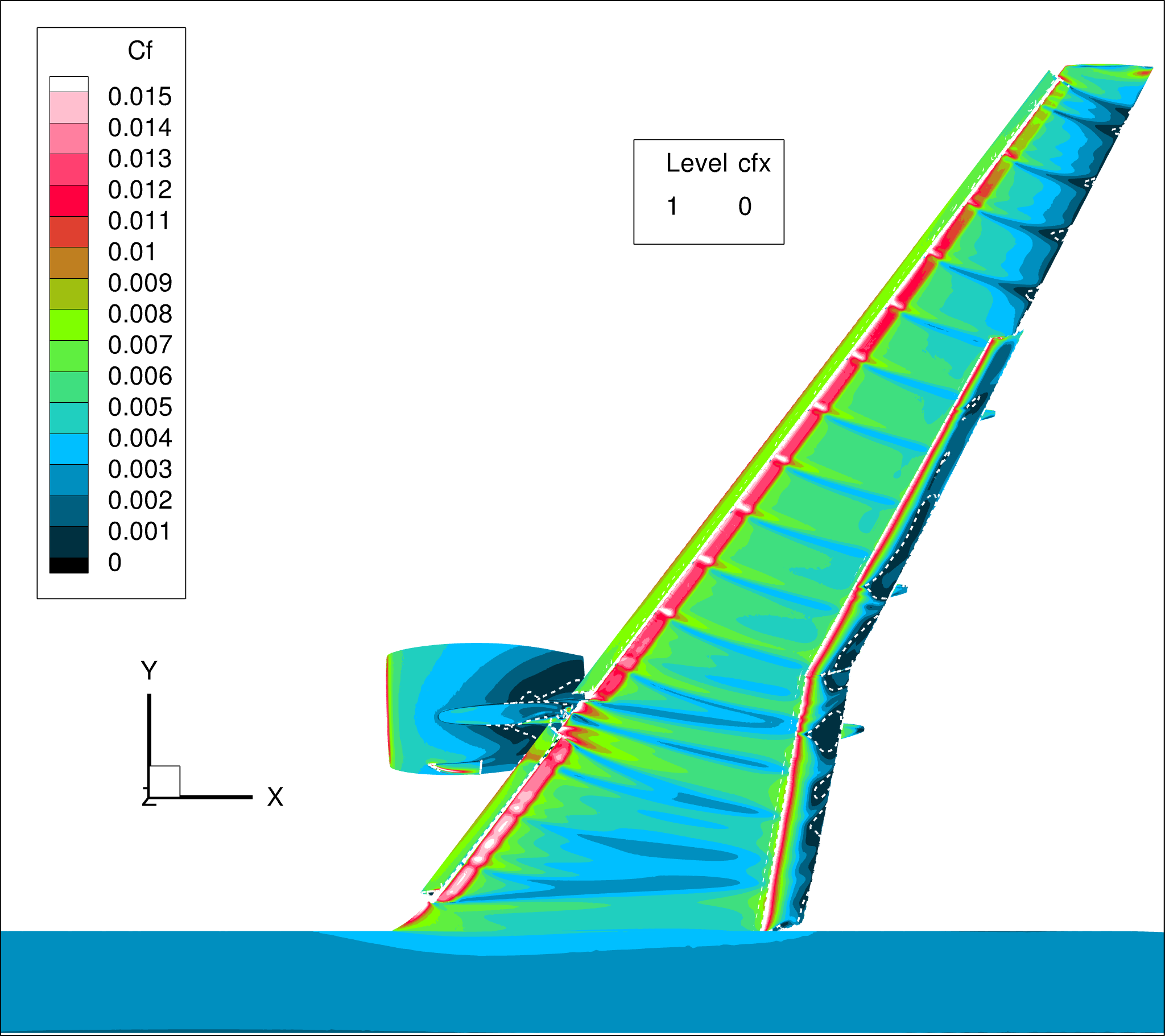}
        %\caption{Subplot 1}
    \end{subfigure}
    \begin{subfigure}[b]{0.33\textwidth}
    	b)\\
        \includegraphics[width=\textwidth, trim=10 10 10 10, clip]{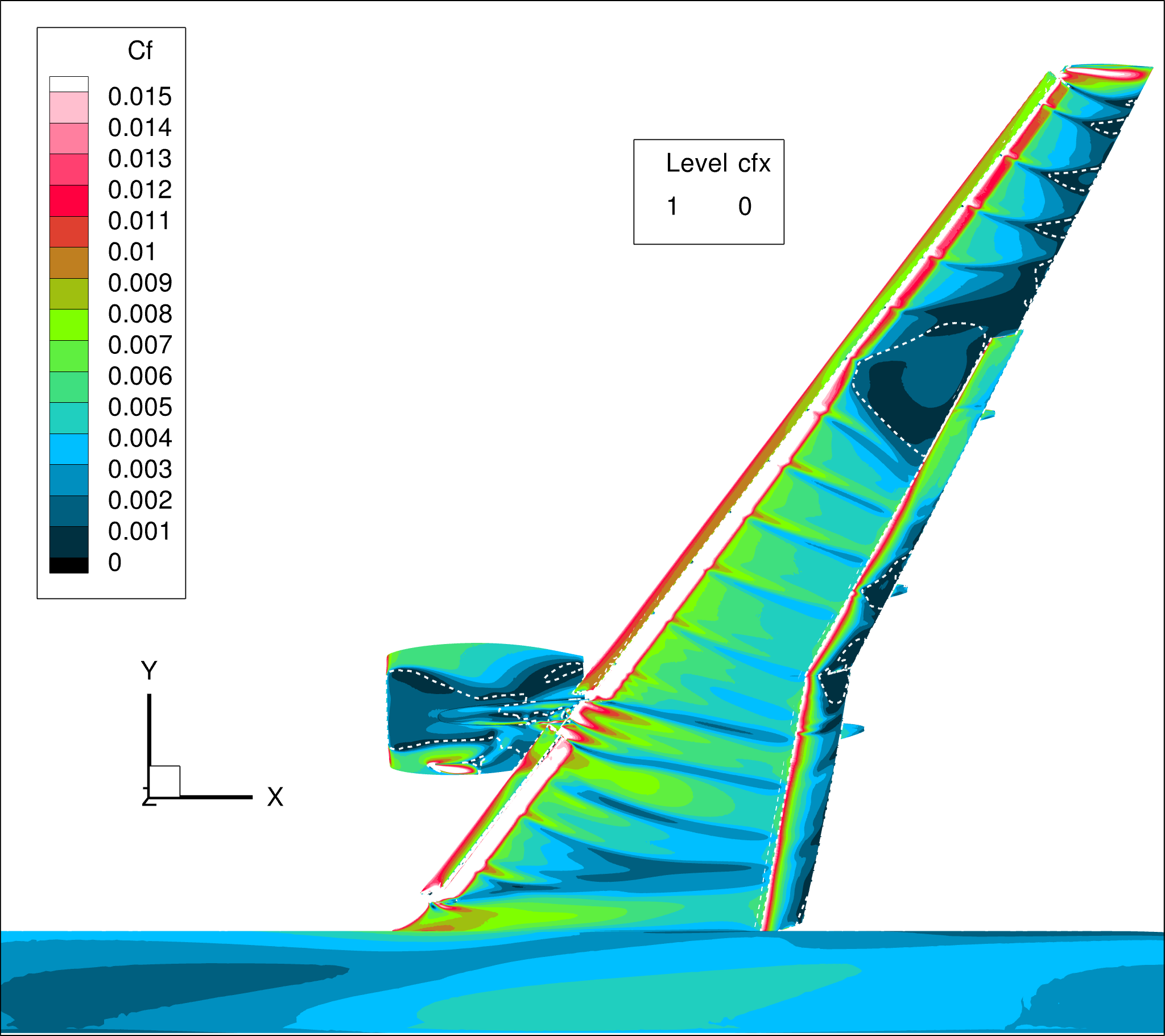}
        %\caption{Subplot 2}
    \end{subfigure}
    \begin{subfigure}[b]{0.33\textwidth}
    	c)\\
        \includegraphics[width=\textwidth, trim=10 10 10 10, clip]{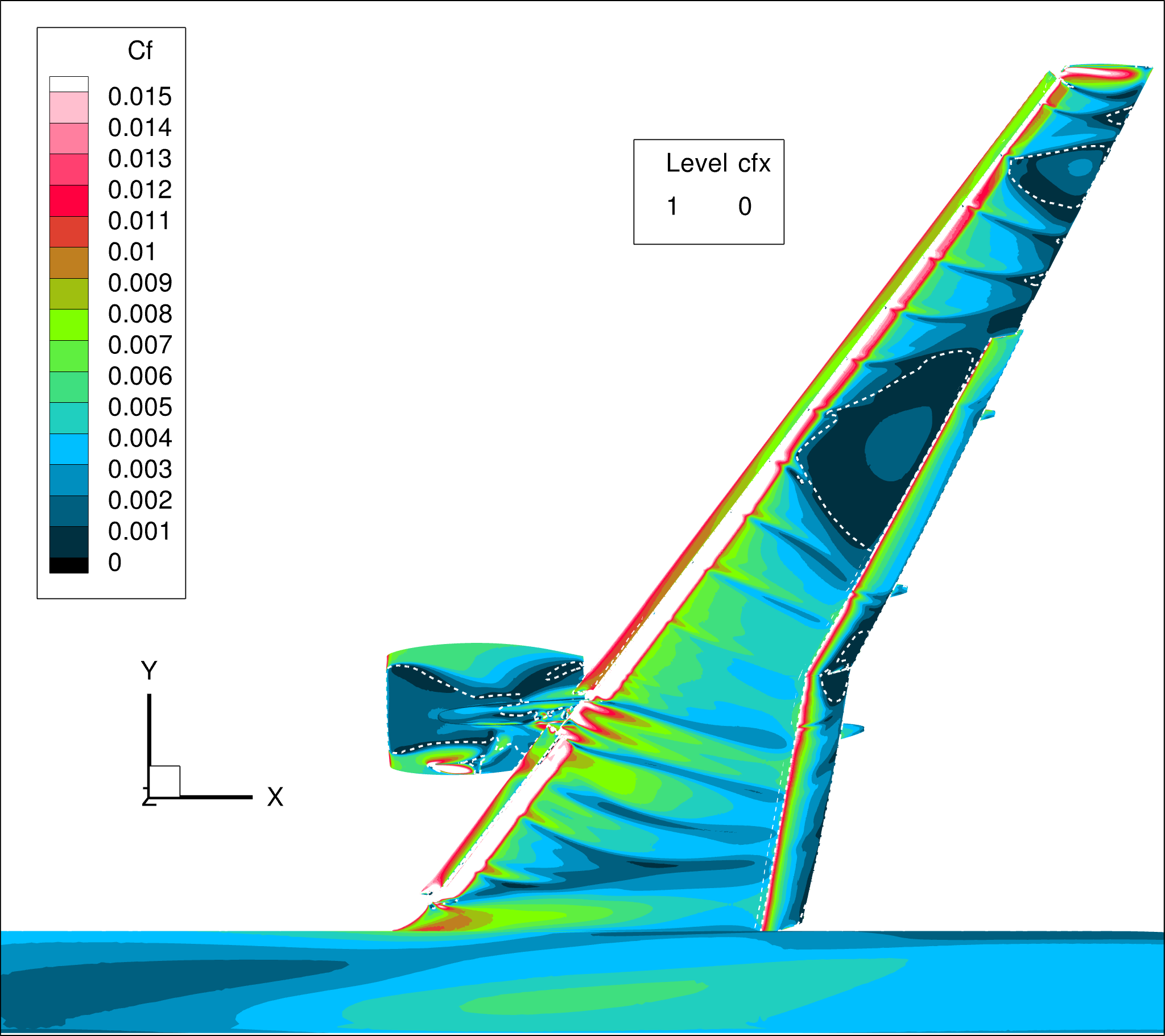}
        %\caption{Subplot 3}
    \end{subfigure}\\
    \vspace{0.1cm} % adjust vertical spacing between rows of subplots
%%%%%%%%%%%%%%%%%%%%%%%%%%%%%%%%%%%%%%%%%%%%%%%%%%%%%%%%%%%%%%%%%%%%%%%%%%%%%%%%%%%%%%%%%%%%%%%%%%
    \begin{subfigure}[b]{0.33\textwidth}
    	d)\\
        \includegraphics[width=\textwidth, trim=10 10 10 10, clip]{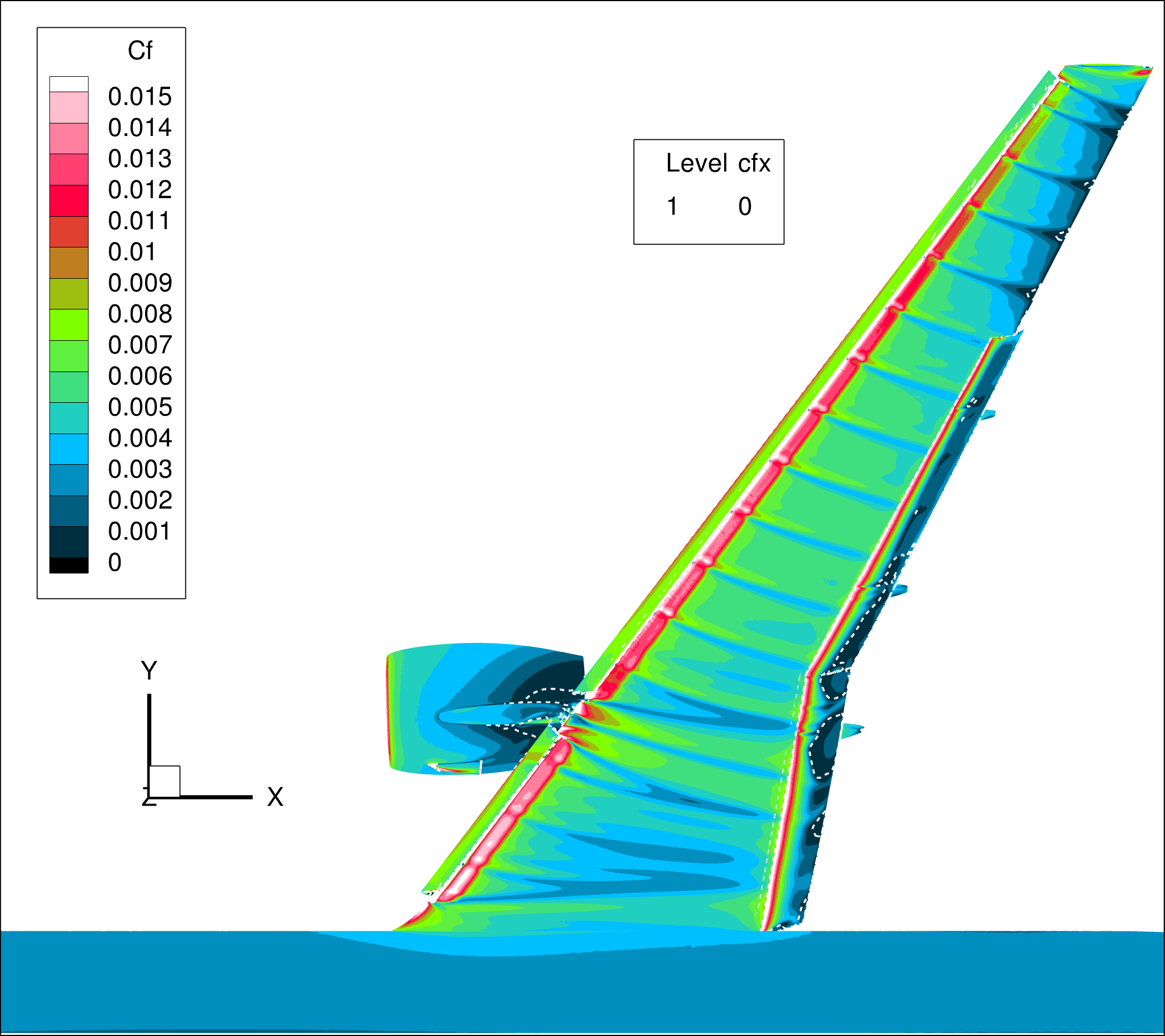}
        %\caption{Subplot 1}
    \end{subfigure}
    \begin{subfigure}[b]{0.33\textwidth}
    	e) Menter SST-V\\
        \includegraphics[width=\textwidth, trim=10 10 10 10, clip]{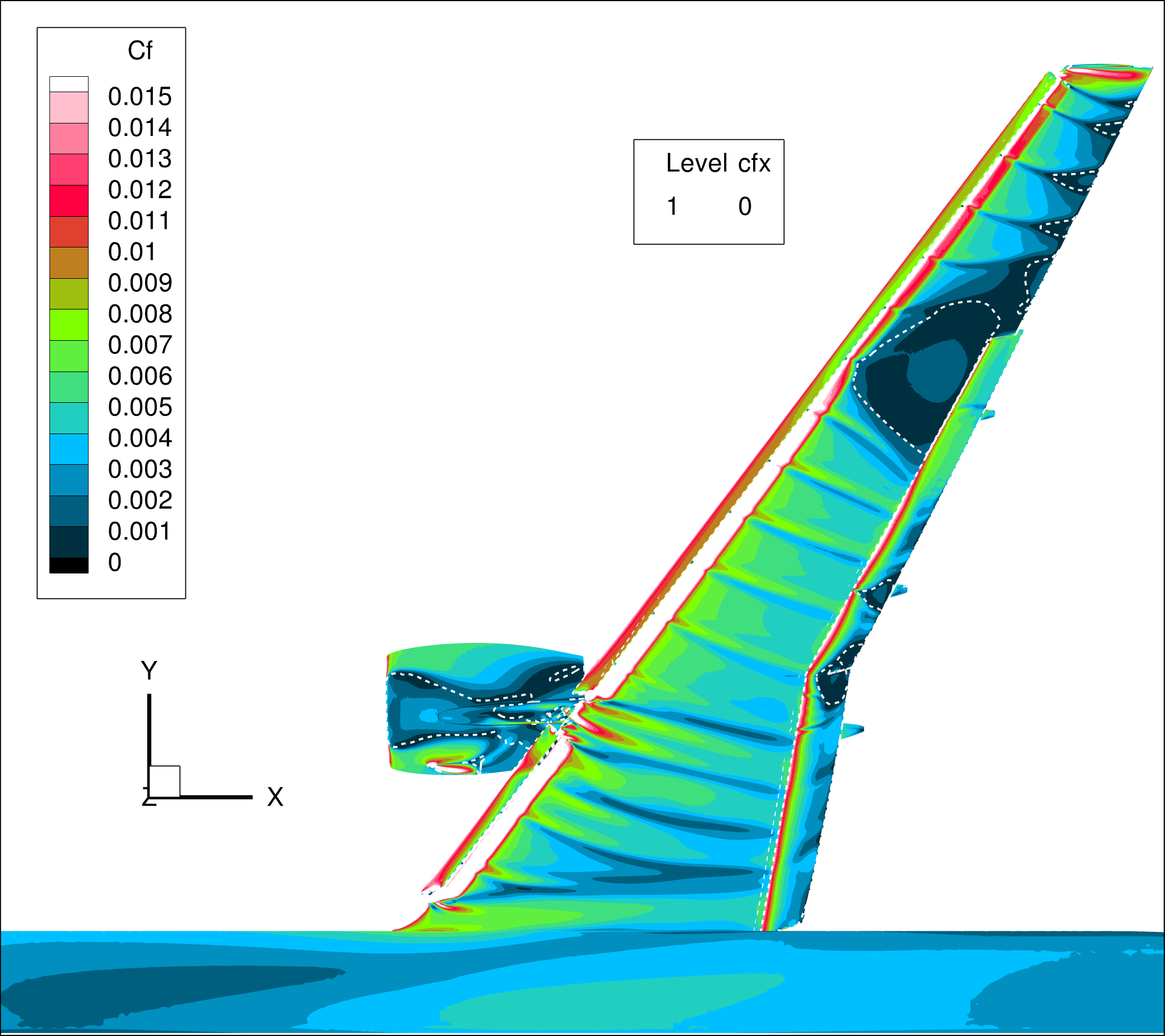}
        %\caption{Subplot 2}
    \end{subfigure}
    \begin{subfigure}[b]{0.33\textwidth}
    	f)\\
        \includegraphics[width=\textwidth, trim=10 10 10 10, clip]{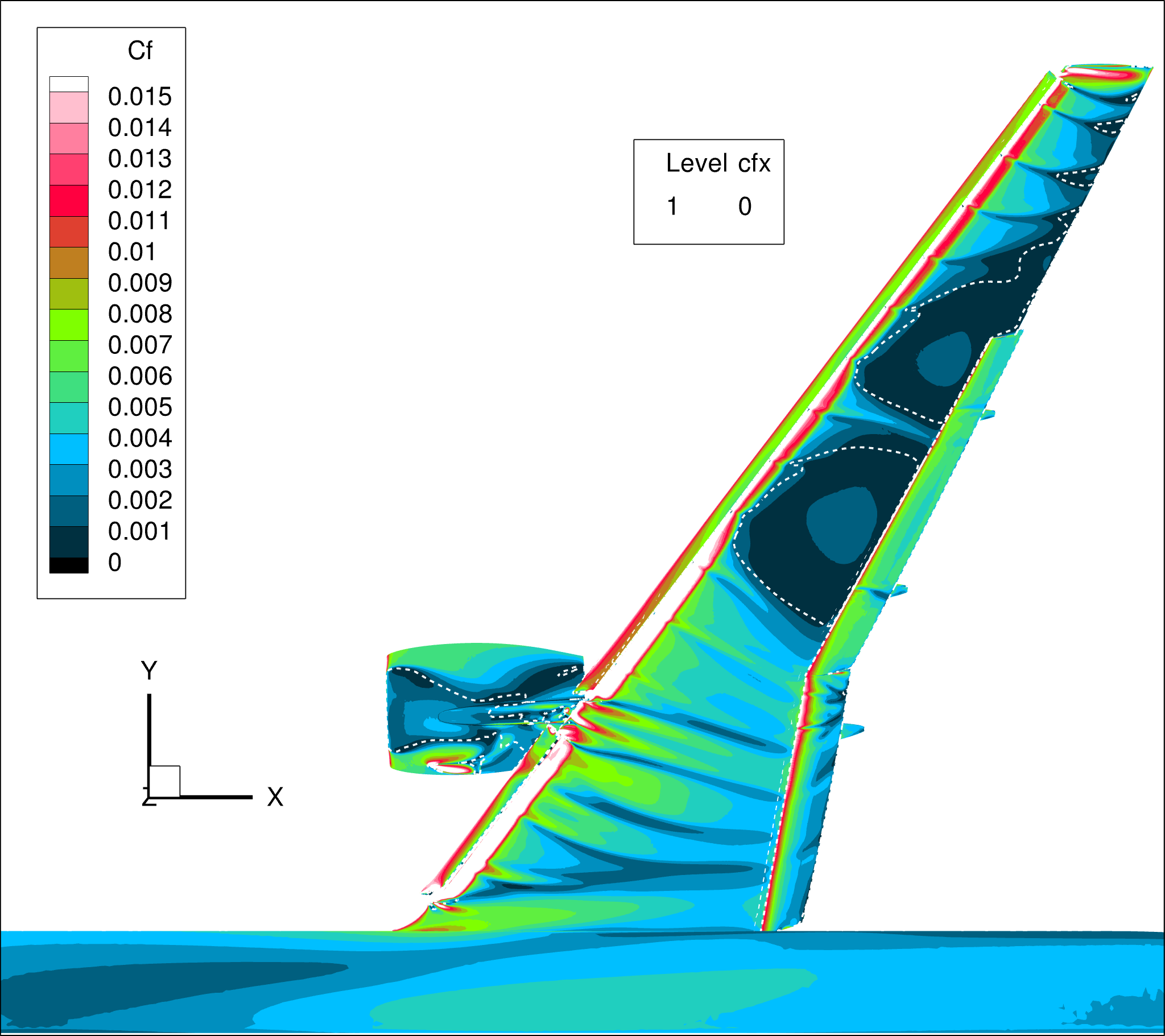}
        %\caption{Subplot 3}
    \end{subfigure}\\
    \vspace{0.1cm} % adjust vertical spacing between rows of subplots
%%%%%%%%%%%%%%%%%%%%%%%%%%%%%%%%%%%%%%%%%%%%%%%%%%%%%%%%%%%%%%%%%%%%%%%%%%%%%%%%%%%%%%%%%%%%%%%%%%
    \begin{subfigure}[b]{0.33\textwidth}
    	g) Menter SST-2003\\
        \includegraphics[width=\textwidth, trim=10 10 10 10, clip]{Figures_FA_RANS/Cf/SST22_cold_07p05.png}
        %\caption{Subplot 1}
    \end{subfigure}
    \begin{subfigure}[b]{0.33\textwidth}
    	h)\\
        \includegraphics[width=\textwidth, trim=10 10 10 10, clip]{Figures_FA_RANS/Cf/SST22_cold_17p05.png}
        %\caption{Subplot 2}
    \end{subfigure}
    \begin{subfigure}[b]{0.33\textwidth}
    	i)\\
        \includegraphics[width=\textwidth, trim=10 10 10 10, clip]{Figures_FA_RANS/Cf/SST22_cold_19p57.png}
        %\caption{Subplot 3}
    \end{subfigure}
    \caption{Skin-friction coefficient contour plots for SST Turbulence models}
    \label{fig:APP_SST_Cf}
\end{figure}

\begin{figure}
    \centering
%%%%%%%%%%%%%%%%%%%%%%%%%%%%%%%%%%%%%%%%%%%%%%%%%%%%%%%%%%%%%%%%%%%%%%%%%%%%%%%%%%%%%%%%%%%%%%%%%%
    \begin{subfigure}[b]{0.33\textwidth}
    	j)  Menter SST-sust\\
        \includegraphics[width=\textwidth, trim=10 10 10 10, clip]{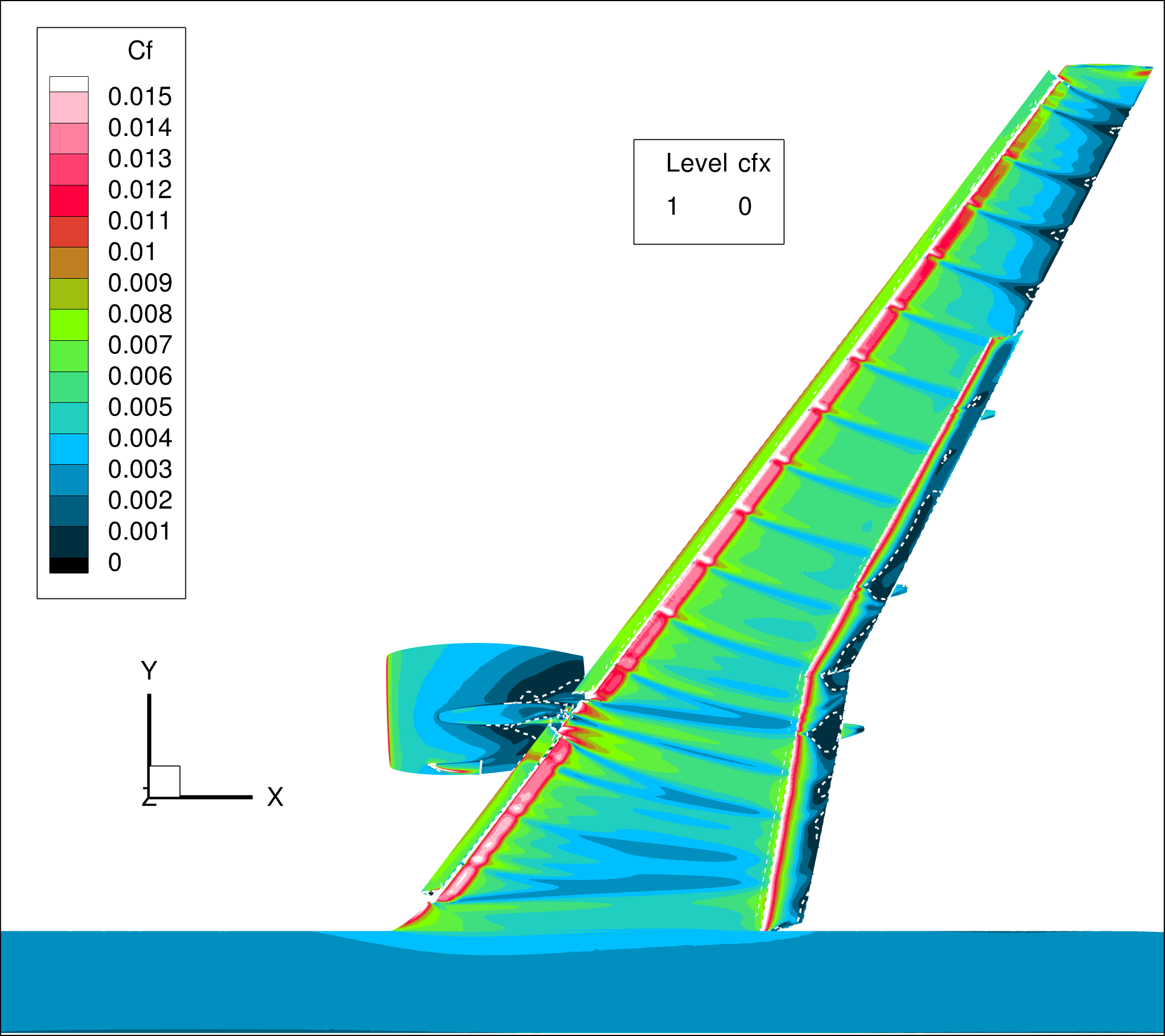}
        %\caption{Subplot 1}
    \end{subfigure}
    \begin{subfigure}[b]{0.33\textwidth}
    	k)\\
        \includegraphics[width=\textwidth, trim=10 10 10 10, clip]{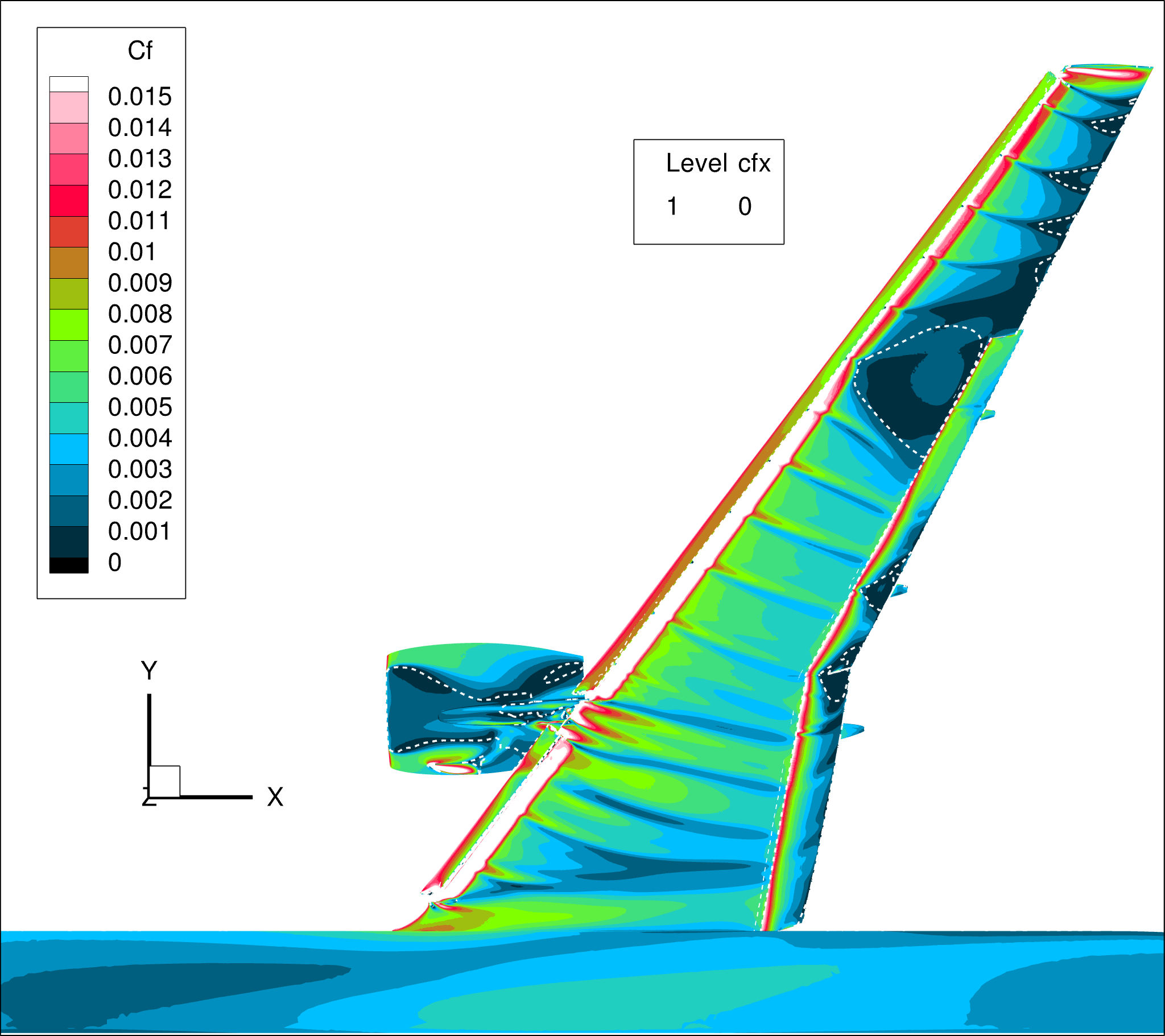}
        %\caption{Subplot 2}
    \end{subfigure}
    \begin{subfigure}[b]{0.33\textwidth}
    	l)\\
        \includegraphics[width=\textwidth, trim=10 10 10 10, clip]{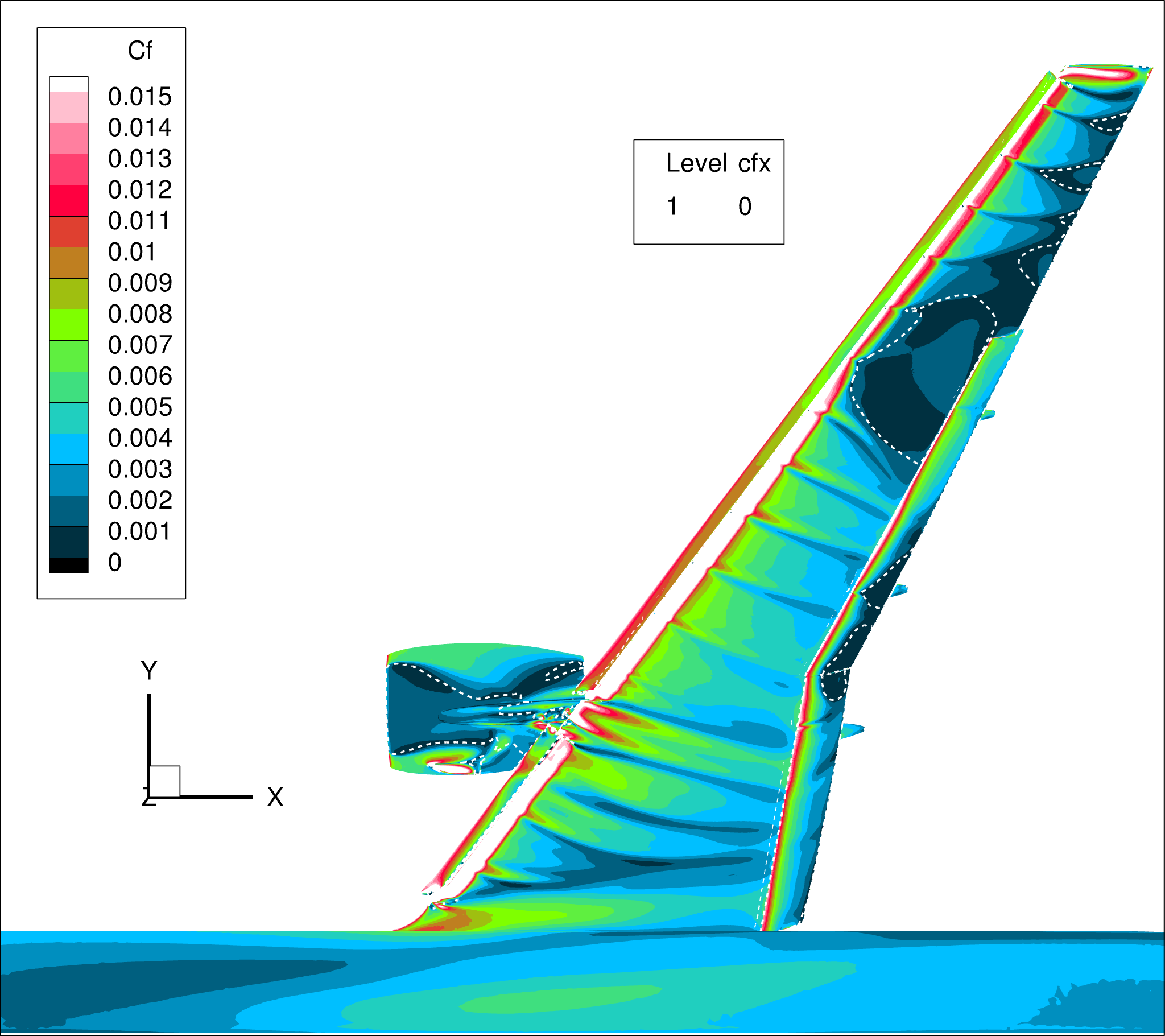}
        %\caption{Subplot 3}
    \end{subfigure}\\
    \vspace{0.1cm} % adjust vertical spacing between rows of subplots
%%%%%%%%%%%%%%%%%%%%%%%%%%%%%%%%%%%%%%%%%%%%%%%%%%%%%%%%%%%%%%%%%%%%%%%%%%%%%%%%%%%%%%%%%%%%%%%%%%
    \begin{subfigure}[b]{0.33\textwidth}
    	j)  Menter SST-Vsust\\
        \includegraphics[width=\textwidth, trim=10 10 10 10, clip]{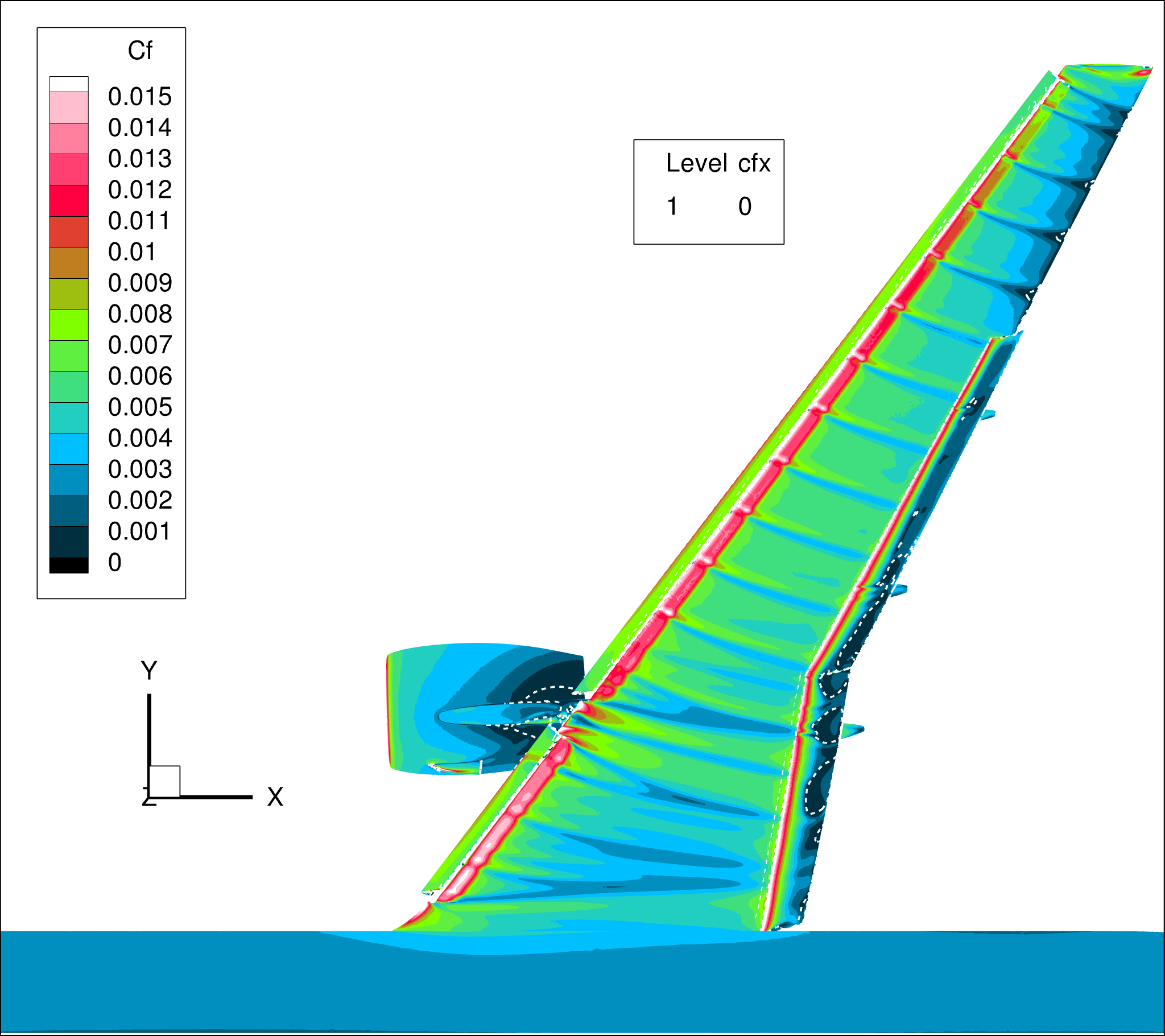}
        %\caption{Subplot 1}
    \end{subfigure}
    \begin{subfigure}[b]{0.33\textwidth}
    	k)\\
        \includegraphics[width=\textwidth, trim=10 10 10 10, clip]{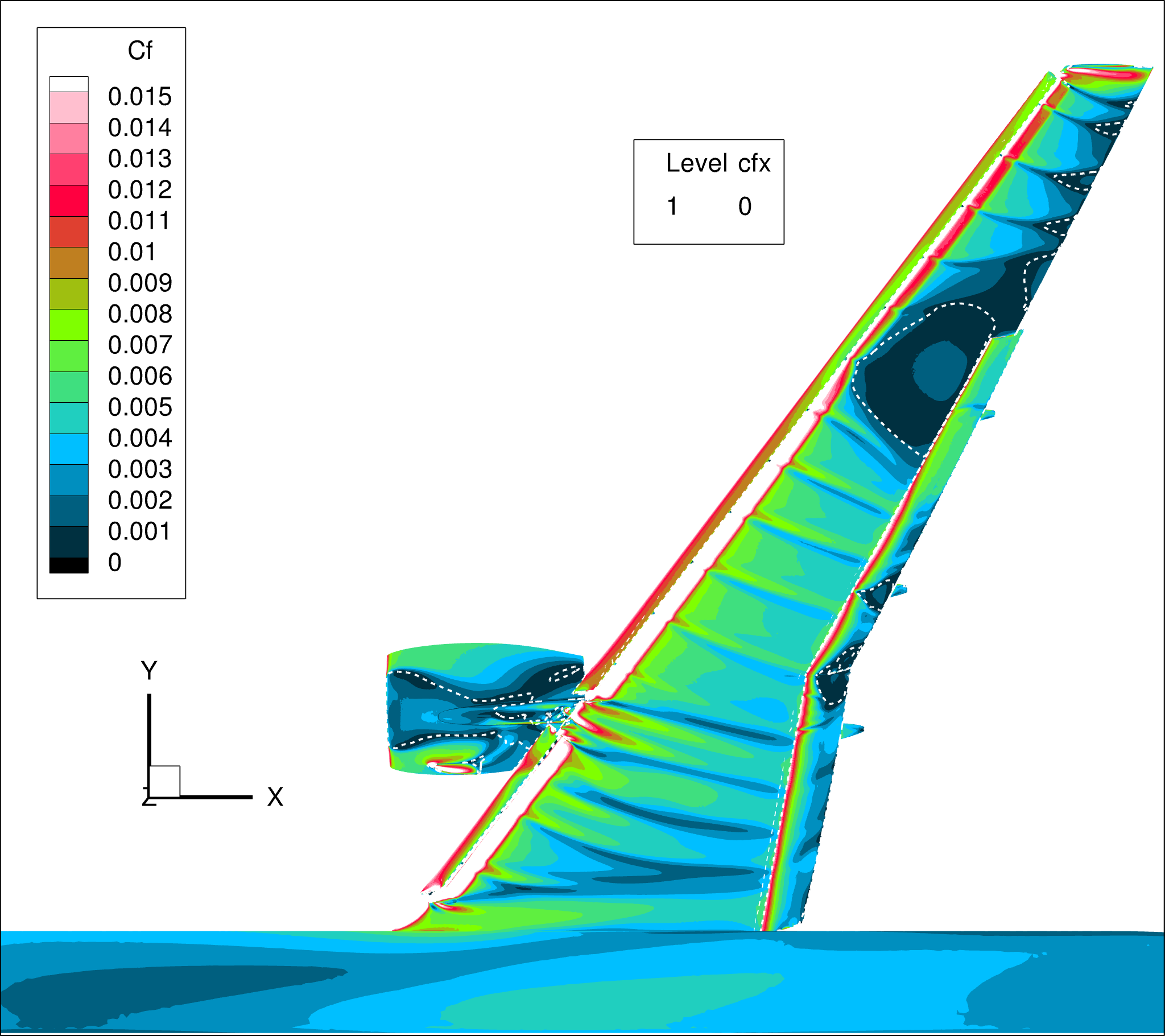}
        %\caption{Subplot 2}
    \end{subfigure}
    \begin{subfigure}[b]{0.33\textwidth}
    	l)\\
        \includegraphics[width=\textwidth, trim=10 10 10 10, clip]{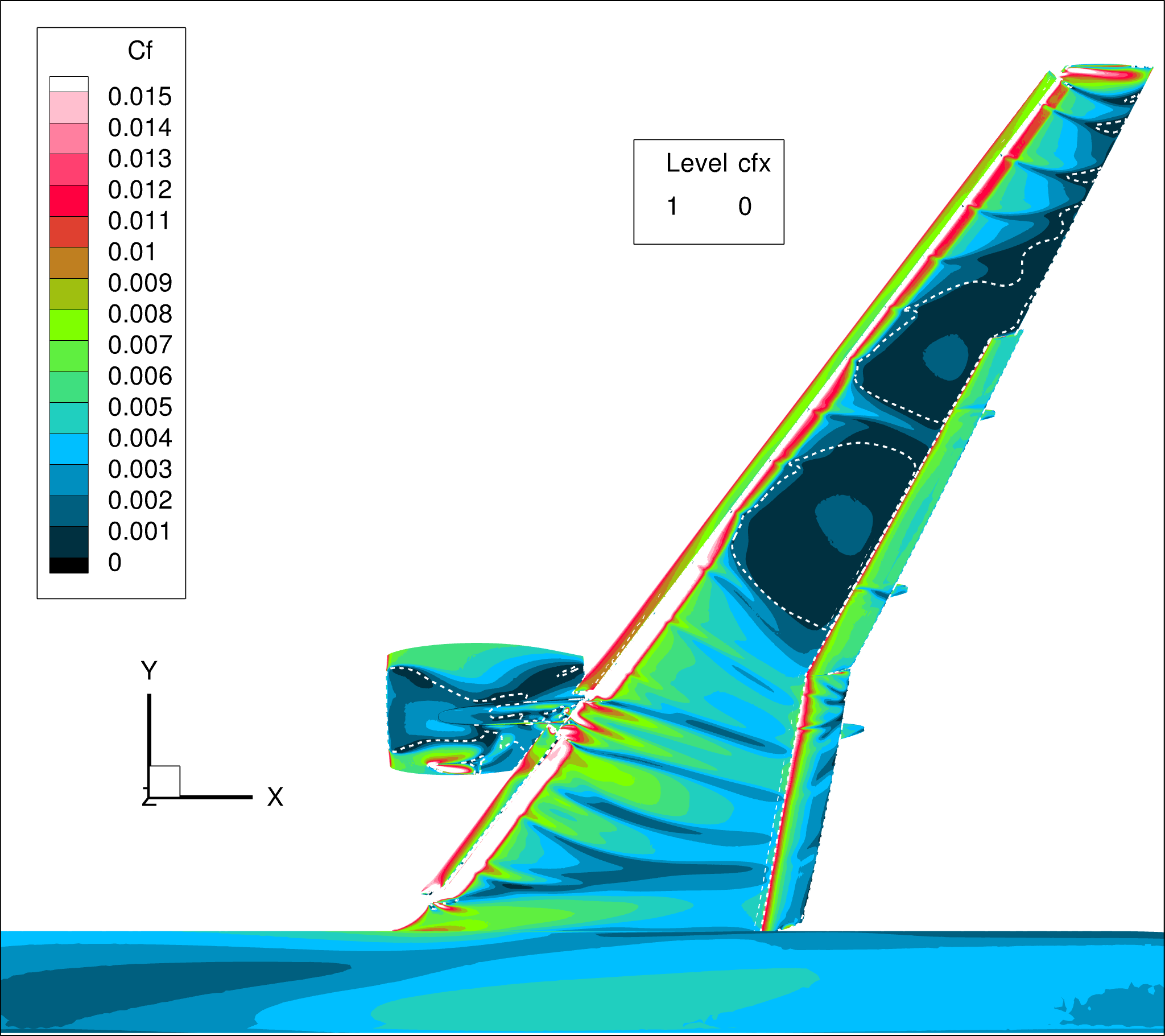}
        %\caption{Subplot 3}
    \end{subfigure}\\
    \vspace{0.1cm} % adjust vertical spacing between rows of subplots
%%%%%%%%%%%%%%%%%%%%%%%%%%%%%%%%%%%%%%%%%%%%%%%%%%%%%%%%%%%%%%%%%%%%%%%%%%%%%%%%%%%%%%%%%%%%%%%%%%
    \begin{subfigure}[b]{0.33\textwidth}
    	j)  Menter SST-2003sust\\
        \includegraphics[width=\textwidth, trim=10 10 10 10, clip]{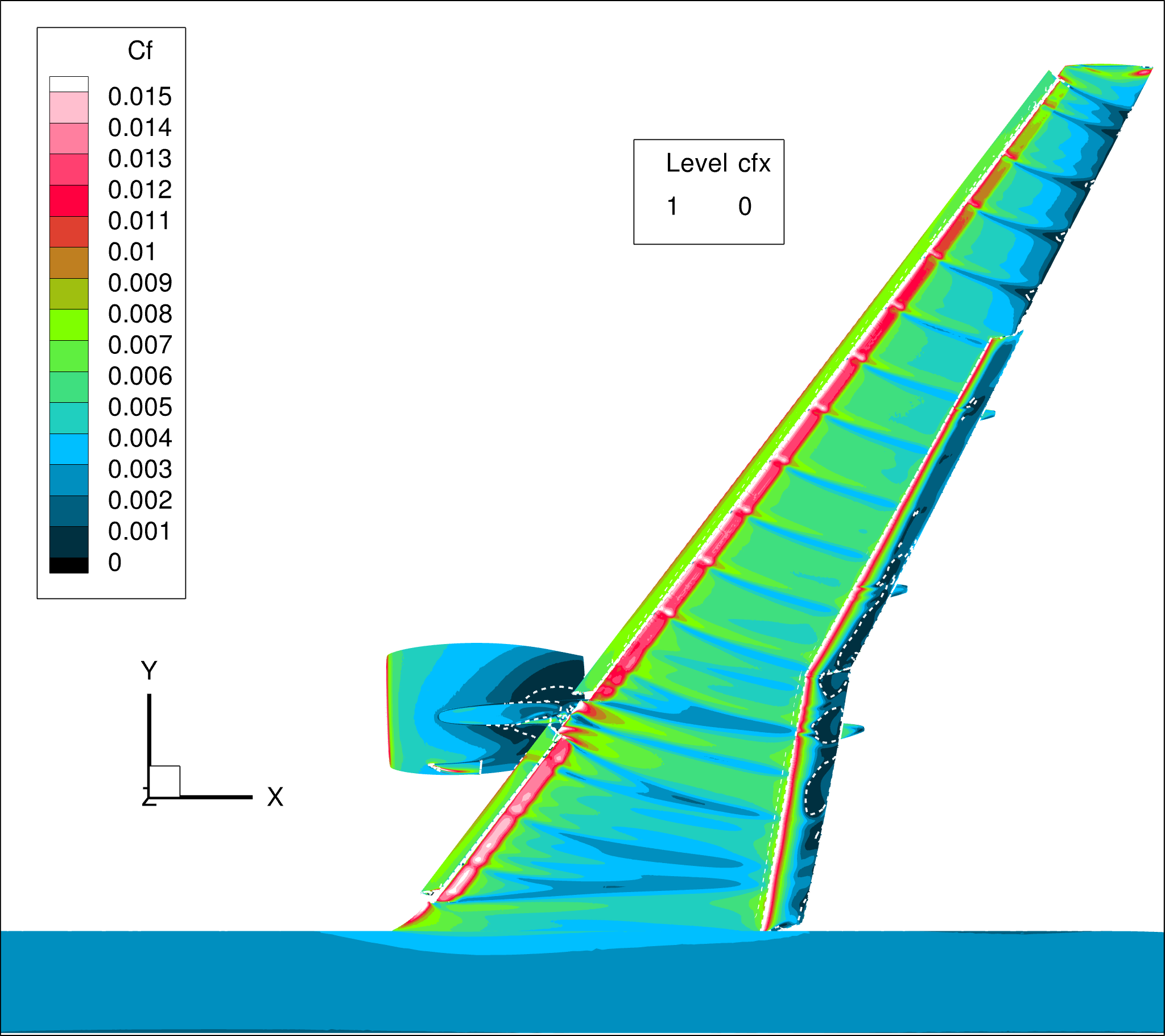}
        %\caption{Subplot 1}
    \end{subfigure}
    \begin{subfigure}[b]{0.33\textwidth}
    	k)\\
        \includegraphics[width=\textwidth, trim=10 10 10 10, clip]{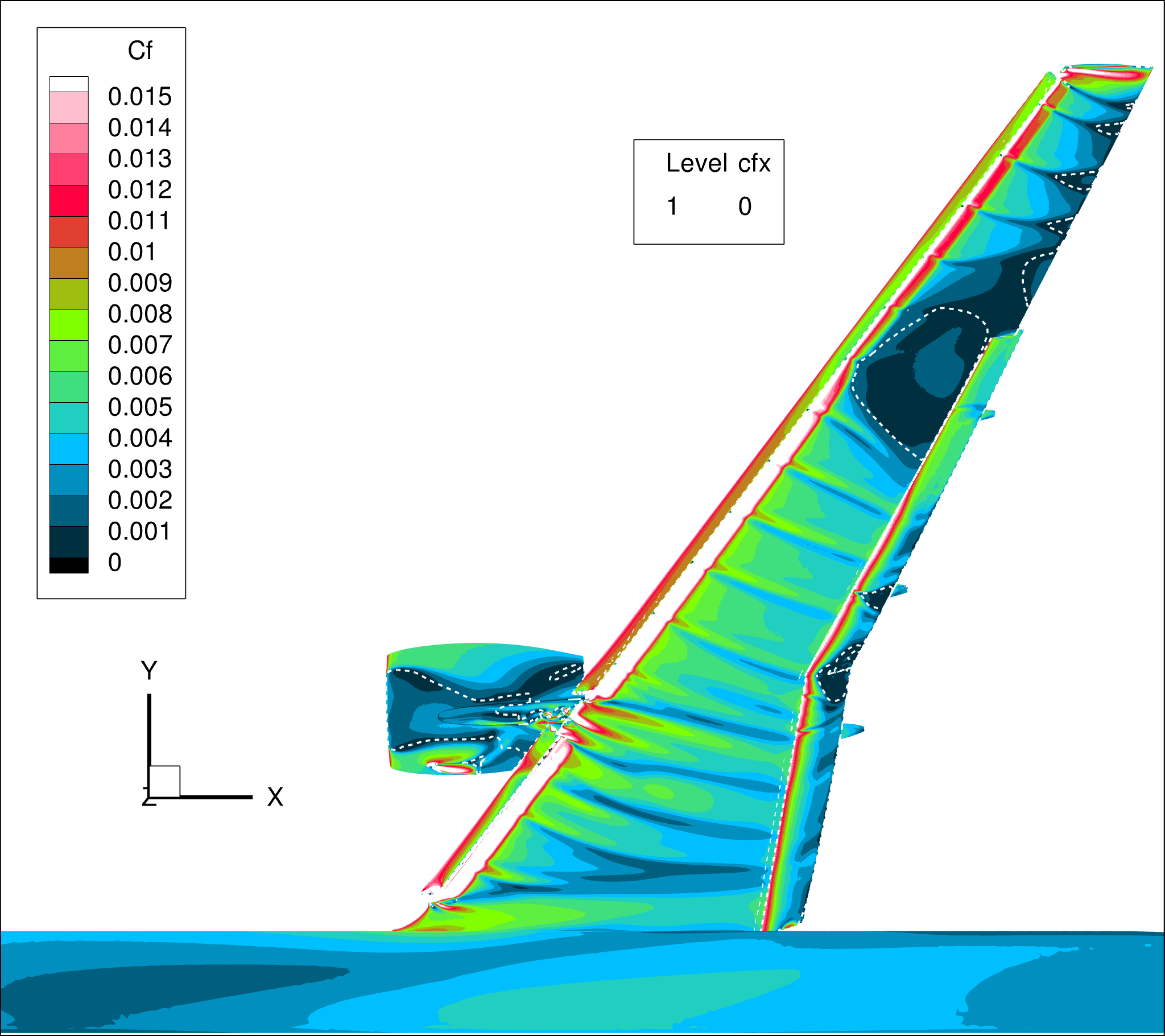}
        %\caption{Subplot 2}
    \end{subfigure}
    \begin{subfigure}[b]{0.33\textwidth}
    	l)\\
        \includegraphics[width=\textwidth, trim=10 10 10 10, clip]{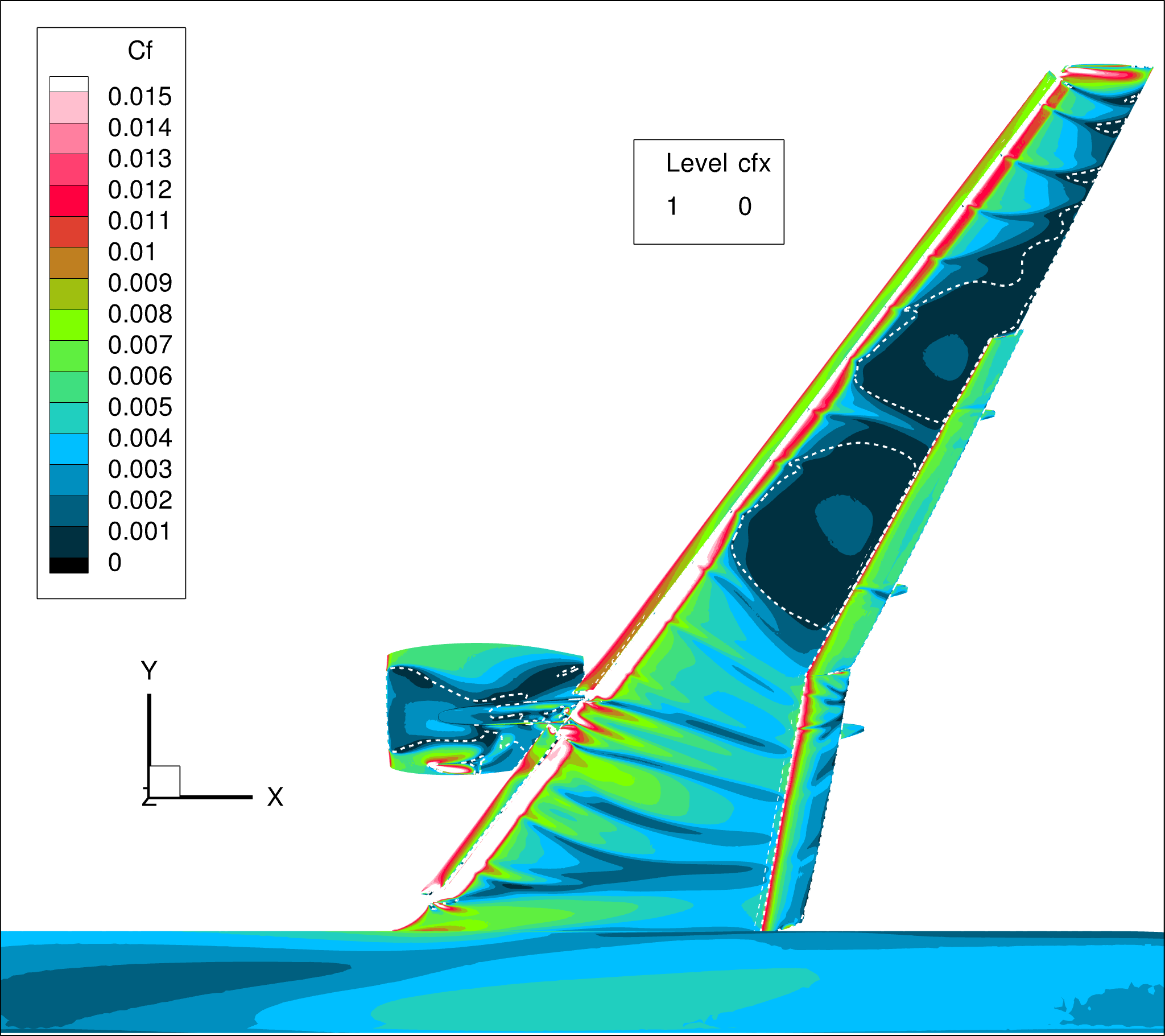}
        %\caption{Subplot 3}
    \end{subfigure}\\
    \vspace{0.1cm} % adjust vertical spacing between rows of subplots
%%%%%%%%%%%%%%%%%%%%%%%%%%%%%%%%%%%%%%%%%%%%%%%%%%%%%%%%%%%%%%%%%%%%%%%%%%%%%%%%%%%%%%%%%%%%%%%%%%
    \caption{Continuation of skin-friction coefficient contour plots for SST Turbulence models}
    \label{fig:APP_SST_Cf_cont}
\end{figure}

\subsection{Steady RANS Grid-refinement Study}\label{sec:AppGridRef}
\begin{figure}
    \centering
    \begin{subfigure}[b]{0.48\textwidth}
        a) \\
        \includegraphics[width=\textwidth]{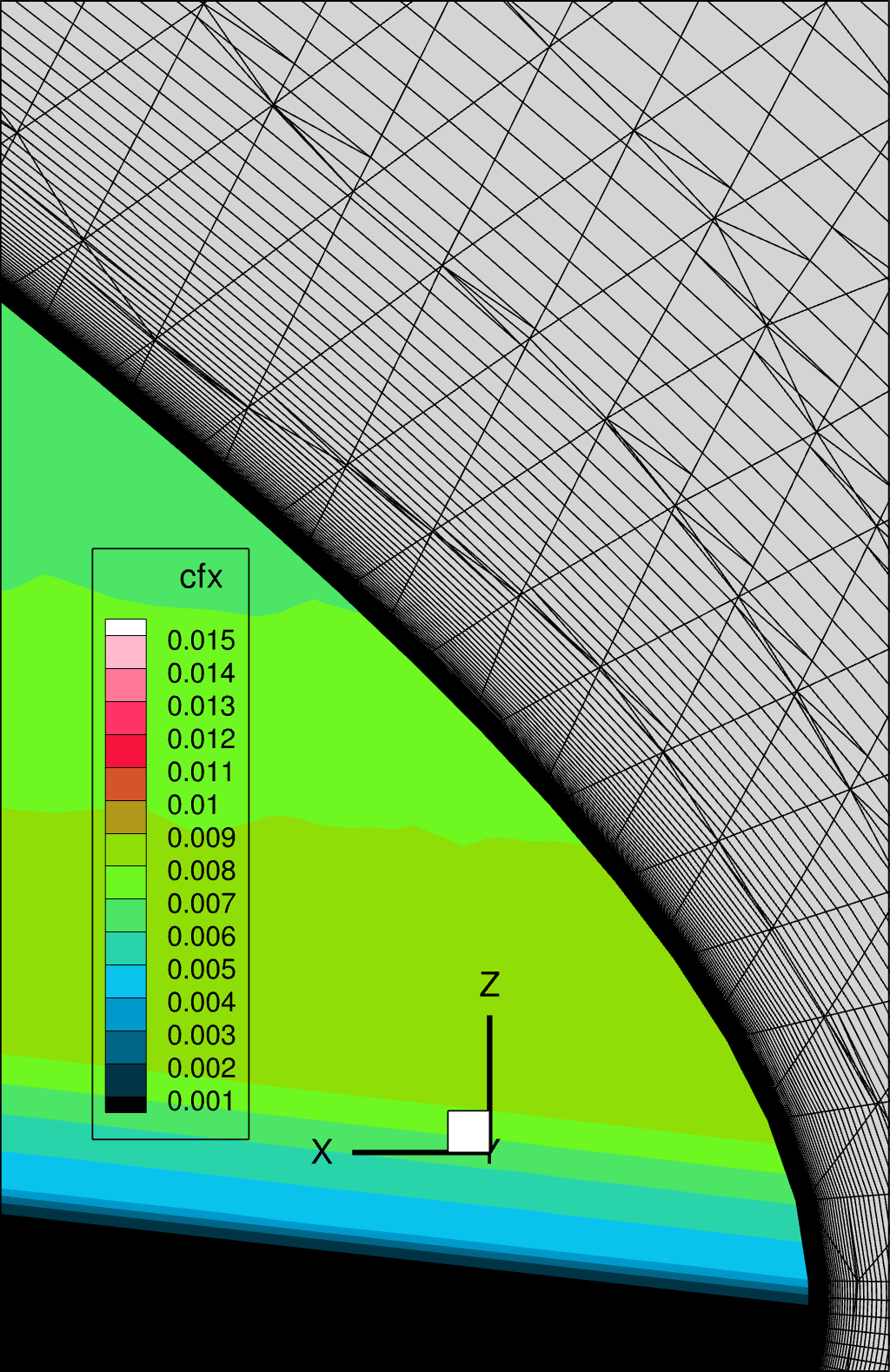}
        %\caption{Subplot 1}
    \end{subfigure}
    \begin{subfigure}[b]{0.48\textwidth}
        b) \\
        \includegraphics[width=\textwidth]{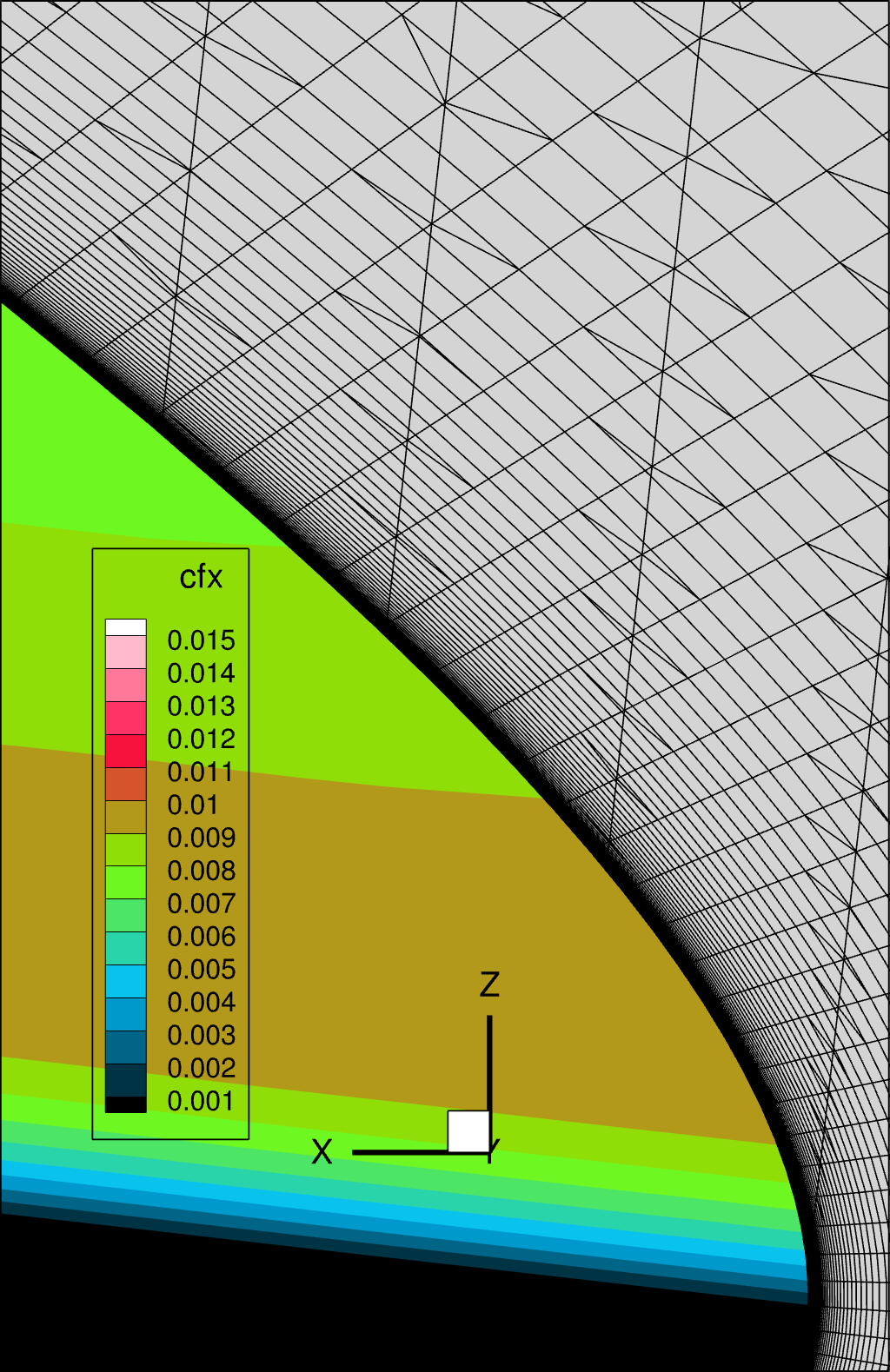}
        %\caption{Subplot 2}
    \end{subfigure}
    \caption{Close-up of (a) refined RANS and (b) DDES grid at a slice indicated by the black line in Fig~\ref{fig:FA_RANS_GridStudy}.}\label{fig:FA_RANS_GridStudy_CU1}
\end{figure}

\begin{figure}
    \centering
    \begin{subfigure}[b]{0.48\textwidth}
        a) \\
        \includegraphics[width=\textwidth]{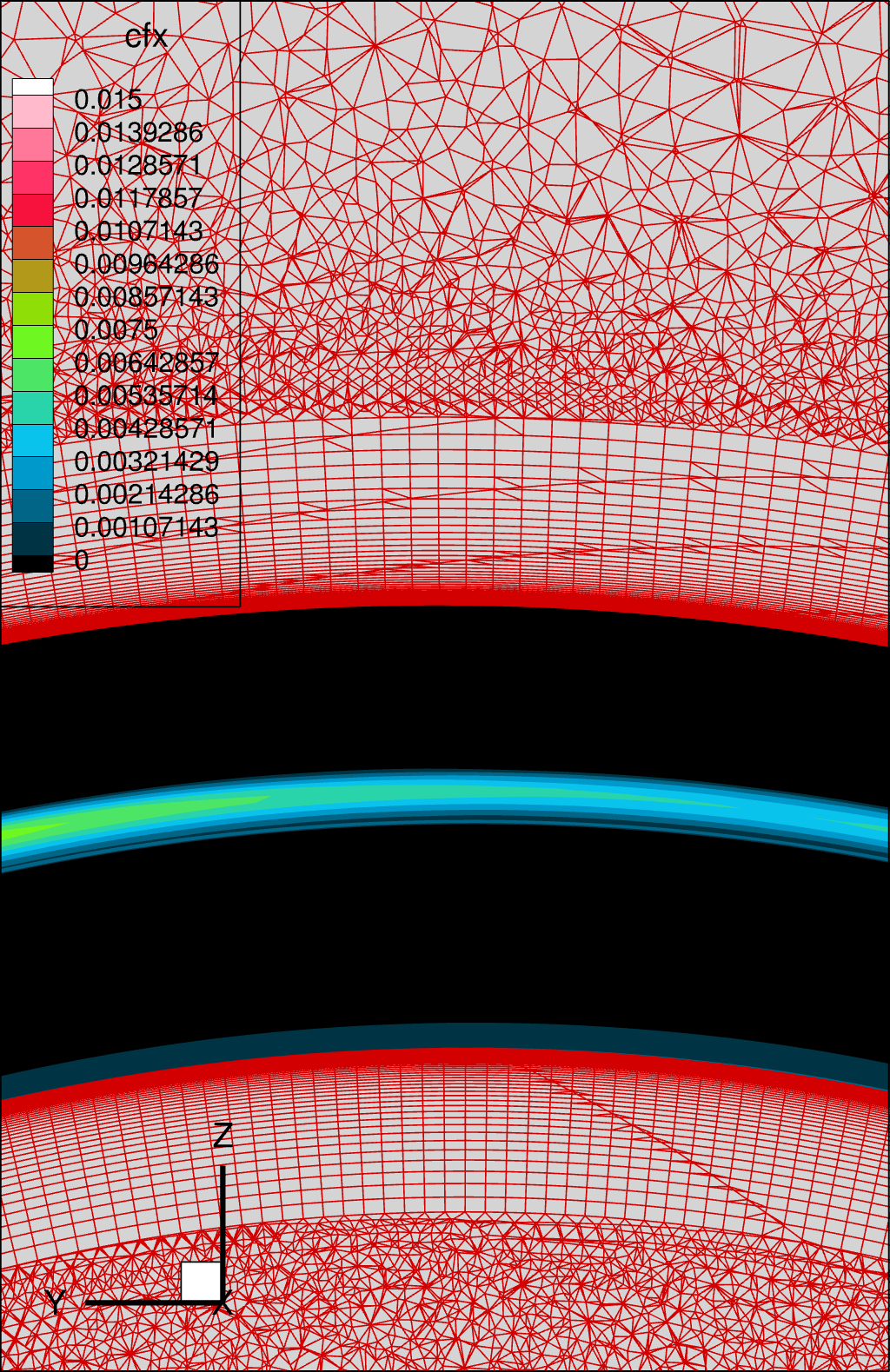}
        %\caption{Subplot 1}
    \end{subfigure}
    \begin{subfigure}[b]{0.48\textwidth}
        b) \\
        \includegraphics[width=\textwidth]{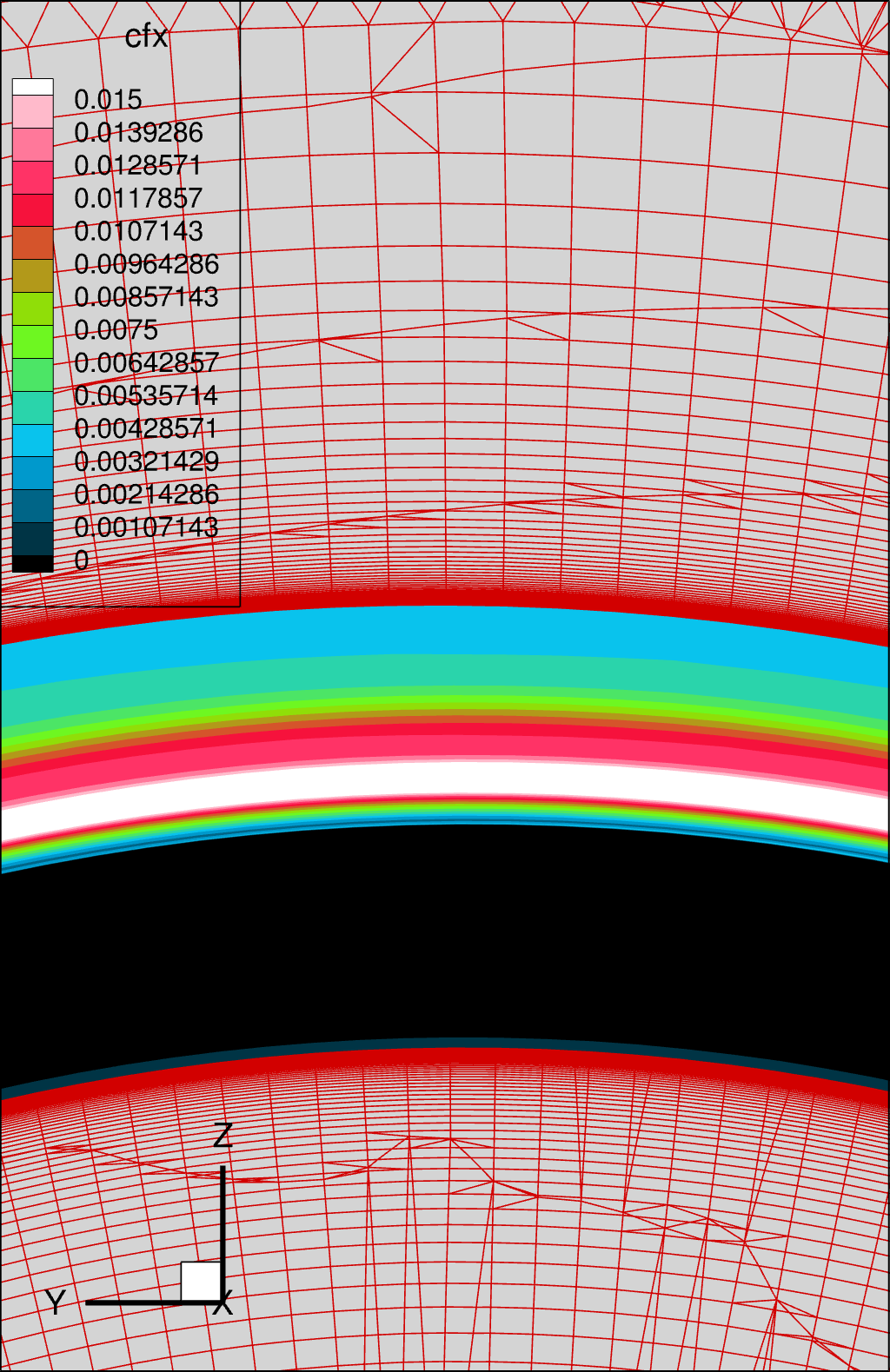}
        %\caption{Subplot 2}
    \end{subfigure}
    \caption{Close-up of (a) standard RANS and (b) DDES grid at a slice indicated by the red line in Fig~\ref{fig:FA_RANS_GridStudy}.}\label{fig:FA_RANS_GridStudy_CU2}
\end{figure}

Figures \ref{fig:FA_RANS_GridStudy_CU1} and \ref{fig:FA_RANS_GridStudy_CU2} show close-ups of the grids for cross-sections respectively indicated by black and red lines in Fig~\ref{fig:FA_RANS_GridStudy}. Fig.~\ref{fig:FA_RANS_GridStudy_CU1} compares refined RANS and DDES grids near the leading edge of the slat in the wing outboard region. Besides different grid topologies, main differences are observed in the streamwise resolution. 
Fig.~\ref{fig:FA_RANS_GridStudy_CU2} shows comparison between standard RANS and DDES grids around the nacelle, where differences are mainly observed in the circumferential resultion.

\section*{Acknowledgments} \label{sec:acknowledgements}
The authors would like to acknowledge Dr. Yasushi Ito (JAXA) for providing grids that were used for RANS simulations. We would also like to thank Dr. Yoimi Kojima (JAXA) and Dr. Kenji Hayashi (Ryoyu Systems) for their great technical support and fruitful discussions. Simulations were carried out on JAXA's high-performance computing facility JSS3. We would like to thank JSS3 for providing resources and technical support.

\bibliography{sample}

\end{document}